\documentclass{book}
\usepackage[dvips]{graphics,color}
\usepackage{bbm,amsmath,amssymb,amsbsy,graphicx,subfigure,psfig,color,amsfonts,appendix,unisaphd}
\begin{document}

\frontmatter

\title{Continuous Variable Teleportation with Non-Gaussian Resources\\in the Characteristic Function Representation}
\author{Lorenzo Albano Far\'{\i}as}
\dept{Universit\`{a} degli Studi di Salerno\\
Dipartimento di Fisica "E.R. Caianiello" } \supervisors{Prof.
Fabrizio Illuminati} \submitdate{November, 2007}
 \phdy{VI Ciclo II Serie (2004-2007)}
\coord{Prof. G. Vilasi}

\maketitle

\mbox{} \vskip13truecm

{\centering \Large \em Faber est suae quisque fortunae.}

%%\clearpage{\pagestyle{empty}\cleardoublepage}

\tableofcontents
\listoffigures
%%\clearpage{\pagestyle{empty}\cleardoublepage}

\chapter*{Acknowledgments}
\addcontentsline{toc}{chapter}{Acknowledgments}

There is only one manner in which I would be able to do justice to everyone involved
in this thesis. Namely, to thank everyone with which I have come into close contact
during these three years.

Of course, my family and friends everywhere; from before I came to Italy, and those I
met in Italy deserve all the thanks that I am able to give. My relatives in Italy who
have done everything I could ask and more when I first came to Italy, I thank from
the heart. May I become a perfect host to you  soon.

The Ciancio and Ruggiero family, I would like to, more than thank, recognize as my
other family here in Italy. Little did I know that I would find myself with a
grandmother and with young brothers and sisters after my grandparents left us and my
brother and sister grew up. I would like to thank further Carmen Ciancio, Raffaele
Ruggiero and Rosa Ruggiero for everything from legal advice to providing me with much
needed assistance in almost every matter.

Carolina, you have shown me that \emph{wonders} are not performed through fraud and
expedient, the impossible conjuring of the non existent or the negation of reality.
But through the traumatic and honest process of upsetting the manner and expectations
we adopt for seeing; of \emph{forcing upon us} new and surprising points of view and
adding them to a wiser perception of reality. Reality is too amazing and too
magnificent to need lies or self deception; too pervasive to allow a lasting place to
the superfluous. For this among many other things, I grant you my love.

My colleagues, PhD students, who constituted my world for these three years, deserve
my most heartfelt thanks, both for being intellectually stimulating and for being
true friends. Specially, Arturo, Francesco, Mauro, "Lupo" and Luca.

I acknowledge and thank my advisors and collaborators Profs. Fabrizio Illuminati and
Silvio de Siena, and Dr. Fabio dell'Anno for the contributions that made most of this
work a going concern. To me, Fabio, you deserve credit as an unofficial advisor. I
would also like to thank Dr. Gerardo Adesso for very fruitful discussions, and for
the huge "bit of assistance" he gave me when I did not know \emph{where to go from
here}.

Acknowledgement is due to the Ministero degli Affari Esteri of Italy, for the
support, financial and otherwise, extended to the author for the duration of his
studies in Italy. I would like to thank specifically Mr. Americo Marrazza and Mrs.
Martina Vardabasso for their continual support and concern for my welfare in these
three years.

\clearpage{\pagestyle{empty}\cleardoublepage}

\chapter*{Introduction}
\addcontentsline{toc}{chapter}{Introduction} \label{INTRO} \markboth{Introduction}{}

Recent theoretical and experimental efforts in quantum optics and quantum information
have been focused on the engineering of highly nonclassical, non-Gaussian states of
the radiation field~\cite{PhysRep}, in order to achieve either enhanced properties of
entanglement or other desirable nonclassical
features~\cite{PhysRep,KimBS,KitagawaPhotsub,DodonovDisplnumb,VanLoock}. It has been
shown that at fixed covariance matrix, some of these properties, including
entanglement and distillable secret key rate, are minimized by Gaussian
states~\cite{ExtremalGaussian}. In the last two decades increasingly sophisticated
schemes for the generation of non-Gaussian states have been proposed, based on
delocalized photon addition or
subtraction~\cite{DeGauss0,DeGauss1,DeGauss2,DeGauss3,DeGauss4}; or on strong
cross-Kerr interactions~\cite{CxKerrKorolkova}. Some of the photon addition and
subtraction schemes have been experimentally implemented to engineer non-Gaussian
photon-added and photon-subtracted states starting from Gaussian coherent or squeezed
inputs~\cite{ZavattaScience,ExpdeGauss1,ExpdeGauss2}.

Through the photon subtraction of a single delocalized photon from a two-mode
entangled initially Gaussian state, it has been possible to realize an state of
enhanced entanglement with negative two-mode Wigner function~\cite{Grangier}.
Remarkably, the photon-addition/subtraction operations, performed on thermal light
fields have led to the demonstration of the commutation relation rules for quadrature
operators; one of the constitutive relations of quantum
mechanics~\cite{BelliniProbing}. Moreover, very recently, a protocol has been
experimentally realized that allows the generation of arbitrarily large squeezed
Schr\"odinger cat states, using homodyne detection and photon number states as
resources~\cite{GrangierCats}. This class of optical cat states is of particular
importance because it is strongly resilient against decoherence~\cite{SerafiniCats}.

Progresses in the theoretical characterization and the experimental production of
non-Gaussian states are being paralleled by the increasing attention on the role and
uses of non-Gaussian entangled resources in quantum information and quantum
computation with continuous-variable systems~\cite{LloydBraunstein}. Concerning
quantum teleportation with continuous variables (\textbf{CV}), the success
probability of teleportation can be greatly increased by using entangled non-Gaussian
resources~\cite{KitagawaPhotsub,Opatrny,Cochrane,Olivares,TelepNonGauss}. In
refs.~\cite{KitagawaPhotsub,Opatrny,Cochrane,Olivares} conditional measurements,
inducing ''degaussification'' through photon-subtraction, are exploited to improve
the efficiency of \textbf{CV} teleportation protocols. Most of these investigations
have used the \emph{transfer
operator}~\cite{NielsenCaves,UnivTeleportTwist,HoffmanTransfer} formalism and Fock
basis representations of the degaussified resources rather than phase-space
representations as used in the original \textbf{CV} protocol~\cite{LloydBraunstein}.
Phase-space and conjugate phase-space representations of operators constitute an
unifying language for the description of quantum optical states and processes;
including the composition of said processes into protocols such as \textbf{CV}
quantum teleportation.

Moreover, non-Gaussian cloning of coherent states has been shown to be optimal with
respect to the single-clone fidelity~\cite{Cerf}. Determining the performance of
non-Gaussian entangled resources  in \textbf{CV} quantum communication protocols can
prove to be useful in a number of concrete applications ranging from hybrid quantum
computation~\cite{Hybrid} to cat-state logic~\cite{Ralph} and in all quantum
computation schemes based on communication that integrate together qubit degrees of
freedom for computation with quantum continuous variables for communication and
interaction~\cite{Communication}.

In the same fashion, but following a more general approach, in
ref.~\cite{TelepNonGauss} the implementation of simultaneous phase matched
multiphoton processes and conditional measurements are used to introduce a general
class of two-mode non-Gaussian entangled states, in the form of squeezed Bell-like
states endowed with a free parameter. The optimization on this free parameter allows
a remarkable increase of the teleportation fidelity for various classes of input
states.

In the present work, we propose a general formalism for the study of Quantum
Teleportation in a \textbf{CV} setting, based on the Wigner's characteristic function
description as a conjugate phase-space representation and on the Weyl's
correspondence~\cite{Weyl} of unitary evolutions and measurements to coordinate
transformations and partial traces over the characteristic function. We show that the
entire quantum teleportation protocol as formulated in~\cite{BraunsteinKimble} for
any combination of entangled resource state and of input state can be represented by
adequate transformations and integrations of the characteristic function of the joint
input-resource physical system on which the protocol is performed. We formulate a
number of complications to the original protocol, intended to simulate imperfections
in the homodyne detection apparatus and the presence of environmental "noise" in the
preparation setup for the resource state~\cite{RealHomodLeonhardt,Marianthermal}.

With the theoretical tools above described, we investigate systematically the
performance of different classes of entangled two-mode non-Gaussian states used as
resources for continuous-variable quantum teleportation. In our approach, the
entangled resources are taken to be non-Gaussian {\em ab initio}, and their
properties are characterized by the interplay between \textbf{CV} squeezing and
discrete, single-photon pumping. Our first aim is to determine the actual properties
of non-Gaussian resources that are needed to assure improved performance compared to
the Gaussian case. At the same time, we carry out a comparative analysis between the
different non-Gaussian cases in order to single out those properties that are most
relevant to successful teleportation. Finally, we wish to understand the role of
adjustable free parameters other than squeezing, in order to \emph{sculpture}
resources to achieve optimized performances within the set of non-Gaussian resources.
With this objective in mind, we propose the squeezed Bell-like
states~\cite{TelepNonGauss} with a single, free superposition parameter which
determines non-Gaussianity. The squeezed Bell-like states are formulated to include
as special cases all the other Gaussian and non-Gaussian resources evaluated in this
work. Then, we optimize the teleportation fidelity using squeezed Bell-like resources
with respect to the superposition parameter. We show that maximal non-Gaussian
improvement of teleportation success depends on the nontrivial relations between
enhanced entanglement, suitably measured level of {\em non-Gaussianity}, and the
presence of a proper Gaussian squeezed vacuum contribution in the non-Gaussian
resources for large values of the squeezing parameter (squeezed vacuum affinity).

The squeezed Bell-like state can be parameterized as a first-order (or two-photon)
truncation of the squeezed vacuum states. A generalization of the squeezed Bell-like
state can be easily constructed by allowing four-photon terms in the (prior to
squeezing) superposition making up the Bell state, thus constructing a more general
superposition of Fock states~\cite{TelNoisyCats}. Further optimization of
teleportation fidelity is made possible by the addition of a new dimension to the
Hilbert space to be explored and the addition of a new free parameter to the set of
parameters over which optimization is carried out. However, optimization reduces
these squeezed superpositions of Fock states to second-order truncations on squeezed
vacuum states. An avenue of exploration of non-Gaussian resources other than
superpositions of few-photon Fock states lies in the formulation of resources
combining two-mode squeezing and the entangled superpositions present in two-mode
Schr\"{o}dinger Cat states. Such two-mode squeezed cat-like states can likewise be
used in \textbf{CV} teleportation and optimized with respect to the free parameter
given by the phase-space distance between the terms of the superposition.

Furthermore, we investigate the effects of the presence of thermal noise on the
performance of two-mode non-Gaussian states used as resources for continuous-variable
quantum teleportation\cite{TelepNoise,TelNoisyCats}. We will consider the
non-Gaussian resources obtained by \emph{superimposing} the classes of squeezed
Bell-like states and squeezed cat-like states over two-mode thermal
states~\cite{Glaubercoher}. Due to the thermal contribution, the state so obtained is
mixed and its correlation properties are modified and deteriorated by the presence of
thermal photons. We limit the discussion to the situation of ideal teleportation
protocol, i.e. ideal Bell measurements and decoherence-free propagation through space
of radiation states. Detailed analysis in the instance of the most general realistic
situation, including various sources of noise, will be discussed elsewhere.

The thesis is organized as follows. In chapter~\ref{Back} we discuss the basics of
\textbf{CV} systems (see section~\ref{Back:QOpt}), i.e. quantum optics including,
most importantly the groups of operations such as displacement and squeezing and
non-unitary "operations" such as homodyne measurement (see section~\ref{Back:Toys}).
We introduce correspondence principles based on the aforementioned groups of
operators and phase-space representations of states of the radiation field, as well
as the characteristic functions of some Gaussian states (see
section~\ref{Back:PhaseSp}). Finally, we discuss quantum teleportation as an
universal procedure and teleportation fidelity as a measure of teleportation success
with a view to the formulation of quantum teleportation in a characteristic
functions' language (see section~\ref{Back:Qtele}).

In chapter~\ref{Form} we derive, using the Wigner's function (see
section~\ref{Form:TeleportWigner}) and Wigner's characteristic function language (see
section~\ref{Form:TeleportIdeal}), the expression for the output state of \textbf{CV}
teleportation for the ideal setup and for several interesting modifications of the
teleportation protocol. Such as the performance of the "homodyne" projective
measurement over a mixed state (see section~\ref{Form:TelepMix}); a resource state
prepared and superimposed over thermal vacuum states (see section~\ref{Form:MixRes})
and "realistic" homodyne detection (see section~\ref{Form:TelepNoise}), whereas
fictitious beam-splitters mix the modes to be measured with external fields and cause
a loss of intensity). We analyze and compare the effect of the complications and
modifications introduced in each case. Finally, we analyze the teleportation fidelity
expression we have chosen~\cite{CriteriaCVTelep,CriteriaCVTelep2} for the
teleportation outputs we have derived in the characteristic function formalism(see
section~\ref{Form:Fide}).

In chapter~\ref{NGau} we study the use of some non-Gaussian states as resources for
an ideal \textbf{CV} teleportation protocol. We introduce and describe relevant
instances of two-mode entangled non-Gaussian resources, including squeezed number
states and typical degaussified states currently considered in the literature, such
as photon-added squeezed and photon-subtracted squeezed states (see
section~\ref{NGau:Chzation}). We compare the relative performances of non-Gaussian
and Gaussian resources in the \textbf{CV} teleportation protocol for different
(single-mode) input states, Gaussian and non-Gaussian, including coherent and
squeezed states, number states, photon-added coherent states, and squeezed number
states (see section~\ref{NGau:Degauss}). We introduce the squeezed Bell-like states
as a generalization including all of the former non-Gaussian resources, as well as
the Gaussian two-mode vacuum and squeezed vacuum (twin-beam) as special cases
(section~\ref{NGau:SqueBell}), and consider the optimization of non-Gaussian
performance in \textbf{CV} teleportation with respect to the extra angular parameter
of squeezed Bell-like states, and show that maximal teleportation fidelity is
achieved in every case using a form of squeezed Bell-like resource \emph{tailored to
the input} that differs both from squeezed number and photon-added/subtracted
squeezed states. We identify some properties that determine the maximization of the
teleportation fidelity (see section~\ref{NGau:Compaprop}) using non-Gaussian
resources; finding that optimized non-Gaussian resources are those that come nearest
to the simultaneous maximization of three distinct properties: the content of
entanglement, the amount of (properly quantified) non-Gaussianity, and the degree of
{\em ''vacuum affinity''}, i.e. the maximum, over all values of the squeezing
parameter, of the overlap between a non-Gaussian resource and the Gaussian twin-beam.
Schemes for the experimental production of optimized squeezed Bell-like resources are
proposed and illustrated (see section~\ref{NGau:ExpGene}).

In chapter~\ref{SSFCat} we introduce a higher-order generalization of the squeezed
Bell-like states: squeezed superpositions of Fock (\textbf{SSSF}) states and a new
class of resources, the squeezed cat-like states. We define, study and optimize the
new class of \textbf{SSSF} states (in section~\ref{SSFCat:Trunk}). We show that all
the squeezed Bell-like states and the optimal \textbf{SSSF} states can be regarded as
"truncations" on Gaussian states. Higher order "truncations", by bestowing an extra
dimension to the Hilbert space on which optimization is performed, further improve
fidelity, over the already optimized fidelity of squeezed Bell-like resources. We
introduce the cat-like resources, two-mode squeezed superpositions of coherent states
(see section~\ref{SSFCat:Cat}). We optimize the cat-like states for fidelity of
teleportation: We find them to be non-Gaussian resources with a teleportation
performance inferior to that of the optimized squeezed Bell-like state; but
nevertheless superior to that of the Gaussian states for equal squeezing. In the two
sections of this chapter, we perform an analysis of the entanglement, non-Gaussianity
and Gaussian affinity for all the resources, similar to the analysis of the same
properties performed in chapter~\ref{NGau}.

chapter~\ref{Noisy} refers to the teleportation protocol using the general class of
squeezed Bell-like states (and thus all the non-Gaussian resources introduced in
chapter~\ref{NGau}), together with the cat-like resources (introduced in
section~\ref{SSFCat:Cat}) in the presence of noise; for the teleportation of coherent
state inputs. The resource has been prepared or propagated, i.e. \emph{superimposed}
in a noisy environment made up of thermal states, resulting in a mixed-state
resource~\cite{Glaubercoher}. First, we compare the performance of the mixed squeezed
Bell-like states and mixed squeezed cat-like states when optimized for maximum
fidelity and a similarly mixed Gaussian resource (see section~\ref{Noisy:Telep}).
Thus we study the simplest instances of teleportation using mixed non-Gaussian
resources. We compare the robustness of the entanglement of squeezed Bell-like
states, squeezed cat-like states and two-mode Gaussian states under noisy conditions
(see section~\ref{Noisy:Insepar}). Firstly, considering the violation of a sufficient
inseparability criterion~\cite{InsepShchukin,InsepDellAnno} for mixed Bell-like and
mixed Gaussian states at given levels of noise. Lastly, by considering the
noise-induced arrival of the teleportation fidelity at the \emph{classical
teleportation threshold} for coherent state inputs as a practical criterion for the
disappearance of the entanglement of teleportation when the resource is noisy: for
squeezed Bell-like, squeezed cat-like and Gaussian resource states alike.

In chapter~\ref{CONCLU}, we present the conclusions of our work and discuss the
possibility of extending the characteristic functions' formalism to more general
teleportation setups; of optimizing the \textbf{CV} protocol itself for non-Gaussian
resources and inputs; and of considering other non-Gaussian resources for the
analysis and optimization of teleportation performance.

\clearpage{\pagestyle{empty}\cleardoublepage}

\mainmatter

\chapter{Initiation}
\label{Back}

In this chapter we will go through some basic and previous concepts that are
necessary to the understanding of the main body of this work. We will also establish
some conventions and definitions that will hold for the next chapters.

In section~\ref{Back:QOpt} we review the basic concepts of the Continuous Variables
(\textbf{CV}) representation of quantum states of the radiation field, which are the
subject matter of Quantum Optics. In section~\ref{Back:Toys} we recall linear
transformations on \emph{Continuous Variables} and the procedure of homodyne
detection together with some quantum states associated with these operations. With
the purpose of making clear the correspondence principle between density operators of
quantum states and phase-space (displacement operator) representations of such; and
of establishing a correspondence between the transformations ( ideally associated
with experimental procedures) on density operators and coordinate transformations on
phase-space representations.

In section~\ref{Back:PhaseSp}, we review the Wigner function and the Wigner
characteristic function; phase-space (and conjugate phase-space) representations that
associate an square-integrable operator (particularly the density operators) with an
analytic function of a complex variable that functions as a pseudo-phase-space
coordinate. An special emphasis will be made on the conjugate phase-space
representation given by the Fourier Transforms of phase-space functions; the
characteristic functions corresponding to observables and quantum states.

We explain in section~\ref{Back:Qtele} the basic concepts of entanglement, maximally
entangled state, and \emph{universal} quantum teleportation; for physical systems
with state vectors belonging to an arbitrary Hilbert space. Lastly we analyze,
briefly, the definition of teleportation fidelity we have chosen for this work.

\section{Quantum optics and continuous variables}
\label{Back:QOpt}

The quantized electromagnetic field has a field operator for the photon particle;
which is the electric field (or the magnetic field, depending on the choice of
phase). For a single frequency $\omega_{k}$ and a single polarization component the
electric field reads

\begin{align}
  \hat{E}_{k}(\vec{r},t)&=\,
  \mathcal{E}_{k}\left(\hat{a}_{k}\,e^{i(\vec{k}\cdot\vec{r}-\omega_{k}t)}+\hat{a}_{k}^{\dag}e^{-i(\vec{k}\cdot\vec{r}-\omega_{k}t)}\right)\notag\\
   &=\hat{E}^{+}_{k}(\vec{r},t)+ \hat{E}^{-}_{k}(\vec{r},t) \label{eq:ElecField}
\end{align}

The correlation functions of the field are the mean values of normally
ordered~\footnote{In the sense that $a_{k}^{\dag}$ is always to the left of $a_{k}$}
products of $\hat{E}^{-}_{k}$ and $\hat{E}^{+}_{k}$ for the appropriate times
$t_{1},t_{2},\ldots$ and positions $\vec{r}_{1},\vec{r}_{2},\ldots$.  For example,
the second-order correlation function for the field (frequency $\omega_{k}$) is given
by~\cite{Glaubercoher,QOptWalls}

\begin{equation}\label{eq:FieldIntensity}
\langle \hat{E}^{-}_{k}(\vec{r}',t')\hat{E}^{+}_{k}(\vec{r},t) \rangle
\end{equation}
where, for $\vec{r}'=\vec{r}$ and $t=t'$, we have the field intensity, or average
number of photons with energy $\hbar\omega_{k}$ for position $\vec{r}$ and time $t$.

The Hamiltonian of the radiation field, which for a classical field is given by
\begin{equation}\label{eq:HamClasField}
  H=2^{-1}\int d\vec{r}\:\left(|\vec{E}|^{2}+|\vec{B}|^{2}\right)
\end{equation}
becomes, for the quantized field and in terms of the annihilation and creation
operators;
\begin{equation}\label{eq:HamHarmOscOne}
\hat{H}_{k}=\sum_{k}\hbar\omega_{k}(\hat{a}_{k}^{\dag}\hat{a}_{k}+\frac{1}{2})
\end{equation}
where the sum is over all the light frequencies allowed by the boundary conditions
established beforehand for the quantization of the field.

The creation and annihilation operators for the field are those of a harmonic
oscillator with bosonic excitations. These have constitutive relations given by their
commutators;
\begin{align}
  [\,\hat{a}_{k}\,,\,\hat{a}_{k'}^{\dag}\,]\,=\,&\delta_{k,k'} \notag \\
  [\,\hat{a}_{k}\,,\,\hat{a}_{k'}\,]\,=\,&[\,\hat{a}_{k}^{\dag}\,,\,\hat{a}_{k'}^{\dag}\,]\,=\,0 \label{eq:AnniCreaComm}
\end{align}

We will limit our review in this work to one frequency only; barring the
\emph{rigorous} study of non-linear quantum optical phenomena such as
down-conversion, the generalization to multiple frequencies is straightforward. The
Hamiltonian of eq.~(\ref{eq:HamHarmOscOne}), limited to one frequency $\omega$, is
linear in the \textit{number} operator of the harmonic oscillator
$\hat{n}\equiv\hat{a}^{\dag}\hat{a}$.

The eigenstates of the number operator have a definite number of photons~\cite{BAYM}.
They are called the \textit{number}, or \textit{Fock} states $|n\,\rangle$;
\begin{align}
 \hat{a}\,|m\,\rangle=&\,\sqrt{m} |m\,\rangle \notag \\
 \hat{a}^{\dag}\,|m\,\rangle=&\,\sqrt{m+1} |m\,\rangle \notag\\
 \hat{n}\,|m\,\rangle=&\,m\,|m\,\rangle \notag \\
 \langle\,n\,|\,m\rangle=&\,\delta_{n,m} \label{eq:Fockstate}
\end{align}
for $m=0,1,2,\ldots$

We can define the position and momentum operators of the harmonic quantum oscillator;
treating it like a particle in a quadratic potential, with \emph{unit mass}. Define
the annihilation and creation operators as

\begin{align}
\hat{a}=\frac{1}{\sqrt{2\hbar\omega}}\left(\,\omega\hat{x}_{k}\,+\,i\hat{p}\,\right)\label{eq:Annihi1}\\
\hat{a}^{\dag}=\frac{1}{\sqrt{2\hbar\omega}}\left(\,\omega\hat{x}\,-\,i\hat{p}\,\right)\label{eq:Creat1}\\
\hat{H}=\frac{1}{2}\left(\hat{p}^{2}+\omega^{2}\hat{x}^{2}\right)\label{eq:HamHarmOscTwo}
\end{align}

The position and momentum operators of the harmonic oscillator of \emph{unit mass}
will correspond to the quadrature operators of the radiation field. The eigenvalues
associated with these Hermitian observables are real numbers; the continuous values
of position and momentum, and the reason for the \textit{Continuous Variables} name
given to systems thus describable.

With the commutation relations of eq.~(\ref{eq:AnniCreaComm}) we can calculate the
commutation relations for the quadrature operators,
\begin{equation}\label{eq:XPComm}
  [\,\hat{x}\,,\,\hat{p}\,]=i\,\hbar\delta
\end{equation}
that define, in turn, the Heisenberg uncertainty relation obeyed by these operators.
Given $\Delta\hat{x}\equiv \hat{x}-\langle\,\hat{x}\,\rangle$ and
$\Delta\hat{p}_{k}\equiv \hat{p}-\langle\,\hat{p}\,\rangle$ we have;
\begin{equation}
\langle\,\Delta \hat{x}^{2}\,\rangle\;\langle\,\Delta
\hat{p}^{2}\,\rangle\,\geq\,\frac{1}{4}\,|\langle[\,\hat{x}\,,\,\hat{p}\,]\rangle|^{2}\,=\,\frac{\hbar^{2}}{4}
\label{eq:HeisUncRel}
\end{equation}

For simplicity, we choose the system of units whereby $\hbar=1/2$ and $\omega=1$. In
this way the quadratures become \emph{dimensionless};

\begin{align}
\hat{x}\,=&\,\frac{1}{2}\left(\hat{a}+\hat{a}^{\dag}\right)=\,\mathrm{Re}[\hat{a}]\notag\\
\hat{p}\,=&\,\frac{1}{2i}\left(\hat{a}-\hat{a}^{\dag}\right)=\,\mathrm{Im}[\hat{a}]
 \label{eq:XPDimless}
\end{align}

In this manner; $\hat{a}=\hat{x}+i{p}$ and the physical quantities associated with
observables $\hat{x}$ and ${p}$ can be made to correspond with a complex phase-space
"coordinate" $\alpha$, for the representation of \textbf{CV} states and
operators~\cite{Weyl}. The Hermitian, dimensionless position and momentum quadratures
would correspond, respectively, to the real and imaginary parts of the coordinate
$\alpha\equiv\,x+\,ip$. For example, the average values of $\langle\hat{x}\rangle$
and $\langle\hat{p}\rangle$ correspond to the average phase-space "coordinate" of the
quantum state.

However, the Heisenberg uncertainty relation (eq.~(\ref{eq:HeisUncRel})) precludes
the joint knowledge of the values of momentum and position; $\hat{x}$ and $\hat{p}$
with arbitrary precision for a given quantum state. This makes it impossible to
define $\alpha$ as a genuine phase-space coordinate with definite values; this is
just not allowed by quantum mechanics. In an analogous manner, having $\alpha$
associated as a physical quantity to an observable operator having an orthonormal
basis of eigenstates is impossible; the corresponding operator $\hat{a}$ is not
Hermitian. The classical radiation field is not subject to such a fundamental
constraint on the precision of the joint knowledge of its phase and amplitude; its
phase-space localization.

We can however, define the eigenstates of the observable operators $\hat{x}$ and
$\hat{p}$. These are the position and momentum eigenstates $|x\,\rangle$ and
$|p\,\rangle$. The position eigenstate $|x'\,\rangle$, for instance, will have a
position $x'$ which is unambiguously defined; $\langle x'|\,\Delta \hat{x}^{2}\,|x'
\rangle=0$. For this state, the Heisenberg uncertainty relation of
eq.~(\ref{eq:HeisUncRel}) forbids any precision in the knowledge of momentum for such
a state; for $\langle x'|\,\Delta \hat{p}^{2}\,|x' \rangle=\infty$. In an analogous
manner, the momentum eigenstates are of known momentum and undefined position.

The position and momentum eigenstates form an orthogonal basis for the representation
of quantum states~\cite{BAYM}. The wave functions of a quantum state of vector $|\psi
\rangle $ can thus be defined as the projections $\psi(x)\,=\langle x|\psi \rangle$
and $\psi(p)\,=\langle p|\psi \rangle$. Given that $\langle
x|p\rangle=\pi^{-1/2}\,e^{2ipx}$, the wave function in the momentum representation is
the Fourier Transform of the wave function in the position representation;

\begin{equation}
\psi(p)=\,\pi^{-1/2}\int_{-\infty}^{\infty}e^{-2ixp}\,\psi(x)\label{eq:WavefunFour}
\end{equation}

The relation between wave functions in position and momentum means that a wave
function narrow in the position representation (small $\langle\,\Delta
\hat{x}^{2}\,\rangle$) will be wide in the momentum representation; following the
Heisenberg uncertainty relation of eq.~(\ref{eq:HeisUncRel}). For the position
eigenstate itself, the position representation wavefunction will be of the form
$\langle x|x'\rangle=\delta(x-x')$.

The position and momentum eigenstates, useful for representing quantum states are,
however, physically unfeasible. The average number of photons of these states is
infinite; an infinite amount of energy would be needed for their preparation. This
can be easily seen, as $\langle\,\Delta \hat{n}^{2}\rangle=\infty$ where
 $\Delta\hat{n}\equiv \hat{n}-\langle\,\hat{n}\,\rangle$ for any
position or momentum eigenstate. The wave functions of the position and momentum
eigenstates are therefore not square integrable; as the energy of a wave is equal to
the integral of the square of its modulus.

However, physically feasible approximates of the position and momentum eigenstates
exist, they are the squeezed states referred to in subsection~\ref{Back:Toys:Squez}.

Define \emph{nonlocal} quadratures for a two-mode state, that are linear combinations
of the quadratures of two (one-mode) states; we have as the eigenstates of such
nonlocal quadratures the well-known Einstein-Podolsky-Rossen (\textbf{EPR})
states~\cite{EPR}.

In section~\ref{Back:Toys:BS} we will see how the \textbf{EPR} state comes about from
the mixing of a position and momentum eigenstate by means of a beam-splitter
transformation. For the same reasons given for the position and momentum eigenstates,
the (\textbf{EPR}) states have infinite energy and are physically unfeasible.

\section{The toy box: beam-splitting, squeezing, displacement and homodyne detection}
\label{Back:Toys}

We will review in this section the unitary transformations, acting on one and two
modes of the radiation field, that constitute a basic toolbox of \textbf{CV}
transformations and a basis for the representation of density matrices~\footnote{and
any bounded operator $\hat{F}$ having a finite Hilbert-Schmidt norm
$\mathrm{Tr}(hat{F}^{\dag}\hat{F})$} of \textbf{CV} states; together with the bases
of pure quantum states directly associated with these transformations. We will also
describe the projective measurement of a quadrature of the radiation field by
homodyne detection.

\subsection{The displacement operator and the coherent states}
\label{Back:Toys:Disp}

The \textit{displacement operators}~\cite{Glaubercoher,CahillGlauber1} and the
coherent states~\cite{Glaubercoher} associated with them constitute the basis of the
conjugate phase-space representations we will use in this work.

Define the displacement operator;
\begin{equation}\label{eq:GlauDispOp}
\widehat{D}(\alpha)=e^{\left(\alpha\,\hat{a}^{\dag}\,-\,\alpha^{*}\hat{a}\right)}
\end{equation}
where $\alpha$ is a complex number.

The displacement operators form an unitary, orthogonal group of transformations under
the operator multiplication map and the form trace of a product of operators. Given
operators $\widehat{A}$ and $\widehat{B}$ with general ordering
identities~\cite{Barnett}
\begin{align}
  e^{\widehat{A}}\widehat{B}e^{-\widehat{A}}&=\widehat{B}+[\,\widehat{A}\,,\,\widehat{B}\:]+\frac{1}{2!}[\widehat{A}\,,\,[\,\widehat{A}\,,\,\widehat{B}\:]]+\frac{1}{3!}[\widehat{A}\,,\,[\widehat{A}\,,\,[\,\widehat{A}\,,\,\widehat{B}\:]]]+\cdots\label{eq:OpTransf}\\
  e^{\widehat{A}}e^{\widehat{B}}&=e^{\widehat{A}+\widehat{B}+1/2[\,\widehat{A}\,,\,\widehat{B}\:]}\;\;\;\;\hbox{for}\;\;\;\;[[\,\widehat{A}\,,\,\widehat{B}\:],\,\widehat{A}\:]=[[\,\widehat{A}\,,\,\widehat{B}\:],\,\widehat{B}\:]=0\label{eq:BCHRels}
\end{align}
and the commutation rules for annihilation and creation operators of
eq.~(\ref{eq:AnniCreaComm}), it can be shown that~\cite{CahillGlauber1}
\begin{align}
\widehat{D}^{\dag}(\alpha)=&\,\widehat{D}^{-1}(\alpha)=\widehat{D}(-\alpha)\label{eq:DispInver}\\
\widehat{D}(\alpha)\widehat{D}(\beta)=&\,e^{\,\frac{1}{2}\,\left(\alpha\beta^{*}-\alpha^{*}\beta\right)}\widehat{D}(\alpha+\beta)\label{eq:DispComp}\\
\mathrm{Tr}(\widehat{D}(\alpha)\widehat{D}^{-1}(\beta))=&\,\pi\delta^{(2)}(\alpha-\beta)\label{eq:DispOrthog}
\end{align}

The states most obviously associated with the displacement operator are the coherent
states~\cite{Glaubercoher} produced by the displacement transformation of
(multiplication by the displacement operator) the vacuum state of the quantum
harmonic oscillator $|0\rangle$. The coherent state $|\,\alpha\,\rangle$ is an
eigenstate of the annihilation operator $\hat{a}$ with complex eigenvalue $\alpha$.
The coherent states form a non-orthogonal, over-complete basis of representation for
one-mode states of radiation;

\begin{align}
 |\alpha\rangle&=\;\widehat{D}(\alpha) \;|0\,\rangle \notag \\
 \hat{a}\;|\alpha\rangle&=\,\alpha\;|\alpha\rangle \notag \\
 |\alpha\rangle&=\,e^{-|\alpha|^{2}/2}\;\sum_{n}^{\infty}\;\alpha^{n}\:n!^{(-1/2)}\:|n\rangle
 & \hbox{where}\;\{\,|n\,\rangle\,\}\hbox{  is the Fock states basis.} \notag \\
 \int&\,d^{2}\alpha\;|\alpha\rangle\:\langle\alpha|=\,\hat{\mathbf{1}}\notag\\
 \langle\beta\,|\,\alpha\rangle&=\,e^{-|\alpha|^{2}/2-|\beta|^{2}/2+\beta^{*}\alpha}\notag\\
\end{align}\label{eq:CoherProp}

 The averages of the \emph{dimensionless} position and momentum
 observables are
$\langle\,x\,\rangle=\mathrm{Re}[\alpha]$  and
$\langle\,p\,\rangle=\mathrm{Im}[\alpha]$ for coherent state $|\,\alpha\rangle$ ,
with the minimum uncertainty allowed by the Heisenberg uncertainty relations
(eq.~(\ref{eq:HeisUncRel})): $\langle\,\Delta \hat{x}^{2}\,\rangle=\langle\,\Delta
\hat{p}^{2}\,\rangle=1/4$. Furthermore, the wave functions of the coherent state on
the momentum and position representations are Gaussian; therefore completely defined
by the average values above.

Coherent states are therefore the closest approximation allowed by quantum mechanics
to definite localization in phase-space; they are defined by their average
"phase-space coordinate" $\alpha$. In the course of this work this "coordinate" will
be given by $\alpha\equiv\,x+ip$; bearing in mind that a \emph{genuine} phase-space
coordinate with definite values has no realization in quantum mechanics.

The coherent states are the closest approximation to a classical radiation field of
known phase and amplitude among the quantum states of radiation: Because of the
near-definite localization in phase-space, which uniquely identifies each coherent
state as it would a coherent, classical field of one mode or a point particle; and
because of their optical coherence properties~\cite{Glaubercoher}.

The transformation effected by the displacement operator on the one-mode annihilation
operator is straightforward to derive, bearing in mind the operator ordering
identities of eqs.~(\ref{eq:OpTransf}),~(\ref{eq:BCHRels}) and the commutation rules
of eq.~(\ref{eq:AnniCreaComm})
\begin{equation}\label{eq:DispOpTransfAni}
 \widehat{D}(\alpha)\:\hat{a}\:\widehat{D}^{\dag}(\alpha)=\,\hat{a}\,+\,\alpha
\end{equation}

Lastly, using the ordering identities of
eqs.~(\ref{eq:OpTransf}),~(\ref{eq:BCHRels}), together with the fundamental identity
\begin{equation}\label{eq:NormComplex}
  \delta^{(2)}(\alpha)=\pi^{-2}\int
  d^{2}\xi\; e^{\alpha\xi^{*}-\alpha^{*}\xi}
\end{equation}
and the completeness properties of the coherent state basis (see
eq.~(\ref{eq:CoherProp})) it can be shown that the displacement
operators~\cite{CahillGlauber1} are a complete, orthogonal basis for the
representation of arbitrary \emph{bounded} operators. Let $\widehat{A}$ be bounded;
such that it's Hilbert-Schmidt norm
$\|\widehat{A}\|=\mathrm{Tr}(\widehat{A}^{\dag}\widehat{A})$ is finite. There exists
an one-to-one correspondence between the bounded operator $A$ and the
square-integrable form $\mathrm{Tr}(\widehat{A}\widehat{D}(\xi))$ such that
\begin{equation}
\widehat{A}=\,\pi^{-1}\int\,d^{2}\xi\;\mathrm{Tr}(\widehat{A}\,\widehat{D}(\xi))\:\widehat{D}^{-1}(\xi)
\label{eq:BasisRepDisp}
\end{equation}

The set of Displacement Operators $\{\widehat{D}(\alpha),\,\forall \alpha \epsilon
\mathbb{C}\}$ form an orthogonal group under multiplication and under the trace of
the product of two operators, and constitute a complete basis of representation for
operators acting on quantum states, while coherent states constitute an over-complete
basis of representation for quantum states. These properties of both Displacement
Operators and coherent states are the mathematical foundation for the
correspondence~\cite{Weyl} of density operators of quantum states (and thus quantum
states) onto functions of phase-space "coordinates" such as the Wigner
Function~\cite{CahillGlauber1,DistFuncWigner} and the Wigner characteristic function.

\subsection{The beam-splitter transformation and nonlocal states}
\label{Back:Toys:BS}

We describe below the transformation effected by an idealized linear optical device;
a lossless, phase-free beam splitter on the two modes entering its ports. The
beam-splitter transformation on the modes' annihilation operators  is unitary; it
preserves the commutation relations (see eq.~(\ref{eq:AnniCreaComm}) and overall
photon number between incoming and outcoming modes.

\begin{figure}[h] \label{fig:BSDiag}
\begin{centering}
\includegraphics[width=12cm]{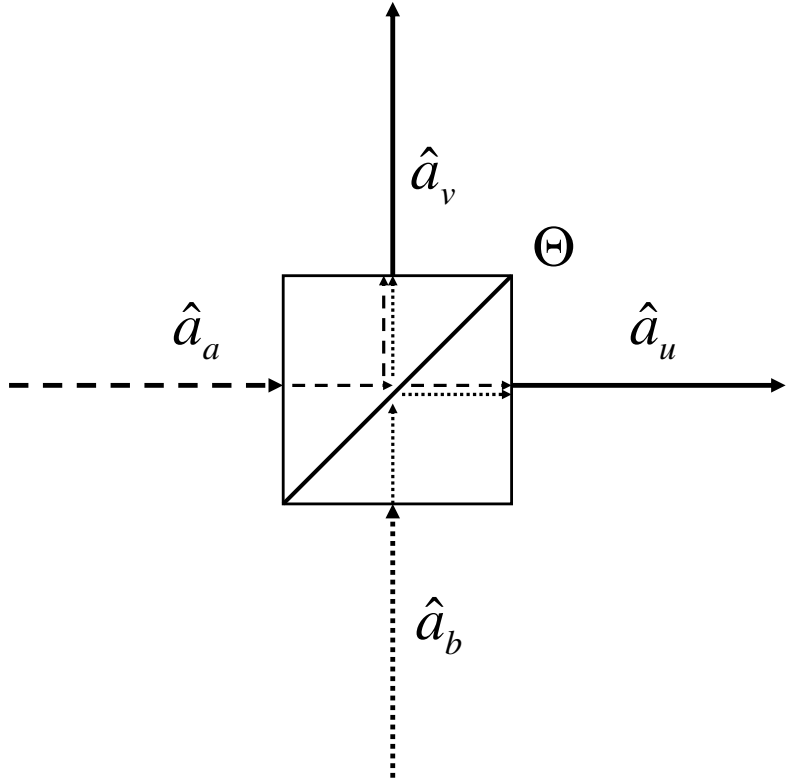}
\end{centering}
\caption[Geometric configuration of an ideal beam-splitter]{Geometric configuration
of an ideal beam-splitter: The incoming modes $\hat{a}_{a}$ (dashed line) and
$\hat{a}_{b}$ (dotted line) are "mixed" by means of a partially reflecting surface of
transmission coefficient $\cos(\Theta)$. The outcoming modes are $\hat{a}_{u}$ and
$\hat{a}_{v}$; each consists in a linear combination of the reflected fraction of one
mode and the transmitted fraction of the other mode.}
\end{figure}

Given two \textit{incoming} modes represented by their annihilation operators in the
Heisenberg interaction picture; $\hat{a}_{a}$ and $\hat{a}_{b}$, and the two
\textit{outcoming} modes' operators $\hat{a}_{u}$, $\hat{a}_{v}$ as illustrated in
fig.~(\ref{fig:BSDiag}); the transformation effected by a lossless, phase-free
beam-splitter is given by~\cite{KimBS,CamposBS}
\begin{align}
\hat{a}_{u}=&\;\widehat{B}_{ab}(\Theta)\;\hat{a}_{a}\;\widehat{B}^{\dag}_{ab}(\Theta)\notag\\
\hat{a}_{v}=&\;\widehat{B}_{ab}(\Theta)\;\hat{a}_{b}\;\widehat{B}^{\dag}_{ab}(\Theta)\notag\\
\widehat{B}_{ab}(\Theta)=&\;e^{\Theta\left(\hat{a}_{b}^{\dag}\,\hat{a}_{a}\,-\,\hat{a}_{a}^{\dag}\,\hat{a}_{b}\right)}
\label{eq:LinOptTransf}
\end{align}
where the transmittance and reflectance coefficients of the beam-splitter apparatus
are, respectively, $\cos^{2}(\Theta)$ and $\sin^{2}(\Theta)$. Given that
$\widehat{B}_{ab}$ is an unitary transformation the commutation relations of
$\hat{a}_{u}$ and $\hat{a}_{v}$ with their Hermitian conjugates and between their
associated quadratures will be those of $\hat{a}_{a}$ and $\hat{a}_{b}$ and
associated quadratures. The overall number of photons is conserved, as
$\hat{n}_{u}+\hat{n}_{v}=\hat{n}_{a}+\hat{n}_{b}$.

Let us recall eq.~(\ref{eq:OpTransf}); it is straightforward to re-state
eq.~(\ref{eq:LinOptTransf}) in the matrix form
\begin{equation}\label{eq:PhaselessBSTransf}
  \begin{pmatrix}
  \hat{a}_{u} \\
  \hat{a}_{v}
\end{pmatrix}  = \begin{pmatrix}
  \cos (\Theta) & - \sin (\Theta) \\
  \sin (\Theta) & \cos (\Theta)
\end{pmatrix} \begin{pmatrix}
  \hat{a}_{a} \\
  \hat{a}_{b}
\end{pmatrix}
\end{equation}

Given that $\hat{a}_{u,v}=\hat{x}_{u,v}+i\hat{p}_{u,v}$, the outcoming modes'
quadratures will be linear combinations of the incoming modes' quadratures with an
analogous relationship to that of eq.~(\ref{eq:PhaselessBSTransf}).

An operator that is an \emph{analytic} function of the modes' operators, such as
$\widehat{F}(\hat{a}_{a},\hat{a}_{b}\,;\,\hat{a}^{\dag}_{a},\hat{a}^{\dag}_{b})$ will
be transformed by eq.~(\ref{eq:PhaselessBSTransf}) onto
\begin{equation}\label{eq:BSTransfOp}
\widehat{F}'(\hat{a}_{u},\hat{a}_{v};\hat{a}^{\dag}_{u},\hat{a}^{\dag}_{v})=\,\widehat{B}_{ab}(\Theta)\:\widehat{F}\:\widehat{B}^{\dag}_{ab}(\Theta)=\widehat{F}(\hat{a}_{a}(\hat{a}_{u},\hat{a}_{v}),\,\hat{a}_{b}(\hat{a}_{u},\hat{a}_{v});\,\hat{a}^{\dag}_{a}(\hat{a}^{\dag}_{u},\hat{a}^{\dag}_{v}),\,\hat{a}^{\dag}_{b}(\hat{a}^{\dag}_{u},\hat{a}^{\dag}_{v}))
\end{equation}
where $\hat{a}_{a,b}(\hat{a}_{u},\hat{a}_{v})$ denotes the \emph{inverse} transform
of eq.~(\ref{eq:PhaselessBSTransf}).

Most importantly, displacement operators for two separate modes
$\widehat{D}_{a}(\alpha_{a})\:\widehat{D}_{b}(\alpha_{b})$ will be transformed to
$\widehat{D}_{a}(\alpha_{a}(\alpha_{u},\alpha_{v}))\:\widehat{D}_{b}(\alpha_{b}(\alpha_{u},\alpha_{v})))=
\widehat{D}_{u}(\alpha_{u})\:\widehat{D}_{v}(\alpha_{v})$. Where
\begin{equation}
 \begin{pmatrix}
  \alpha_{u} \\
  \alpha_{v}
\end{pmatrix}  = \begin{pmatrix}
  \cos (\Theta) & - \sin (\Theta) \\
  \sin (\Theta) & \cos (\Theta)
\end{pmatrix} \begin{pmatrix}
  \alpha_{a} \\
  \alpha_{b}
\end{pmatrix}
\label{eq:XUPVUnitary}
\end{equation}

A separable two-mode state of radiation with a density matrix
$\hat{\rho}_{a}\otimes\hat{\rho}_{b}$ entering the beam-splitter will be have, after
the beam-splitter, a density matrix depending on modes' operators $a_{u}$ , $a_{v}$
and their Hermitian conjugates. The resulting state will be usually~\cite{KimBS}
entangled, as the density matrix will not be factorized into two separate density
matrices for the outcoming modes $u$ and $v$.

Consider for simplicity's sake the separate wave function for a pure state
$\psi_{a}(x_{a})\psi_{b}(x_{b})$; after the beam-splitter transformation it becomes
(according to eq.~(\ref{eq:PhaselessBSTransf})) equal to
 $\psi_{a}(x_{a}(x_{u},x_{v}),x_{b}(x_{u},x_{v}))$. The limit case
for the entangled states achievable by beam-splitter interaction in quantum optics
illustrates the point nicely; assume $\psi{a}(x_{a})=e^{2ix_{a}p'}$, a position
eigenstate, and $\psi_{b}(x_{b})=\delta(x_{b}-x')$ a momentum eigenstate. After the
beam-splitter transformation the joint wave function is
\begin{equation}\label{eq:EPRWavefunc}
  \delta\left((\cos(\Theta)\,x_{v}\,-\,\sin(\Theta)\,x_{u})\,-\,x'\,\right)\;e^{2i\left(\cos(\Theta)\,x_{u}\,+\,\sin(\Theta)\,x_{v}\right)\,p'}
\end{equation}
which, for $\Theta=\pi/4$ and for the modes $u$ and $v$ is the wave function of the
maximally entangled state in a \textbf{CV} setting, the \textbf{EPR}
state~\cite{EPR}. That this state is maximally entangled can be seen easily; it is a
joint eigenstate \footnote{the associated observable operators commute:
 $2^{-1}[\hat{x}_{v}-\hat{x}_{u},p_{u}+p_{v}]=0$} of the
continuous, nonlocal quadratures $2^{-1/2}(x_{v}-x_{u})$ and $2^{-1/2}(p_{u}+p_{v})$,
with eigenvalues $x'$ and $p'$, respectively. While the measurements effected on one
of the local quadratures, say $x_{u}$, will yield a random result $x^{(m)}_{u}$, with
a constant probability for all the values of $x_{u}$; this same measurement will fix
the value of $x_{v}=2^{1/2}x'-x^{(m)}_{u}$.

It has been shown that the \textbf{EPR} state is physically unfeasible, unless for an
infinitesimal normalization constant, the wavefunction of eq.~(\ref{eq:EPRWavefunc})
is also not square-integrable.

\subsection{Homodyne detection}
\label{Back:Toys:Homo}

The procedure for homodyne detection of quadratures of the radiation field is the
basic detection scheme of \textbf{CV} quantum information protocols; such as quantum
teleportation~\cite{BraunsteinKimble,Furusawa1,Furusawa2,Bowen,DeMartini} and quantum
tomography~\cite{DArianoMacchiParis,DArianoLPTomo}. We will describe the experimental
procedure for homodyne measurement and the simple projective measurement that
(ideally) projects the mode thus detected onto an eigenstate of the measured
quadrature.

\begin{figure}[h] \label{fig:idealhomod}
\begin{centering}
\includegraphics[width=12cm]{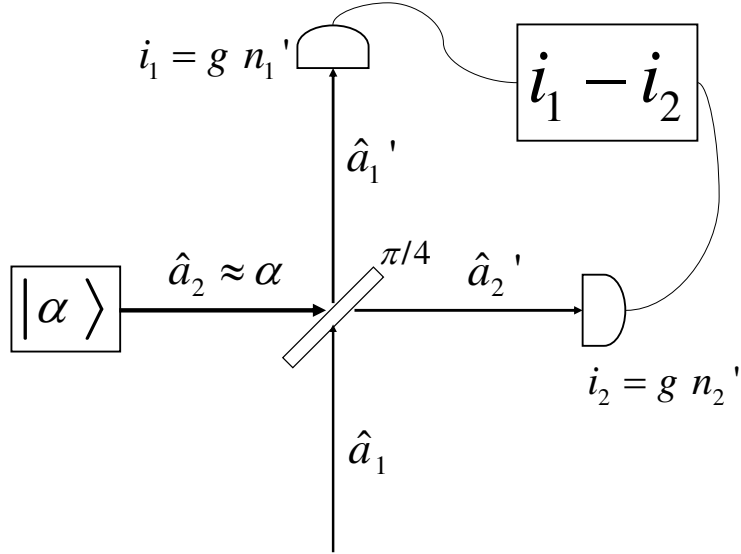}
\end{centering}
\caption[Idealized Homodyne Detection Scheme]{Idealized Homodyne Detection Scheme:
Mode $\hat{a}_{1}$, to be measured, is mixed by means of a symmetric beam-splitter
with mode $\hat{a}_{2}$, prepared in a coherent state $|\alpha\,\rangle$ of high
intensity $|\alpha|^{2}$. Photo-detection is performed on modes $\hat{a}'_{1}$ and
$\hat{a}'_{2}$. The photocurrents $i_{1}$ and $i_{2}$, proportional to the measured
intensities, are subtracted to obtain a difference photocurrent that is proportional
to the quadrature's value.}
\end{figure}

The experimental scheme for balanced homodyne detection~\cite{YuenChanHomod} is
illustrated in fig.~(\ref{fig:idealhomod}).

Mode $\hat{a}_{1}$ is mixed with a reference mode $\hat{a}_{2}$ by means of a
symmetric beam-splitter. Mode $\hat{a}_{2}$ is in a coherent state of a very high
average photon number (light intensity), approximating a classical coherent source of
light; it's behavior can therefore be described approximately by it's complex
amplitude, thus $\hat{a}_{2}\approx\alpha=|\alpha|e^{i\Phi}$. The outcoming modes of
the beam-splitter read
\begin{align}
\hat{a}'_{1}=&\,2^{-1/2}(\,|\alpha|e^{i\Phi}\,+\,\hat{a}_{1}) \notag \\
\hat{a}'_{2}=&\,2^{-1/2}(\,|\alpha|e^{i\Phi}\,-\,\hat{a}_{1})\label{eq:OutModHomod}
\end{align}

The intensity of the two outcoming modes is measured by photo-detectors (see
fig.~(\ref{fig:idealhomod})), resulting in the photo-currents
$i_{1}=\,g\langle\hat{a}_{1}^{'\dag}\hat{a}'_{1}\rangle$ and
$i_{2}=\,g\langle\hat{a}_{2}^{'\dag}\hat{a}'_{2}\rangle$; which are proportional
(with a known gain factor $g$) to the number of photons detected. Subtracting the
photo-currents, we obtain
\begin{equation}\label{eq:DeltaCuHomo}
\delta\,i=\,i_{1}-i_{2}=\,g|\alpha|\langle\,e^{-i\Phi}\hat{a}_{1}+e^{i\Phi}\hat{a}_{1}^{\dag}\,\rangle
\end{equation}

The observable measured in eq.~(\ref{eq:DeltaCuHomo} is equal to $\hat{x}_{1}$ for
$\Phi=0$ and is equal to $\hat{p}_{1}$ for $\Phi=\pi/2$. We can measure, controlling
the phase of the reference coherent state $|\alpha\rangle$, a generalized quadrature
of mode $1$;
\begin{equation}\label{eq:GenQuadHomod}
\hat{x}_{1}^{(\Phi)}\equiv
\,2^{-1}\left(e^{-i\Phi}\,\hat{a}_{1}\,+\,e^{i\Phi}\,\hat{a}_{1}^{\dag}\right)
\end{equation}
which would be identical to $\hat{x}_{1}$, for a phase shifted mode $1$ where
$\hat{a}_{1}\rightarrow\hat{a}_{1}e^{-i\Phi}$. The conjugate quadrature to
$\hat{x}_{1}$, satisfying the commutation relations in eq.~(\ref{eq:AnniCreaComm} is
simply $\hat{x}_{1}^{\Phi+\pi/2}$.

Ideally, the measurement of the quadrature $\hat{x}_{1}^{(\Phi)}$ on an arbitrary
state $\rho_{1}$ of mode $1$ will obtain a random measurement result $x^{(\Phi)}$ and
"collapse" the state into the pure eigenstate $|x^{(\Phi)}\rangle_{1}$ of
$\hat{x}_{1}^{(\Phi)}$. With a probability equal to
\begin{equation}\label{eq:ProjProbDens}
\mathrm{Tr}(\hat{\rho}_{1}|\,x^{(\Phi)}\rangle_{1}\langle\,x^{(\Phi)}|_{1})
\end{equation}
thus, the system after the measurement can be said to be in a mixed
state~\footnote{According to the Copenhagen Intepretation of Quantum Mechanics.}; The
pure components of the mixture being the $|x^{(\Phi)}\rangle_{1}$ eigenstates, with
probability given by eq.~(\ref{eq:ProjProbDens}).

To illustrate the concept of a projective measurement further, and particularly for
two-mode states, let us consider a state with a density matrix $\hat{\rho}_{1,3}$.
The state of mode $3$ after a \emph{projective} measurement on mode $1$ obtaining a
result $x^{(\Phi)}$ can be represented by a partial trace of the form
\begin{equation}\label{eq:ProjMeasDens}
\mathrm{Tr}_{1}(\hat{\rho}_{1,3}|\,x^{(\Phi)}\rangle_{1}\langle\,x^{(\Phi)}|_{1})=\mathrm{P}(x^{(\Phi)})\hat{\rho}_{3}(x^{(\Phi)})
\end{equation}
where $\mathrm{P}(x^{(\Phi)})$ is the probability density for the measurement result
$x^{(\Phi)}$ when the quadrature $\hat{x}_{1}^{(\Phi)}$ is measured and
$\hat{\rho}_{3}(x^{(\Phi)})$ is the state of mode $3$ after measurement. The operator
of eq.~(\ref{eq:ProjMeasDens} is not a normalized, proper density operator, because
the "projection" is not an unitary operation. It is easy to see also that the state
of mode $3$ will not be affected by the measurement process on mode $1$ if the
$\hat{\rho}_{1,3}$ state is \emph{separable} into two density matrices;
$\hat{\rho}_{1,3}=\hat{\rho}_{1}\otimes\hat{\rho}_{3}$. If $\hat{\rho}_{1,3}$ is an
entangled, \emph{inseparable} state; the state of mode $3$ after the measurement
depends on the random outcome $x^{(\Phi)}$ of the measurement performed on mode $1$,
with a probability $\mathrm{P}(x^{(\Phi)})$ for the state
$\hat{\rho}_{3}(x^{(\Phi)})$. Such a state of "classical ignorance" produced by
measurement is a mixture of pure states. Therefore, we choose a properly normalized
state for mode $3$, accounting for the random outcome of a projective measurement;

\begin{equation}\label{eq:ProjMeasMix}
\hat{\rho}_{3}=\int d x^{(\Phi)} P(x^{(\Phi)})\hat{\rho}_{3}(x^{(\Phi)})=\int d
x^{(\Phi)}\mathrm{Tr}_{1}(\hat{\rho}_{1,3}|\,x^{(\Phi)}\rangle_{1}\langle\,x^{(\Phi)}|_{1})
\end{equation}

\subsection{The one and two-mode squeezing operator and the squeezed states}
\label{Back:Toys:Squez}

Producing physically feasible states approximating as an asymptotic limit the
position (or momentum) eigenstates, that can then be entangled by a beam-splitter
requires interactions of a finite energy that cause these states to have a narrower
position (or momentum) wave function and a wider momentum (or position) wave
function. Though any state that has a different variance for position and momentum
can be thought of as \textit{squeezed}, we will generally term \textit{squeezed}
states those that have been transformed by a particular kind of unitary evolution
named the \textit{squeezing} transformation.

To effect such an evolution in the laboratory requires the use of nonlinear optical
elements and optical pumping~\cite{FourWallsMix,DownCWuKimble}. The simplest example
of an squeezing evolution involves the use of a nonlinear medium down-converting
photons of a given frequency to two photons of half this
frequency~\cite{QOptWalls,DownCWuKimble}. Let $\hat{a}$ be the annihilation operator
for the mode to be squeezed, of frequency $\hbar\omega$. Let $\hat{b}$ be the
annihilation operator for an intense coherent field; the \emph{pump}, of frequency
$2\hbar\omega$. Let $\Xi^{(2)}$ be the strength of the coupling between the two
modes, and the nonlinear coefficient of the medium inside which the squeezing
evolution occurs. For a lossless setup the Hamiltonian for the interaction is given
by
\begin{equation}\label{eq:SqueHam}
  \widehat{H}_{I}=2^{-1}\:i\:\hbar\:\Xi^{(2)}\,(\hat{b}^{\dag}\,\hat{a}^{2}\,-\,\hat{b}\,\hat{a}^{\dag\,2}\,)
\end{equation}
For each photon of the pump annihilated, two photons of the mode are created, and
viceversa. Given the high intensity of the pump field and its quasi-classical
character, it's annihilation operator can be approximated by a complex amplitude,
$\hat{b}\approx\beta$.

The unitary evolution operator, $U(t)=e^{-i\hbar^{-1}\widehat{H}_{I}\,t}$ for the
time independent Hamiltonian in eq.~(\ref{eq:SqueHam}) will therefore be
\begin{equation}\label{eq:OneModSqOp}
  \widehat{S}(\zeta) \,=\, e^{\,\frac{1}{2}\,\left(\zeta^{*}\, \hat{a}^{2}\,-\,\zeta
\,\hat{a}^{\dag 2}\right)}
\end{equation}
with $\zeta=\Xi^{(2)}\,\beta\,t$, the phase of argument of the operator being that of
the pump field. This evolution operator is unitary, therefore
$\widehat{S}(-\zeta)=\widehat{S}^{\dag}(\zeta)=\widehat{S}^{-1}(\zeta)$.

$\widehat{S}(\zeta)$ has been named the \emph{squeezing
operator}~\cite{QOptWalls,Barnett}, having a complex argument
$\zeta=|r|\,e^{i\varphi}$. The modulus~\footnote{We will usually refer to a signed,
real quantity $r$ instead of the non-negative modulus $|r|$.} of this argument,
$|r|$, usually called the \emph{squeezing coefficient} is a characteristic
interaction time and a logarithmic measure of the \emph{degree} of squeezing. The
squeezing phase $\varphi$, equal to that of the pump field, will determine the
phase-space orientation of the squeezing transformation.

Using the ordering identities of eqs.~(\ref{eq:AnniCreaComm})
and~(\ref{eq:OpTransf}), the results of the squeezing transformation on annihilation
and creation operators, and on the displacement operator can be derived,
\begin{align}
\widehat{S}(\zeta)\,\hat{a}\,\widehat{S}^{\dag}(\zeta)=&\,\hat{a}\,\cosh(r)\,-\,\hat{a}^{\dag}\,\sinh(r)\,e^{i\varphi}\notag\\
\widehat{S}(\zeta)\,\hat{a}^{\dag}\,\widehat{S}^{\dag}(\zeta)=&\,\hat{a}^{\dag}\,\cosh(r)\,-\,\hat{a}\,\sinh(r)\,e^{-i\varphi}\notag\\
\widehat{S}(\zeta)\,\widehat{D}(\,\alpha)\,\widehat{S}^{\dag}(\zeta)=&\,\widehat{D}(\alpha\,\cosh(r)\,-\,\alpha^{*}\,\sinh(r)\,e^{i\varphi}\,)\label{eq:SqueeAnniCreDisp}
\end{align}

Given eqs.~(\ref{eq:GenQuadHomod}) and~(\ref{eq:SqueeAnniCreDisp}), the result of the
squeezing transformation of the generalized quadrature $\hat{x}^{(\Phi)}$ can be
calculated;
\begin{equation}\label{eq:SqueeGenQuad}
\widehat{S}(\zeta)\;\hat{x}^{(\Phi)}\;\widehat{S}^{\dag}(\zeta)=\,-e^{r}\,\sin(\Phi-\varphi/2)\:\hat{x}^{(\varphi/2+\pi/2)}\,+\,e^{-r}\,\cos(\Phi-\varphi/2)\:\hat{x}^{(\varphi/2)}
\end{equation}

Taking $\Phi=\varphi/2$; the expression in eq.~(\ref{eq:SqueeGenQuad}) is simplified
as the phase orientation of the quadrature $\hat{x}^{(\Phi)}$ coincides with the
phase orientation of the squeezing operator. Thus we have,
\begin{align}
\widehat{S}(r\,e^{2i\Phi})\;\hat{x}^{(\Phi)}\;\widehat{S}^{\dag}(r\,e^{2i\Phi})=&\,e^{-r}\,\hat{x}^{(\Phi)} \notag \\
\widehat{S}(r\,e^{2i\Phi})\;\hat{x}^{(\Phi+\pi/2)}\;\widehat{S}^{\dag}(r\,e^{2i\Phi})=&\,e^{r}\,\hat{x}^{(\Phi+\pi/2)}\notag\\
\Delta(\widehat{S}(r\,e^{2i\Phi})\;\hat{x}^{(\Phi)}\;\widehat{S}^{\dag}(r\,e^{2i\Phi}))^{2}=&\,e^{-2r}\,\Delta(\hat{x}^{(\Phi)})^{2}\notag\\
\Delta(\widehat{S}(r\,e^{2i\Phi})\;\hat{x}^{(\Phi+\pi/2)}\;\widehat{S}^{\dag}(r\,e^{2i\Phi}))^{2}=&\,e^{2r}\,\Delta(\hat{x}^{(\Phi+\pi/2)})^{2}\label{eq:SqueeQuad}
\end{align}

For $r>0$, the variance of $\hat{x}^{(\Phi)}$ is scaled by a factor of $e^{-2r}$ and
the associated wavefunction is made narrower; this is termed \emph{squeezing} in the
language of quantum optics. While the opposite happens to the conjugate quadrature
$\hat{x}^{(\Phi+\pi/2)}$, "expanded" by the inverse factor $e^{2r}$. Thus the
squeezing transformation preserves the product of variances in Heisenberg's
uncertainty relation, eq.~(\ref{eq:HeisUncRel}). A quantum state for which that
product is equal to the lower bound of $\frac{1}{16}$, a \emph{minimum uncertainty}
state, will continue to be of minimum uncertainty after being "squeezed".

With the purpose of simplifying calculations, we will take the squeezing operator's
argument to be real; $\zeta=r$, where $r$ can be negative. This is equivalent to the
choice made for eq.~(\ref{eq:SqueeQuad}) of
 squeezing phase-orientation; on quadratures  $\hat{x}^{(0)}$ and $\hat{p}=\hat{x}^{(\pi/2)}$.

The states usually associated with the squeezing operator $\widehat{S}(\zeta)$ are
the \textit{coherent squeezed states}~\cite{Barnett,YuenTwoPhot}, produced by the
squeezing transformation of a vacuum state; on which is then performed a displacement
operation. Consider the case where $\Phi=\varphi=0$;
\begin{equation}\label{eq:SqueCohStat}
 |\alpha;r\,\rangle=\widehat{D}(\alpha)\:\widehat{S}(r)|0\,\rangle
\end{equation}

The coherent squeezed states are minimum uncertainty states like the coherent states,
with the variance $\Delta\hat{x}^{2}=e^{-2r}/4$ reduced and
$\Delta(\hat{p}^{2}=e^{2r}/4$ increased when $r>0$. When $r<0$, which corresponds to
$\varphi=\pi$, the $\hat{p}$ quadrature is the one "squeezed" and it's variance is
reduced. The wave function in either basis of representation is Gaussian, and the
averages of position and momentum are, respectively, the real and imaginary part of
$\alpha$.

The infinite squeezing (and coupling~$\times$~intensity of pump~$\times$~time, as
$r=|\beta|\xi^{(2)}t$) limit for the squeezed states $\widehat{S}(re^{i\varphi})$ are
the $\hat{x}^{(\varphi/2)}$ quadrature eigenstates. In such a way a physically
feasible approximation for a quadrature eigenstate can be obtained by appropriate
squeezing of an initial coherent state where $\alpha$ is wholly real (position
eigenstate, $r>0$) or imaginary (momentum eigenstate, $r<0$), the limitation being
technological.

Quantum state preparation \emph{approximating} the maximally entangled \textbf{EPR}
states is possible given strong nonlinear interactions, control of the phase of the
individual modes, and beam-splitters (see
eqs.~(\ref{eq:PhaselessBSTransf}),~(\ref{eq:BSTransfOp}) and~(\ref{eq:EPRWavefunc})).
Two coherent squeezed states, squeezed in position and in momentum can be mixed by a
beam-splitter obtaining a Gaussian state "squeezed" in nonlocal quadratures. Which,
in the limit $r\,\rightarrow\,\infty$ becomes an \textbf{EPR} state.

The operation described above can be represented by the \textit{Two-mode squeezing}
operator~\cite{Barnett}. Two-mode squeezing is the transformation preparing an
entangled, symmetric Gaussian state state; from a two mode vacuum state vector $|\,0
\rangle\_{1},\otimes\,|\,0\rangle_{2}$:
\begin{equation}\label{eq:TwoModSqOp}
\widehat{S}_{12}(\zeta)= e^{ -\zeta \hat{a}_{1}^{\dag}\hat{a}_{2}^{\dag} + \zeta
\hat{a}_{1}\hat{a}_{2}} \;\;\;\;\;\;\;\zeta\equiv\,r\,e^{i\phi}
\end{equation}

It is straightforward to show that the two-mode squeezing operator can be written as
the product of two squeezing operators for different modes and a beam-splitter
transformation;
\begin{equation}\label{eq:TwoModSqPre}
\widehat{S}_{12}(r)\equiv\,\widehat{B}_{12}(\pi/4)\:\widehat{S}_{1}(r)\:\widehat{S}_{2}(-r)
\end{equation}
where the two squeezing operations represent the nonlinear interaction (see
eqs.~(\ref{eq:SqueHam}) and~(\ref{eq:OneModSqOp})) transforming each of the initial
vacuums into states squeezed, respectively, in the quadratures $\hat{x}_{1}$ and
$\hat{p}_{2}$. The symmetric beam-splitter transformation mixes the two squeezed
states (see eq.~(\ref{eq:EPRWavefunc}) and discussion above) into an approximate
\textbf{EPR} state, the \textit{two-mode squeezed} vacuum; $|\zeta\rangle_{AB}$.

Recalling the beam-splitter transformation's operation on a two-mode quantum state
wavefunction, it is easily seen that the two mode-squeezed vacuum is Gaussian in
wavefunction. The two-mode squeezed vacuum is an entangled state, and correlations
arise in the measurement of observables other than the nonlocal quadratures. In the
Fock state basis representation;
\begin{equation}\label{eq:TwoModSqVacRFock}
|r\,\rangle_{12}=\,(\cosh(r))^{-1}\sum_{n=0}^{\infty}\left(\,\tanh(r)\,\right)^{n}\;|n\rangle_{1}\;|n\rangle_{2}
\end{equation}

The state vector is symmetric under exchange of modes $1$ and $2$, having an even
overall number of photons. A measurement of photon number must give the same result
for modes $1$ and $2$. For the two-mode squeezed state
$\langle\,\hat{n}_{1}-\hat{n}_{2}\rangle=0$, even if the overall photon number
$n_{1}+n_{2}$ is not unambiguously defined.

The factor $(\tanh(r))^{n}$ in eq.~(\ref{eq:TwoModSqVacRFock}) makes for a
probability of measuring (overall) $2\;n$ photons that decreases exponentially with
$2n$; and ensures the convergence of the infinite sum for finite $r$. The infinite
squeezing limit $r\rightarrow\infty$ of eq.~(\ref{eq:TwoModSqVacRFock}) is the
\textbf{EPR} state of eq.~(\ref{eq:EPRWavefunc}). In this limit the infinite sum does
not converge and the normalization constant $(\cosh{r})^{-1}$ becomes infinitesimal.

To study the effect of two-mode squeezing on non-Gaussian states' wave functions and
phase-space representations; it suffices to study the transformation of modes'
operators $\hat{a}_{1}$ and $\hat{a}_{2}$;

\begin{equation}\label{eq:TwoModBogCom}
 \widehat{S}_{12}(\zeta)\, \hat{a}_{i} \, \widehat{S}_{12}^{\dag}(\zeta)=\cosh(r) \, \hat{a}_{i}
-e^{i\phi}\sinh(r) \, \hat{a}_{j}^{\dag} \;\;\;\; \hbox{for}\;\; i\neq j=1,2
\end{equation}
a Bogoliubov transformation mixing mode operators.

Analytic functions of the operators $\hat{a}_{1,2},\hat{a}_{1,2}^{\dag}$ will be
transformed in an analogous manner to that of eq.~(\ref{eq:PhaselessBSTransf}); with
the Bogoliubov transformation of eq.~(\ref{eq:TwoModBogCom}) substituting for the
beam-splitter transformation. An interesting and useful (for the purposes of this
work) example is that of a product of displacement operators in two modes,
\begin{align}
&\widehat{S}_{12}(\zeta)\;\widehat{D}_{1}(\alpha_{1})\;\widehat{D}_{2}(\alpha_{2})\;\widehat{S}^{\dag}_{12}(\zeta)=\widehat{D}_{1}(\alpha'_{1})\;\widehat{D}_{2}(\alpha'_{2})\label{eq:TwoModBogDisp}\\
&\alpha'_{i} =\: \cosh(r)\, \alpha_{i}\, +\,e^{i\phi} \sinh(r)\, \alpha_{j}^{*}
\;\;\;\hbox{for}\;\; i\neq j=1,2 \label{eq:TwoModBogChar}
\end{align}

\section{The Wigner and the Wigner characteristic function}
\label{Back:PhaseSp}

In this section, we introduce phase-space representations; of quantum states' density
matrices, and of any square-integrable operator~\cite{CahillGlauber1}. In particular,
we introduce the Wigner function~\cite{Wigner1} and the Wigner characteristic
function~\cite{CahillGlauber1} as particularly useful for the representation of
\textbf{CV} systems because they are explicit, analytic functions of phase-space
"coordinates" that transform in the same manner as the wave functions of quantum
states~\cite{DistFuncWigner}. Though the Wigner function is not a genuine probability
distribution in momentum and position, the two conjugate functions allow the
calculation of quantum mechanical averages, including partial traces, to take the
form of integrals over the complex plane of phase-space. Unitary evolutions and
measurements over a multi-mode quantum state likewise take the form of simple unitary
transformations on the arguments of the functions, and projections on adequate
eigenstates.

\subsection{Phase-space representations: mainly Wigner function and characteristic function}
\label{Back:PhaseSp:Def}

The Wigner function was proposed and chosen to be an Hermitian form (real scalar) of
the density operator of a quantum state fulfilling a number of desirable conditions;
to be real and bounded, to transform according to the same rules for a classical
distribution of probability; to produce the quantum mechanical averages pertaining to
the density operator~\cite{DistFuncWigner}; to give the appropriate probability
distributions when integrated. Such a function was proposed in ref.~\cite{Wigner1} in
the form
\begin{equation}\label{eq:WignerFunc}
 W(\alpha \,=\, x+ i p )=\int dy\; \langle\, x-\frac{y}{2}|\,\hat{\rho}\,|x+\frac{y}{2}\,\rangle\: e^{2 i p y}
\end{equation}
with the unit system defined so $\hbar=\frac{1}{2}$. The Wigner function is real, and
can be demonstrated to exist and be square-integrable for any density
operator~\cite{CahillGlauber1}. It has been termed a pseudo-distribution and as such,
a conjugate function, it's Fourier transform, the Wigner characteristic function has
been defined~\cite{DistFuncWigner};

\begin{align}\label{eq:FourierWig}
\chi(\xi)=&\,\pi^{-1}\,\int d^{2}\,\alpha\: e^{\xi\alpha^{*}-\xi^{*}\alpha}
\;W(\alpha)\\
W(\alpha)=&\,\pi^{-1}\,\int d^{2}\,\xi\: e^{\alpha\xi^{*}-\alpha^{*}\xi}\;\chi(\xi)
\end{align}
with $\xi=w+iz$ being a conjugate phase-space "coordinate". The characteristic
function for a density operator, being the Fourier transform of the analytic, real
Wigner's function is thus an analytic function. We remark here that "Wigner"
representations of suitable operators other than the density operators can be
calculated, and are necessary to the calculation of physicalle relevant quantum
mechanical averages using this representation.Let us review the \emph{forms} (in a
strict mathematical sense) corresponding to quantum mechanical averages and
normalization conditions.

The normalization condition on density operators; $\mathrm{Tr}(\hat{\rho})=1$ is
equivalent, in the Wigner and Characteristic Function descriptions to
\begin{equation}\label{eq:TrWigChar}
\mathrm{Tr}(\hat{\rho})=\pi^{-1}\, \int d^{2}\, \xi
  \;\chi_{\rho}(\xi)= \pi^{-1}\, \int d^{2}\, \alpha
 \;W_{\rho}(\alpha) \,=\, 1
\end{equation}

Finite Wigner function and characteristic functions can be derived, and used to
calculate quantum mechanical averages for any operator fulfilling the condition that
the Hilbert-Schmidt norm $\mathrm{Tr}(\widehat{F}^{\dag}\widehat{F})$ be finite, in
other words that the operator be \textit{bounded}~\cite{CahillGlauber1}. The density
operators are, furthermore, \textit{trace-class} operators with trace equal to $1$;
therefore $\mathrm{Tr}(\hat{\rho}^{2}) \leq 1 $, and
\begin{equation}\label{eq:TrSqWigChar}
\mathrm{Tr}(\hat{\rho}^{2})\,=\,\pi^{-1}\: \int d^{2}\, \xi
  \;|\chi_{\rho}(\xi)|^2\,=\, \pi^{-1} \:\int d^{2}\, \alpha
  \;|W_{\rho}(\alpha)|^2 \,\leq\, 1
\end{equation}

The quantity $\mathrm{Tr}(\hat{\rho}^{2})$ is a quantitative measure of the
\emph{purity} of the quantum state, and is thus named.

Quantum mechanical averages or expectation values of the product of at least one
trace-class operator and a bounded operator are finite and are given by the form
$\mathrm{Tr}(\widehat{F}\widehat{G})$;
\begin{align}
  \mathrm{Tr}(\widehat{F}\,\widehat{G})\,=\,&\pi^{-1}\: \int d^{2}\, \xi
  \;\chi_{F}(\xi) \;\chi_{G}(-\xi) \label{eq:TraceTwoChar}\\
=\,&\pi^{-1}\, \int d^{2}\, \alpha\;
  W_{F}(\alpha)\; W_{G}(\alpha)\label{eq:TraceTwoWig}
\end{align}
where $W_{F},W_{G}$ and $\chi_{F},\chi_{G}$ are the Wigner and characteristic
functions corresponding to $\widehat{F},\widehat{G}$ respectively.

The Wigner function can be easily shown not to be a genuine probability distribution
for density operators, even though it produces expectation values and is normalized
to $1$ like a classical probability distribution. Let
$\hat{\rho}_{\Phi}=\,|\Phi\rangle\,\langle\Phi|$ and
$\hat{\rho}_{\Psi}=\,|\Psi\rangle\,\langle\Psi|$, density operators for two pure
states. If the states are orthogonal, then
$\langle\,\Phi\,|\,\Psi\,\rangle=\,\mathrm{Tr}(\hat{\rho}_{\Phi}\:\hat{\rho}_{\Psi})\,=\,0$.
The integral of the product of two nonzero Wigner functions in
eq.~(\ref{eq:TraceTwoWig}) is equal to zero. Hence, the Wigner function of at least
one of the density operators must have negative values. The Wigner function cannot be
a genuine probability distribution for all quantum states, and is thus termed a
pseudo-probability. Conversely, quantum states with a positive Wigner function that
is a genuine probability distribution cannot be orthogonal with each other.

Quantum states of a nonclassical character have a Wigner
function~\cite{DeGauss4,ZavattaScience} with negative values in parts of its domain;
the \textit{negative volume} of the Wigner function has been proposed as a measure of
the nonclassical character of a quantum state~\cite{KenfackNegWig}.

The Wigner characteristic function is the trace of the product of the density
operator and the displacement operator~\cite{CahillGlauber1,DistFuncWigner}; the mean
value of the displacement operator for the density $\hat{\rho}$. Recall that
displacement operators form an orthogonal basis for the representation of
square-integrable operators (see eq.~(\ref{eq:BasisRepDisp}) and that any bounded
operator may be represented by a linear combination of displacement
operators~\cite{CahillGlauber1}. The characteristic function is the coefficient for
this representation, and is necessarily square-integrable (see
eq.~(\ref{eq:TrSqWigChar})) when the operator itself is trace-class. The
characteristic function is also given by
\begin{equation}
\chi(\xi)=\,\mathrm{Tr}(\,\hat{\rho}\,\widehat{D}(\xi)\,) \label{eq:CharTracDisp}
\end{equation}
which for a pure state $\hat{\rho}=\,|\psi\rangle\,\langle\psi|$ becomes
\begin{equation}
\chi(\xi)=\,\langle\,\psi\,|\:\widehat{D}(\xi)\:|\,\psi\,\rangle\label{eq:CharPurDisp}
\end{equation}

Generalizing the characteristic function to more than one mode is straightforward;
the product of operators to be traced over in eq.~(\ref{eq:CharTracDisp}) being that
of the density operator and the \emph{commuting} displacement operators
$\widehat{D}_{1}(\xi_{1})\:\widehat{D}_{2}(\xi_{2})\ldots\widehat{D}_{N}(\xi_{N})$,
for $N$ modes.

A most important property of the characteristic function, is that transformations on
density operators of quantum states such as those we have reviewed in the preceding
section will correspond, in the language of the characteristic functions, to
transformations on the arguments $\xi_{i}$ of the displacement operators (see
eqs.~(\ref{eq:XUPVUnitary}),~(\ref{eq:DispComp}),~(\ref{eq:SqueeAnniCreDisp})
and~(\ref{eq:TwoModBogChar})); which are the arguments of the characteristic
function.

The Wigner function and Wigner characteristic function are associated to a symmetric
ordering of operators~\cite{CahillGlauber1,DistFuncWigner}; where \emph{ordering}
means the left to right ordering of the non-commuting $\hat{a}$ and $\hat{a}^{\dag}$
operators inside other operators such as density matrices and observables. For
example: the number operator $\hat{n}=\hat{a}^{\dag}\hat{a}$ is symmetrically ordered
as $2^{-1}(\hat{a}^{\dag}\hat{a}+\hat{a}\hat{a}^{\dag})$, normally ordered as
$\hat{a}^{\dag}\hat{a}$, anti-normally ordered as $\hat{a}\hat{a}^{\dag}$.

Other phase-space representations exist based on alternative orderings. In the manner
discussed above, the $P(\alpha)$~\cite{Glaubercoher} representation and the
$Q(\alpha)$~\cite{Husimi} function are conjugate with characteristic functions
associated with the normal and anti-normal \emph{ordering} of operators,
respectively. Particularly, with \emph{the ordering of the displacement operator} of
which the characteristic function is a coefficient of representation;
\begin{align}
\{\widehat{D}(\xi)\}=\,&e^{\xi\hat{a}^{\dag}-\xi^{*}\hat{a}}
\notag \\
\widehat{D}(\xi,s)=\,&e^{s|\xi^{2}|/2}\{\widehat{D}(\alpha)\}\notag\\
\widehat{D}(\xi,1)=\,&e^{\xi\hat{a}^{\dag}}\:e^{-\xi^{*}\hat{a}}\notag\\
\widehat{D}(\xi,-1)=\,&e^{-\xi^{*}\hat{a}}\:e^{\xi\hat{a}^{\dag}}
\label{eq:DispOrder}
\end{align}
where the index takes the values $s=1$ for normal ordering, $s=0$ for symmetric
ordering, also denoted by the brackets $\{\ldots\}$, and $s=-1$ for antinormal
ordering. It is obvious from inspection of eq.~(\ref{eq:CharTracDisp}) and
eq.~(\ref{eq:DispOrder}) that the transformation between characteristic functions
associated with different orderings is just a power of $e^{|\xi^{2}|/2}$.

The characteristic function of the Wigner distribution, similarly to a classical
characteristic function, is the \emph{generating function} for the symmetrically
ordered moments of annihilation and creation operators~\cite{Barnett};
\begin{equation}\label{eq:CharGenMom}
\langle\left\{\,(\hat{a}^{\dag})^{m}\,\hat{a}^{n}\,\right\}\rangle=\,\mathrm{Tr}(\hat{\rho}\,\left\{(\hat{a}^{\dag})^{m}\,\hat{a}^{n}\,\right\})\,=\,
\left.\left(\,\frac{\partial}{\partial\xi}\right)^{m}\left(-\frac{\partial}{\partial\xi^{*}}\right)^{n}\,\chi(\xi)\;\right|_{\:\xi\,=\,\xi^{*}\,=\,0}
\end{equation}
simplifying the calculation of statistical moments; specially quantum mechanical
averages and covariances.

\subsection{Characteristic functions of the two-mode squeezed vacuum and EPR states}
\label{Back:PhaseSp:Ent}

The characteristic function for a Gaussian, two-mode squeezed vacuum state
$|0\rangle_{a}\otimes|0\rangle_{b}$ is given by
\begin{equation}\label{eq:CharGaussVac}
  \chi_{S}(\xi_{a},\xi_{b})\,=\,e^{-1/2\left(|\xi'_{a}|^{2}\,+\,|\xi'_{b}|^{2}
  \right)}
\end{equation}
where the relation of the pair of variables $\xi'_{a},\xi'_{b}$ to the pair of
variables $\xi_{a},\xi_{b}$ is the Bogoliubov transformation described in
eq.~(\ref{eq:TwoModBogChar}). Other, displaced two-mode coherent states have a
characteristic function similar to the one in eq.~(\ref{eq:CharGaussVac}) with a
multiplying phase factor. For an overall phase-space displacement of, say
$x'\,+\,ip'$ on $A$ mode this factor is given by $e^{2i(z_{a}\,x'\:-\:w_{a}\,p')}$.

The infinite squeezing limit of the Gaussian two-mode squeezed state is the
\textbf{EPR} state. In a \textbf{CV} setting, it is the maximally entangled
state~\cite{EPR} of two nonlocal quadratures $x_{u},p{v}$, as defined by the unitary
transformation of eq.~(\ref{eq:XUPVUnitary}). Given the wave function (see
eq.~(\ref{eq:EPRWavefunc})), the Wigner function of the state can be derived easily
from it's definition in eq.~(\ref{eq:WignerFunc}),
\begin{equation}\label{eq:WignerEPR}
W_{x'_{u},p'_{v}}(\alpha_{a}\,,\,\alpha_{b}) \, = \, \mathcal{C} \;\delta
(\cos(\Theta)\,x_{a}\,-\,\sin(\Theta)\,x_{b}\,-\,x'_{u})
\;\delta(\sin(\Theta)\,p_{a}\,+\,\cos(\Theta)\,p_{b}\,-\,p'_{v})
\end{equation}
where $\mathcal{C}$ is a normalization factor that becomes infinitesimally small as
$r\rightarrow\infty$. The characteristic function of the EPR state is easy to
calculate as the Fourier transform (eq.~(\ref{eq:FourierWig})) of
eq.~(\ref{eq:WignerEPR});
\begin{equation}\label{eq:CharEPR}
\chi_{x'_{u},p'_{v}}(\xi_{a},\xi_{b}) \,=\, \mathcal{C} \; \delta (z_{a}\,\sin
(\Theta)\, +\, z_{b}\,\cos (\Theta)) \;\delta(w_{a}\,\cos (\Theta)\,-\,w_{b}\,\sin
(\Theta))\:e^{2i(z_{a}\,x'_{u}\,\sec (\Theta)\,-\,w_{a}\,p'_{v}\,\csc(\Theta))}
\end{equation}
which is, again, not square-integrable. However, the product of this characteristic
function and a square-integrable characteristic function can be traced over
(integrated) in the manner of eq.~(\ref{eq:TraceTwoChar}).

\section{Universal quantum teleportation (in Continuous Variables)}
\label{Back:Qtele}

This section will review both the universal protocol for Quantum Teleportation of a
quantum state (and apply it to a \textbf{CV} setting) and the fidelity of
teleportation in the language of Wigner characteristic functions.

%%\subsection{Pure entangled resource states: measures} \label{Back:Qtee:EntPure}

\subsection{Quantum teleportation}
\label{Back:Qtele:UnivProt}

The quantum teleportation protocol involves the transcription of the \emph{unknown}
quantum state of a physical system $in$ (named \textit{input}) onto the quantum state
of another, similar physical system $B$~\footnote{belonging to a party named
\textit{Bob}.} that is \emph{remote} with respect to $in$. The protocol itself is a
projective measurement of the maximally entangled states for that
setting~\cite{NielsenCaves,UnivTeleportTwist,DArianoBell}, the Bell states~(in
Continuous Variables the \textbf{EPR} states) on the joint system consisting on state
$in$ and another, similar system $A$~\footnote{belonging to a party named
\textit{Alice}, also in possession of the $in$ system.}.

Quantum teleportation is not "cloning" the quantum state, as this operation is
impossible for arbitrary states~\cite{NoCloningTheo}. It is no scheme of quantum
measurements on the input state or of repeated measurements on identical instances of
preparation as is done in quantum tomography~\cite{VogelRisTomo,DArianoTomoRev}. The
state itself is unknown and is to remain unknown through teleportation; in principle
there is only one copy available for teleportation.

The transcription of an unknown state from $in$ to physically separate system $B$ by
measurements on systems $A$ and $in$ is only possible if a quantum correlation exists
between the states of the two systems $A$ and $B$ (the joint, entangled state of both
being named \textit{resource}) and if a correlation is forced on the $A,in$ state by
interaction between the two systems before measurement; there is no Bell state
measurement without previous interaction~\cite{VaidmanYoranTel}.

An obvious maximally entangled or Bell state for a quantum system Hilbert space
$\mathcal{H}$ of dimensionality $d$ can be formulated with
ease~\cite{UnivTeleportTwist}. Let $\{|n\rangle\}$ be a complete basis of
representation of $\mathcal{H}$; the simplest Bell state has a state vector
\begin{equation}\label{eq:FirstBellState}
  |\Psi_{0}\,\rangle_{AB}\,=\,d^{-1/2}\:\sum_{n}\:e^{-i\phi_{n}}\;|\,n\,\rangle_{A}\;|\,n\,\rangle_{B}
\end{equation}
belonging to the joint Hilbert space $\mathcal{H}\otimes\mathcal{H}$. This state has
the same form when written in terms of any complete basis of representation of
$\mathcal{H}$. Let the basis $\{|n\rangle\}$ be made of eigenstates of an observable
operator $\hat{n}$; \textit{Alice's} measurement of $\hat{n}_{A}$ will determine the
result for \textit{Bob's} measurement of $\hat{n}_{B}$. It follows, on inspection,
that the Bell state of eq.~(\ref{eq:FirstBellState}) is an eigenstate of nonlocal
observable $\hat{n}_{A}-\hat{n}_{B}$ with eigenvalue $0$. nonlocal observables having
as eigenstates the maximally entangled Bell states are termed \textit{Bell
observables}; their measurement is taken to be a \emph{projective measurement} of the
Bell state~\footnote{Following the Copenhagen Interpretation of Quantum Mechanics.}.
Note that the squeezed vacuum of eq.~(\ref{eq:TwoModSqVacRFock}) and its infinite
squeezing limit, the \textbf{EPR} state of \textbf{CV} systems~\footnote{Also termed
infinite dimensional} are eigenstates of a nonlocal observable (the difference in
photon number of the two systems).

Other maximally entangled eigenstates, associated to other eigenvalues of Bell
observables can be produced by a local unitary transformation on
eq.~(\ref{eq:FirstBellState});
\begin{equation}\label{eq:BellStatesTra}
  |\Psi_g\rangle_{AB}=\,\widehat{U}_{A}(g)\;|\,\Psi_{0}\,\rangle_{AB}
\end{equation}
In this case \textit{Alice} has performed the operation $\widehat{U}(g)$. To produce
all the possible Bell states, $\widehat{U}(g)$ must belong to a group of unitary
transformations $\{\widehat{U}(g)\}$ having an irreducible unitary representation in
the space of operators acting on $\mathcal{H}$. Schur's lemma for the density
operators of the Bell states takes the form:
\begin{equation}\label{eq:BellPOVM}
  \int dg\:\widehat{U}_{A}(g)\:|\,\Psi_{0}\,
  \rangle_{AB}\:\langle\, \Psi_{0}\,|_{AB}
  \:\widehat{U}_{A}^{\dag}(g)\,=\,d^{-1}\;\hat{\mathbf{1}}_{A}\,\otimes\,\hat{\mathbf{1}}_{B}
\end{equation}
with the integration over the argument $g$ being replaced by a sum whenever the group
is discrete. The factor $d^{-1}$ becomes infinitesimal for \textbf{CV} systems.

The identity in eq.~(\ref{eq:BellPOVM}) is a proof that the \emph{group} consisting
of density operators of Bell states form a Positive Operator Valued
Measure~\cite{NielsenCaves} (abbreviated \textbf{POVM}). It is also a proof that the
Bell states form a complete, orthonormal basis for the Hilbert space
$\mathcal{H}\otimes\mathcal{H}$; eq.~(\ref{eq:BellPOVM}) is a completeness
relationship for Bell basis states.

For the $2$-dimensional (spin $1/2$) Hilbert space of the original teleportation
proposal~\cite{Bennett}, the unitary transformation group consists in the set of the
Pauli operators
$\left\{\,\hat{\mathbf{1}}\,,\,\hat{\sigma}^{(x)}\,,\,\hat{\sigma}^{(y)}\,,\,\hat{\sigma}^{(z)}\,\right\}$.
The Bell basis is given by
\begin{align}
 &|\uparrow\,\rangle_{A}\,|\uparrow\,\rangle_{B}\;+\;e^{i\Theta}\,|\downarrow\,\rangle_{A}\,|\downarrow\,\rangle_{B}
 \notag \\
 &|\uparrow\,\rangle_{A}\,|\downarrow\,\rangle_{B}\;+\;e^{i\Theta}\,|\downarrow\,\rangle_{A}\,|\uparrow\,\rangle_{B}
\label{eq:BellStatesDV}
\end{align}
with $\Theta=0,\pi$. The Bell observables are given by
\begin{align}
 &\hat{\sigma}^{(z)}_{A}\,-\,\hat{\sigma}^{(z)}_{B}\notag\\
 &\hat{\sigma}^{(z)}_{A}\,+\,\hat{\sigma}^{(z)}_{B} \label{eq:BellObsDV}
\end{align}

In a \textbf{CV} setting a $2$-dimensional Hilbert space (of bosonic excitations,
though) can be defined by a truncation of the Fock basis to the first two states
$\{|0\rangle,|1\rangle\}$. The most general superposition state for this truncated
space is of the form
$\cos(\varepsilon)\,|\,0\,\rangle\:+\:e^{i\Theta}\,\sin(\varepsilon)\,|\,1\,\rangle$.
The Bell observables are $\hat{n}_{A}-\hat{n}_{B}$ and $\hat{n}_{A}+\hat{n}_{B}$ and
their eigenstates have a form similar to that of eq.~(\ref{eq:BellStatesDV}).

For a full \textbf{CV} setting the group of unitary transformations producing the
Bell states is that of the displacement operators and the Bell basis is made of the
\textbf{EPR} states of eq.~(\ref{eq:EPRWavefunc}). The state in
eq.~(\ref{eq:FirstBellState}) being the infinitely squeezed two-mode vacuum (see
eq.~(\ref{eq:TwoModSqVacRFock})); having nonlocal quadratures' (Bell observables)
eigenvalues $\hat{p}_{A}+\hat{p}_{B}=0$ and $\hat{x}_{A}-\hat{x}_{B}=0$. The
displacement operation, when performed by either \textit{Alice} or \textit{Bob} will
produce all the other possible values for the nonlocal quadratures. The \textbf{POVM}
for the Bell-basis measurement of Bell operators will be the homodyne measurement
\textbf{POVM}~\cite{DArianoBell}.

For the quantum teleportation protocol, we have a three-party joint system of the
entangled resource $AB$, and the input $in$ with a density operator
$\hat{\rho}_{in}\otimes\hat{\rho}_{AB}$. The teleportation protocol consists
 in the measurement of the Bell observables (projective measurement of the Bell states) over the
modes $A,in$ in \textit{Alice's} possession.

Given the \textbf{POVM} defined in eq.~(\ref{eq:BellPOVM}), a measurement with a
definite result $g$ would result in a state, after measurement (see
eqs.~(\ref{eq:ProjProbDens}),~(\ref{eq:ProjMeasDens}) and~(\ref{eq:ProjMeasMix}) and
the associated discussion of projective measurements);
\begin{equation}\label{eq:DensPOVMMeas}
\mathrm{Tr}\left(\,\hat{\rho}_{in}\,\otimes\,\hat{\rho}_{AB}\:|\,\Psi_{g}
  \,\rangle_{in,A}\:\langle\, \Psi_{g}\,|_{in,A}\,\right)_{A,in}
\end{equation}
which is generally not normalized. To simplify and restrict exposition to Bell
resource teleportation, let the input state be the pure state $|\psi\rangle_{in}$;
let the resource be $|\Psi_{0}\rangle_{AB}$ of eq.~(\ref{eq:FirstBellState}); lastly,
let the basis of representation for the states $\{|\,n\,\rangle\}$ be orthonormal.
The outcome of the projective measurement of eq.~(\ref{eq:DensPOVMMeas}) with a
result $g$ is the state
\begin{equation}\label{eq:PurePOVMMeas}
  d^{\,-1}\:\sum_{n,n'}\,e^{i\,\phi_{n'}\,-\,i\,\phi_{n}\,}\:\langle\, n'\,|_{A}\:\langle
  \,n'\,|_{in}\;\widehat{U}_{in}^{\dag}(g)\;\;|\,n\,\rangle_{A}\:|\,n\,\rangle_{B}\:|\,\psi\,\rangle_{in}=\,d^{\,-1}\:\sum_{n}\:\langle
  \,n\,|\:\widehat{U}^{\dag}(g)\:|\,\psi\,\rangle\:|\,n\,\rangle_{B}
\end{equation}

Given the complete and orthonormal nature of the basis $\{|\,n\,\rangle\}$; the state
in eq.~(\ref{eq:PurePOVMMeas}) is the original input state (now "living" in system
$B$ instead of $in$), on which the transformation $\widehat{U}^{\dag}(g)$ has been
performed.

Note that the Bell state for the outcome of measurement $g$ and the resource state
are related by the transformation
$|\Psi_{g}\rangle_{AB}=\widehat{U}_{A}(g)|\Psi_{0}\rangle_{AB}$. Moreover, the
generalization to a teleportation protocol where the resource state is a Bell state
different from
 $|\Psi_{0}\rangle_{AB}$ is straightforward; as $\{\widehat{U}(g)\}$ is a group, with a group composition law:
\begin{equation}\label{eq:UnitGroupComp}
\widehat{U}(g')\:\widehat{U}(g'')\,=\,\widehat{U}(g'\cdot g'')\:e^{i\Phi(g',g'')}
\end{equation}
where $e^{i\Phi(g',g'')}$ is a phase factor (f.e. the phase in
eq.~(\ref{eq:DispComp}) for displacements); and the operation indicated by $g'\cdot
g''$ is algebraic. The phase factor and the  arguments $g$ under the operation
$\cdot$ are associative, with an inverse element and an identity element. Thus, the
transformations on a Bell state not equal to $|\Psi_{0}\rangle_{AB}$ can be
compounded with ease.

To finish the teleportation protocol, \textit{Bob} must apply the transformation
$\widehat{U}_{B}(g)$ onto the state $B$ in his possession. The result of measurement
is random in a \emph{fundamental} manner; to know which transformation to apply
\textit{Bob} has to receive a communication from \textit{Alice} via a classical
communication channel.

Owing to the random nature of projective measurement, \textit{Bob's} output state
should in reality be an ensemble of all the possible outputs corresponding to
possible results of such measurement (as in eq.~(\ref{eq:ProjMeasMix})). We have
explained teleportation with a Bell state resource, entailing perfect transcription
of the input state for \emph{every measurement result}. For this special case, the
ensemble nature or "mixedness" of the output induced by teleportation does not exist
as all results produce the same output. In a \textbf{CV} setting, and in any other
setting where the protocol does not rely on a Bell state resource and produces
different outputs according to different measurement results, the mixed nature of the
output state is relevant and is to be taken into account. In
refs.~\cite{BraunsteinKimble,VanLoockBraunstein} and~\cite{ChizhovKnollWelsch}
(dealing with \textbf{CV} teleportation) integration over the possible outcomes of
homodyne measurement is performed on the un-normalized output state of
eq.~(\ref{eq:DensPOVMMeas})).

In the \textbf{CV} protocol~\cite{BraunsteinKimble} \textit{Alice} communicates the
results of homodyne measurements of nonlocal quadratures to \textit{Bob}; he applies
this same displacement to system $B$ in his possession. For a review of the
\textbf{CV} teleportation protocol as a projective measurement of pure state wave
functions; see also refs.~\cite{VaidmanTelep,VaidmanAnotherLook,ErezTeleProOp}.

\subsection{The fidelity of teleportation for Continuous Variables}
\label{Back:Qtele:Fide}

The most widely accepted definition~\cite{CriteriaCVTelep} for the \textit{fidelity}
between two quantum states; $\hat{\rho}_{out}$, the output state of quantum
teleportation and $\hat{\rho}_{in}$, the input state is given by the trace of the
product of both density operators. In the the phase-space representations used in
this work, such a fidelity is given by

\begin{align}
  \mathcal{F}\,=&\, \mathrm{Tr}(\hat{\rho}_{in}\,\hat{\rho}_{out}) \label{eq:Fidedens} \\
=&\,\pi^{-1}\,\int d^{2}\alpha\;W_{in}(\alpha)\:W_{out}(\alpha)\label{eq:FideWig}\\
=&\,\pi^{-1}\,\int d^{2}\xi\;\chi_{in}(\xi)\,\chi_{out}(-\xi)\label{eq:FideChar}
\end{align}

The positivity and normalization properties of all density operators (see
eq.~(\ref{eq:TrSqWigChar})) determine that the fidelity is bounded from above and
below: $0\,\leq\,\mathcal{F}\,\leq\,1$. Supposing that at least \emph{one} of the
states is pure; say, $\hat{\rho}_{in}$ (see ref.~\cite{CriteriaCVTelep}), a fidelity
of $1$ means that $\hat{\rho}_{in}=\hat{\rho}_{out}$. A fidelity of $0$ would mean
that the two states are orthogonal.

Generally, the "practical" lower bound to fidelity is taken to be the classical
teleportation threshold or \textit{classical fidelity}, for a given input and a
separable two-party resource; considered a \emph{worst}, limit case of a class of
possible entangled resources. For the classical teleportation of an arbitrary
coherent state by separate two-mode vacuum states~\cite{CriteriaCVTelep}, it is
$\mathcal{F}_{cls}=\frac{1}{2}$.

Such a classical threshold can be calculated for a given input, and a class of
resource states. First, define a separable, "limit" resource state, with density
operator $\hat{\rho}_{A}\otimes\hat{\rho}_{B}$ and an input state $\rho_{in}$ (or an
ensemble of likely inputs). To perform the teleportation protocol a projective
measurement of a Bell state is performed by \textit{Alice} on
$\hat{rho}_{A}\otimes\hat{\rho}_{in}$. Probability for result $g$ given by
\begin{equation}\label{eq:ClassTelepBell}
  P_{class}(g)=\,\mathrm{Tr}(\hat{\rho}_{A}\otimes\hat{\rho}_{in}\;|\,\Psi_{g}
  \,\rangle_{in,A}\:\langle\, \Psi_{g}\,|_{in,A})
\end{equation}
The ensemble state of all the possible "corrections" performed by \textit{Bob} on
state $\hat{\rho}_{B}$ in his possession will be the output state for the
teleportation protocol;
\begin{equation}\label{eq:ClassTelepensemble}
\rho_{out}=\,\sum_{g}\: P_{class}(g)\,
\widehat{U}_{B}(g)\:\hat{\rho}_{B}\:\widehat{U}^{\dag}_{B}(g)
\end{equation}

The \textit{classical fidelity}  is given by eq.~(\ref{eq:Fidedens}); calculated for
the input state $\rho_{in}$ and the output state of
eq.~(\ref{eq:ClassTelepensemble}). This fidelity can be modulated and optimized by
changing constraints on the likely input states' ensemble and on the character of the
limit, separable resource.

%%In every case, the upper bound to the fidelity is given by the
%%purity of the purest state~\cite{????}

\chapter{Characteristic function formalism for \textbf{CV} teleportation}
\label{Form}

The universal procedure for teleportation (see section~\ref{Back:Qtele}) of input
quantum state (designated $in$) using an entangled resource shared by parties
\textit{Alice} and \textit{Bob} (designated $A\,,\,B$) consists on a projective
measurement by \textit{Alice} of $A$ and $in$ joint system onto a Bell state.
Followed by a local unitary operation on $B$ by \textit{Bob}, based on measurement
results communicated by \textit{Alice}.

In this chapter, we will describe quantum teleportation using characteristic
functions as a means of representation. In this representation it is straightforward
to obtain an elegant expression for the teleported state; for \emph{all} (one-mode)
input states and \emph{all} (two-mode) resource states in a \textbf{CV} context, and
to introduce simple modifications~\footnote{Mostly related to realistic, noisy
measurement or preparation of the resource} of the teleportation protocol which
obtain general, elegant expressions for the output state. For these reasons, we speak
of a characteristic function formalism for \textbf{CV} teleportation.

Mixing in external modes by means of beam-splitters, making projective measurements
on the modes arising from such beam-splitters, slightly modifying the nonlocal
measurement that constitutes the basis of quantum teleportation are a few examples of
the possible modifications on the basic \textbf{CV} teleportation protocol that can
be studied with ease with the characteristic function formalism.

\section{Teleportation in the language of Wigner functions}
\label{Form:TeleportWigner}

The maximally entangled states in a \textbf{CV} setting are the \textbf{EPR} states
see (eq.~(\ref{eq:EPRWavefunc})). These are the eigenstates of the nonlocal
quadratures $x_{u}$ and $p_{v}$ defined in eq.~(\ref{eq:XUPVUnitary}). Entangled
states of these quadratures are achieved in \textbf{CV} settings from separable
states by means of beam-splitters (as illustrated in section~\ref{Back:Toys:BS}).

\begin{figure}[bt] \label{fig:telesetup}
\begin{centering}
\includegraphics[width=15cm]{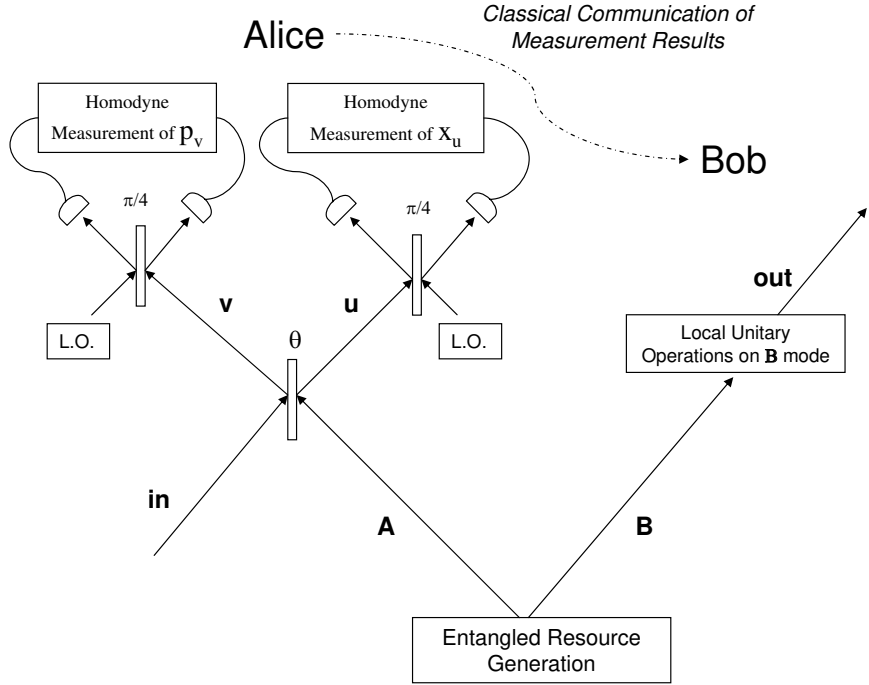}
\end{centering}
\caption[The Experimental Setup for \textbf{CV} Teleportation]{The Experimental Setup
for Continuous Variable Teleportation. \textit{Alice} mixes input mode $in$ with mode
$A$ of the entangled state that is used as a teleportation resource by means of the
beam-splitter "$\theta$". An homodyne measurement of nonlocal, commuting $x_{u}$ ,
$p_{v}$ quadratures is performed, "collapsing" the joint state of modes $A$ and $in$
into an \textbf{EPR} state and transferring the initial state of mode $in$ onto mode
$B$; save for a phase-space displacement corresponding to the actual values of the
quadratures measured. This displacement is corrected for by \textit{Bob} on receiving
a communication of the measurement results from \textit{Alice}.}
\end{figure}

To perform quantum teleportation in this setting~\cite{BraunsteinKimble},
\textit{Alice} must produce an entangled state of modes $A$ and $in$ by means of a
beam-splitter; then she is able to perform measurements of nonlocal
quadratures~\cite{VaidmanYoranTel} by homodyne detection. The experimental setup in
fig.~(\ref{fig:telesetup}) illustrates such a scheme.

Initially, the joint system of input and resource is in a state given by the Wigner
function
\begin{equation}\label{eq:Wignerinitial}
  W_{AB}(\alpha_{A};\,\alpha_{B})\;W_{in}(\alpha_{in})
\end{equation}

The "$\theta$" beam-splitter, with transmission coefficient $\cos(\theta)$
(fig.~(\ref{fig:telesetup})) mixes the modes incoming modes $A$ and $in$, producing
the outcoming modes $u$ and $v$ corresponding to nonlocal phase-space variables
$\alpha_{u}$ and $\alpha_{v}$. The state of the system after the beam-splitter
operation is given by
\begin{equation}\label{eq:WignerBS}
  W_{AB}(\alpha_{A}(\alpha_{u},\alpha_{v}),\:\alpha_{B})\;W_{in}(\alpha_{in}(\alpha_{u},\alpha_{v}))
\end{equation}
where variables
$\alpha_{A}(\alpha_{u},\alpha_{v})\:,\;\alpha_{in}(\alpha_{u},\alpha_{v})$ are given
by the inverse of the transformation in eq.~(\ref{eq:XUPVUnitary}) (see also
eq.~(\ref{eq:BSTransfOp})).

On such a entangled state of $A$ and $in$ modes homodyne measurements of the
nonlocal, commuting quadratures $x_{u}$ and $p_{v}$ are performed; thus a Bell
observable measurement is realized. The state of part $B$ of the system after a
projective measurement with results $x'_{u}$ and $p'_{v}$ is represented by the
partial trace over $A$ and $in$ (see eq.~(\ref{eq:TraceTwoWig}) of the Wigner
function of the system (eq.~(\ref{eq:WignerBS}) and the Wigner function
$W_{x'_{u},p'_{v}}$ (eq.~(\ref{eq:WignerEPR})) of the respective
element~\footnote{Namely, the density operator of the \textbf{EPR} state with
eigenvalues $x'_{u},p'_{v}$.} of the Bell measurement;
\begin{align}
W_{measured}(\alpha_{B}\,;\,x'_{u},p'_{v}) = \pi^{-2}\:\int d^{2}\alpha_{in}\,
d^{2}\alpha_{A}\;
W_{AB}(\alpha_{A}(\alpha_{u},\alpha_{v}),\:\alpha_{B})\;W_{in}(\alpha_{in}(\alpha_{u},\alpha_{v}))\notag \\
\times \, \mathcal{C} \:\delta (\cos(\theta)x_{in}\,-\,\sin(\theta)x_{A}\,-\,x'_{u})
\;\delta(\sin(\theta)p_{in}\,+\,\cos(\theta)p_{A}\,-\,p'_{v})
\label{eq:WignerProjMeas}
\end{align}

$W_{measured}(\alpha_{B};\:x'_{u},p'_{v})$ in eq.~(\ref{eq:WignerProjMeas}) is not a
normalized Wigner function (as is the case in eqs.~(\ref{eq:ProjMeasDens})
and~(\ref{eq:DensPOVMMeas})). It is the product of the probability
$\mathrm{P}(x'_{u},p'_{v})$ of a measurement result $x'_{u}\,,\,p'_{v}$ and the
Wigner function of the state of the $B$ part of the system after a measurement giving
such a result; given by the conditional pseudo-probability $W_{measured}(
\alpha_{B}\,|\:x'_{u},p'_{v})$. The probability can be obtained from
eq.~(\ref{eq:WignerProjMeas}) by tracing out the $B$ mode;
\begin{equation}\label{eq:WignerProbMeas}
\mathrm{P}(x'_{u},p'_{v})\,=\,\pi^{-1}\:\int
d^{2}\alpha_{B}\:W_{measured}(\alpha_{B};\:x'_{u},p'_{v})
\end{equation}

Perform the integral over $\alpha_{A}$ in eq.~(\ref{eq:WignerProjMeas}), yielding a
convolution integral of the resource and the input states;
\begin{align}
W_{measured}(\alpha_{B};\:x'_{u},p'_{v})\,=\,\frac{\pi^{-2}\mathcal{C}}{|\sin(\theta)|\:|\cos(\theta)|}\:\int
d^{2}\alpha_{in}\:W_{in}(\alpha_{in})\notag \\
\times
W_{AB}(x_{in}\cot(\theta)\,-\,x'_{u}\csc(\theta)\,+\,i\,\left(p_{v}\sec(\theta)\,-\,p_{in}\tan(\theta)\right);\:\alpha_{B})
\label{eq:WignerProjMeas2}
\end{align}

Up to this point, we have not discussed the operations that \textit{Bob} must perform
on $B$ mode. Only to see clearly what these might be, we will consider an ideal
resource, the \textbf{EPR} state of Wigner function
\begin{equation}\label{eq:WignerABEPRSimp}
  W_{AB}(\alpha_{A};\:\alpha_{B})=\,2 \mathcal{C}\:
  \delta(x_{A}\,-\,x_{B})\;\delta(p_{A}\,+\,p_{B})
\end{equation}

Substituting this resource in eq.~(\ref{eq:WignerProjMeas2} and integrating yields
the resulting, un-normalized Wigner function;
\begin{equation}\label{eq:WignerProjMeasSimp}
  \frac{2\,\pi^{-2}\,\mathcal{C}^{2}}{|\sin(\theta)|\:|\cos(\theta)|}
  W_{in}(x_{B}\tan(\theta)\,+\,x'_{u}\sec(\theta)\,+\,i\,(p_{B}\cot(\theta)\,+\,p'_(v)\csc(\theta)))
\end{equation}
where the probability of obtaining the arbitrary result $x'_{u},p'_{v}$ of
measurement is easily seen to be the vanishing constant
$2\,\mathcal{C}^{2}\,\pi^{-2}\;(|\sin(\theta)|\:|\cos(\theta)|)^{-1}$.

To recover the input state at his end, \textit{Bob} must perform two unitary
operations on the $B$ mode; one of them \emph{after} knowing the results of the
measurement, communicated by \textit{Alice} via a classical channel. Namely,
\begin{itemize}
\item A displacement in phase-space of $(x'_{u}\sec(\theta)\,+\,i\,
p'_{v}\csc(\theta))$. For homodyne measurement apparatus with gain coefficients
$0\,<\,g_{x}\,\leq\,1$ and $0\,<\,g_{p}\,\leq\,1$ for $x_{u}$ and $p_{v}$,
respectively; this displacement will be $(g_{x} x'_{u}\sec(\theta)+i\,g_{p}
p'_{v}\csc(\theta))$. Thus the operation is $\widehat{D}((g_{x}
x'_{u}\sec(\theta)+i\,g_{p} p'_{v}\csc(\theta)))$.

\item A transformation $x_{B}\,+\,i\,p_{B}\longrightarrow a x_{B}\,+\,i \, a^{-1}
p_{B}$, with $a\equiv\,\cot(\theta)$. Therefore, is is a squeezing operation
$\widehat{S}(r)$, where $e^{-r}\,=\,\cot(\theta)$. Obviously, $\cot(\theta)$ is part
of the description of the beam-splitter inside our experimental setup and can be
known beforehand by \textit{Alice} and \textit{Bob}.
\end{itemize}

The complete unitary operation applied by Bob would be given by the operator
\begin{align}
  \widehat{U}_{B}=&\,\widehat{S}_{B}(-\ln(\cot(\theta)))\;\widehat{D}_{B}((x'_{u}\sec(\theta)\,+\,i\,
p'_{v}\csc(\theta))) \notag \\
=&\,\widehat{D}_{B}((x'_{u}\csc(\theta)\,+\,i\,
p'_{v}\sec(\theta)))\;\widehat{S}_{B}(-\ln(\cot(\theta))) \label{eq:BobUnitary}
\end{align}

The squeezing operation performed by \textit{Bob} can be made unnecessary. Consider a
teleportation resource of the form
\begin{equation}\label{eq:WignerABEPRSquee}
  W_{AB}(\alpha_{A};\:\alpha_{B})=\,\mathcal{C}\,
  \delta(x_{A}\tan(\theta)\,-\,x_{B})\delta(p_{A}\cot(\theta)\,+\,p_{B})
\end{equation}
instead of that of eq.~(\ref{eq:WignerABEPRSimp}). This would give the end result
\begin{equation}\label{eq:WignerProjMeasSquee}
  \frac{\pi^{-2}\,\mathcal{C}^{2}}{|\sin(\theta)|\:|\cos(\theta)|}
  W_{in}(x_{B}\,+\,x'_{u}\sec(\theta)\,+\,i\,(p_{B}\,+\,p'_(v)\csc(\theta)))
\end{equation}
thus, in this case, \textit{Bob} must perform \emph{only} the displacement above
described to recover the input state and realize teleportation.

Note that a squeezing transformation $x_{A}+i\,p_{A}\longrightarrow \cot(\theta)
x_{A}+i \, \tan(\theta) p_{A}$ (identical to \textit{Bob's} squeezing "correction")
on mode $A$ of the resource in eq.~(\ref{eq:WignerABEPRSimp}) would result in the
resource of eq.~(\ref{eq:WignerABEPRSquee}). A displacement of
$(x'_{u}\csc(\theta)+i\, p'_{v}\sec(\theta))$ on mode $A$ of the resource in
eq.~(\ref{eq:WignerABEPRSquee}) would yield the \textbf{EPR} state
$W_{x'_{u},p'_{v}}$, identical to the state of the $A$ and $in$ modes \emph{after}
the Bell measurement with results $x'_{u},p'_{v}$. The unitary transformations on
mode $A$ described above; transforming the \textbf{EPR} state in
eq.~(\ref{eq:WignerABEPRSimp}) into the \textbf{EPR} state $W_{x'_{u},p'_{v}}$ of
eq.~(\ref{eq:WignerEPR}) are identical to those applied by \textit{Bob} to mode $B$
to realize teleportation (in eq.~(\ref{eq:BobUnitary}).

We have just shown that teleportation using an asymmetric beam-splitter of angle
$\theta$ together with the simplest maximally entangled resource
(eq.~(\ref{eq:WignerABEPRSimp}), will result in added squeezing on the output state.
That squeezing will have to be corrected by the application of the inverse squeezing
transformation. Previously applying this inverse transformation to mode $A$ of the
resource will eliminate the need for such a correction by \textit{Bob}.

%%Equivalently, if the entangled resource is produced by mixing two
%%separate, squeezed modes $a$ (squeezing parameter $r_{a},
%%r_{a}\geq 0$) and $b$ (squeezing parameter $-r_{b},r_{b}\geq 0$)
%%with a symmetric beam-splitter:

%%\begin{equation}\label{}
%%  W_{AB}(\alpha_{A};\alpha_{B})=W_{a}
%%\end{equation}
%%incomplete

The use of an asymmetric beam-splitter for mixing the $A$ and $in$ modes would be
desirable only if the teleportation protocol were intended to produce an squeezed
output. The effect of having an asymmetric beam-splitter in the experimental setup
can be mimicked by appropriate local transformations on the resource state.

In order to simplify exposition, and assuming that we have no further use for
\emph{additional squeezing} of the output state we will take the beam-splitter
"$\theta$" to be symmetric ($\theta=\pi / 4$) and without a phase. The use of a
symmetric beam-splitter with a phase would only rotate in phase-space the
displacement to be performed by \textit{Bob}.

Using a symmetric beam-splitter, and after \textit{Bob's} correction (the
displacement in eq.~(\ref{eq:BobUnitary})) we have for the output state of the system
(see eq.~(\ref{eq:WignerProjMeas2}));

\begin{align}
  W_{out}(\alpha_{B};\:x'_{u},p'_{v})=\, 2 \, \pi^{-2}\,\mathcal{C}\:\int
d^{2}\alpha_{in}\: W_{in}(\alpha_{in}) \notag \\
\times\,W_{AB}(x_{in}\,-\,2^{1\!/2}\, x'_{u}\,+\,i\,(2^{1\!/2}\,
p'_{v}\,-\,p_{in});\:\alpha_{B}\,-\,(2^{1\!/2}\,g_{x}\,
x'_{u}\,+\,i\,2^{1\!/2}\,g_{p}\,p'_{v})) \label{eq:WignerOutNotMix}
\end{align}

This output of teleportation is un-normalized, and so far dependent on the outcome
$x'_{u},p'_{v}$ of the Bell measurement. It is the product of a probability for such
an outcome and a conditional Wigner function (see eq.~(\ref{eq:WignerProjMeas}).
Given
\begin{itemize}
\item the fundamental randomness of the results of quantum measurement,
particularly Bell measurement of $x'_{u},p'_{v}$.
\item that teleportation is, in principle performed in an absence of \emph{any} knowledge (by \textit{Alice} and \textit{Bob})
of the input state; thus in the absence of any knowledge, even statistic, of
measurement results.
\item that teleportation is performed in an automatic manner by \textit{Alice} and \textit{Bob}, without change
to the experimental setup due to the knowledge of particular set of measurement
results.
\item that for a realistic fidelity of teleportation, it is necessary to consider all the random outcomes of measurement, modulated by their probability;
a fidelity coefficient depending on a single random result is not acceptable because
it cannot be repeated reliably.
\item that a conditional output state, on a random result is not an acceptable answer for a teleportation output, as it
comes about randomly.
\end{itemize}
it becomes evident that a final output state that is acceptable is given by a mixture
of conditional states (see eq.~(\ref{eq:WignerProjMeas})) with the probability of
measurement in eq.~(\ref{eq:WignerProbMeas}). Therefore, integration over $x'_{u}$
and $p'_{v}$ of eq.~(\ref{eq:WignerOutNotMix}) will yield a normalized Wigner
function that is an ensemble of conditional output states corresponding to individual
measurement results,
\begin{align}
  W_{out}(\alpha_{B})=&\, \mathcal{C}^{-1}\:\int
dx'_{u} \, dp'_{v}\; W_{out}(\alpha_{B};\:x'_{u},p'_{v}) \notag \\
=& \, 2\, \pi^{-2}\, \int
d^{2}\alpha_{in}\;  W_{in}(\alpha_{in})\notag \\
\times\, \int dx'_{u} \, dp'_{v}&\; W_{AB}(x_{in}\,-\,2^{1\!/2}\, x'_{u}\,+\,
i\,(2^{1\!/2}\, p'_{v}\,-\,p_{in});\:\alpha_{B}\,-\,(g_{x}\,2^{1\!/2}\,
x'_{u}\,+\,i\,g_{p}\,2^{1\!/2}\, p'_{v}))\label{eq:WignerOutMix}
\end{align}

%%where $\mathcal{C}^{-1}$ is the infinitesimal constant, equal to the area of the phase-space support of
%%an \textbf{EPR} state.

The Wigner function $W_{out}(\alpha_{B})$ is the outcome of the convolution of the
Wigner function of the input $W_{in}$ and a bipartite (on $in$ and $B$ modes) Wigner
function given by
\begin{equation}\label{eq:WignerTeleKernel}
  K(\alpha_{in};\alpha_{B})=2\,\pi^{-1} \int dx'_{u} \, dp'_{v}\; W_{AB}(x_{in}\,-\,2^{1\!/2}\,
x'_{u}\,+\,i\,(2^{1\!/2}\, p'_{v}\,-\,p_{in});\:\alpha_{B}\,-\,(g_{x}\,2^{1\!/2}\,
x'_{u}\,+\,i\,g_{p}\,2^{1\!/2}\,p'_{v}))
\end{equation}
named the \emph{teleportation kernel} in the
literature~\cite{ChizhovKnollWelsch,Furasek}. It is easily seen that the kernel is
the Wigner function of an ensemble (with constant, flat probability) of
\emph{Transfer Operators}~\cite{UnivTeleportTwist,TelepFormal3} (one for each value
of $x'_{u}\,,\,p'_{v}$), having the Wigner function
\begin{equation}\label{eq:WigTransfer}
  W_{AB}(x_{in}\,-\,2^{1\!/2}\,
x'_{u}\,+\,i\,(2^{1\!/2}\, p'_{v}\,-\,p_{in});\:\alpha_{B}\,-\,(g_{x}\,2^{1\!/2}\,
x'_{u}\,+\,i\,g_{p}\,2^{1\!/2}\,p'_{v}))
\end{equation}

Let the characteristic function $\chi_{out}(\xi_{B})$ be the Fourier transform of
$W_{out}(\alpha_{B})$ (see eq.~(\ref{eq:FourierWig});
\begin{align}
  \chi_{out}(\xi_{B})=\, 2\, \pi^{-3}\:\int d^{2}\alpha_{B}\,\int d^{2}\alpha_{in}\,\int dx'_{u} \, dp'_{v}
  \:e^{2i(x_{B}z_{B}\,-\,p_{B}w_{B})} \notag  \\
  \times W_{in}(\alpha_{in})\; W_{AB}(x_{in}\,-\,2^{1\!/2}\,
x'_{u}\,+\,i\,(2^{1\!/2}\, p'_{v}\,-\,p_{in});\:\alpha_{B}-(g_{x}\,2^{1\!/2}\,
x'_{u}\,+\,i\,g_{p}\,2^{1\!/2}\, p'_{v})) \label{eq:WignerCharOut1}
\end{align}

Making the substitution $x''_{u}\equiv\,2^{1\!/2}\,x'_{u}$,
$p''_{v}\equiv\,2^{1\!/2}\,p'_{v}$, and multiplying the integrand in
eq.~(\ref{eq:WignerCharOut1}) by $e^{\pm\,2\,i\,(g_{x}\,z_{B}\,x''_{u}\,-\,
g_{p}\,w_{B}\,p''_{v} )}$ we obtain
\begin{align}
\chi_{out}(\xi_{B})=&\,\pi^{-1}\,\int d^{2}
\alpha_{in}\;W_{in}(\alpha_{in})\;\pi^{-1}\,\int dx''_{u}
\,dp''_{v}\: e^{2\,i\,(z_{B}\,g_{x}\,x''_{u}\,-\,w_{B}\,g_{p}\,p''_{v})} \notag \\
&\times\, \pi^{-1}\,\int d^{2} \alpha_{B}\:
e^{2\,i\,(z_{B}\,(x_{B}\,-\,g_{x}\,x''_{u})\,-\,w_{B}\,(p_{B}\,-\,g_{p}\,p''_{v}))}
\notag
\\
&\times\,
W_{AB}(x_{in}\,-\,x''_{u}\,+\,i\,(p''_{v}\,-\,p_{in});\:\alpha_{B}\,-\,(g_{x}\,x''_{u}+\,i\,g_{p}\,p''_{v}))
\label{eq:WignerCharOut2}
\end{align}

This is an explicit Fourier transformation over the $\alpha_{B}$,
$x''_{u}\,+\,i\,p''_{v}$ variables; as well as a convolution over $\alpha_{in}$.
Therefore, the characteristic function of the output state is straightforward to
calculate:
\begin{equation}\label{eq:WignerCharOutFin}
\chi_{out}(\xi_{B})=\,\chi_{AB}(g_{p}\,w_{B}\,-\,i\,g_{x}\,z_{B};\:\xi_{B})\;\chi_{in}(g_{p}\,w_{B}\,+\,i\,g_{x}\,z_{B})
\end{equation}
A result equivalent to that obtained in ref.~\cite{MarianCVTelep} for a symmetric
beam-splitter using the transfer operator formalism.

Keeping the beam-splitter asymmetric, and having \textit{Bob} perform the squeezing
operation described in eq.~(\ref{eq:BobUnitary}) will result in output state
\begin{equation}\label{eq:WignerCharOutAsym}
\chi_{out}(\xi_{B})=\,\chi_{AB}(g_{p}\,w_{B}\,-\,i\,g_{x}\,z_{B}\,;\:\tan(\theta)\,w_{B}\,+\,i\,\cot(\theta)\,z_{B})\;\chi_{in}(g_{p}\,\tan(\theta)\,w_{B}\,+\,i\,g_{x}\,\cot(\theta)\,z_{B})
\end{equation}

\section{Teleportation in the language of characteristic functions}
\label{Form:TeleportIdeal}

The derivation of the \textbf{CV} teleportation output made in the previous section
will now be repeated in the characteristic function representation and will be shown
to be much simpler. The experimental setup for teleportation is illustrated, as
before, in fig.~(\ref{fig:telesetup}). Let the initial state of the joint system
$A,B,in$ be described by their characteristic functions
\begin{equation}\label{eq:Charinitial}
  \chi_{AB}(\xi_{A};\:\xi_{B})\;\chi_{in}(\xi_{in})
\end{equation}

There is a (symmetric) beam-splitter transformation from this initial state into the
modes $u$ and $v$ and their associated quadratures, which, it can be seen easily (see
eq.~(\ref{eq:FourierWig})), transforms the conjugate phase-space variables
$\xi_{in},\xi_{A}$ in a similar manner to eq.~(\ref{eq:XUPVUnitary}) to
$\xi_{u}\,\,xi_{v}$. We will consider the \textbf{EPR} state's characteristic
function (eq.~(\ref{eq:CharEPR})) in terms of the variables $\xi_{in},\xi_{A}$. And
realize the partial trace, or projection into the \textbf{POVM} element (for results
$x'_{u},p'_{v}$) thus,
\begin{align}
\chi_{measured}(\xi_{B};\:x'_{u},p'_{v})=\,&\pi^{-2}\,\int\,
d^{2}\xi_{in}\,d^{2}\xi_{A}\;\chi_{AB}(\xi_{A};\:\xi_{B})\;\chi_{in}(\xi_{in})\;\chi_{x'_{u},p'_{v}}(-\xi_{in}\,,\,-\xi_{A})
\notag \\
=&\pi^{-2}\,\int\,d^{2}\xi_{in}\,d^{2}\xi_{A}\;\chi_{AB}(\xi_{A};\:\xi_{B})\;\chi_{in}(\xi_{in})\notag
\\
&\times\, 2\,
\mathcal{C}\,\delta(w_{in}\,-\,w_{A})\:\delta(z_{in}\,+\,z_{A})\:e^{-2\,i\,\left(z_{in}\,2^{1\!/2}\,x'_{u}\,-\,w_{in}\,2^{1\!/2}\,p'_{v}\right)}
\label{eq:CharProjMeas}
\end{align}

This characteristic function is, like its Wigner function equivalent in
eq.~(\ref{eq:WignerProjMeas}), un-normalized. It is the product of the probability of
measurement $\mathrm{P}(x'_{u},p'_{v})$ of a result $x'_{u},p'_{v}$ and the
\emph{conditional} characteristic function of the system $\chi_{measured}(
\xi_{B}|x'_{u},p'_{v})$, on the aforementioned results. Tracing out the $B$ mode in
eq.~(\ref{eq:CharProbMeas}) will give us the probability of measurement;
\begin{equation}\label{eq:CharProbMeas}
\mathrm{P}(x'_{u},p'_{v})=\,\pi^{-1}\,\int
d^{2}\xi_{B}\;\chi_{measured}(\xi_{B};\:x'_{u}\,,\,p'_{v})
\end{equation}

With the purpose of having a look into \textit{Bob's} part in the teleportation
protocol, we will consider the resource to be in a simple \textbf{EPR} state of the
form
\begin{equation} \label{eq:CharEPRSimp}
\chi_{AB}(\xi_{A};\:\xi_{B})=\,2\,\mathcal{C}\:\delta(w_{A}-w_{B})\:\delta(z_{A}+z_{B})
\end{equation}

Performing integration of eq.~(\ref{eq:CharProjMeas}) with the aforementioned
resource yields
\begin{equation}\label{eq:CharProjMeasSimp}
  \chi_{measured}(\xi_{B};\:x'_{u},p'_{v})=4\,\pi^{-2}\,\mathcal{C}^{2}\,\chi_{in}(\xi_{B})\,
  e^{\,2\,i\,\left(w_{B}\,2^{1\!/2}\,p'_{v}\,-\,z_{B}\,2^{1\!/2}\,x'_{u}\right)}
\end{equation}

which is a product of the characteristic function of the input state $\chi_{in}$ (in
mode $B$), with an additional phase-space displacement of
$-2^{1\!/2}x'_{u}\,-\,i2^{1\!/2}p'_{v}$; and a constant probability for every single
result of measurement of $4\pi^{-2}\mathcal{C}^{2}$. This result is consistent with
that of eq.~(\ref{eq:WignerProjMeasSimp}) (for $\theta=\pi/4$). Thus, \textit{Bob}
must perform that which, to his knowledge (given the non-unit gain of the apparatus)
is the opposite displacement operation,
$\widehat{D}_{B}(g_{x}\,2^{1\!/2}\,x'_{u}\,+\,i\,g_{p}\,2^{1\!/2}p'_{v})$ to recover
the input state;
\begin{equation}\label{eq:CharOutNotMix}
\chi_{out}(\xi_{B};\:x'_{u},p'_{v})\,=\,e^{-2\,i\,\left(w_{B}\,g_{p}\,2^{1\!/2}p'_{v}\,-\,z_{B}\,g_{x}\,2^{1\!/2}x'_{u}\right)}\;\chi_{measured}(\xi_{B};\:x'_{u},p'_{v})
\end{equation}

For the same reasons and on the same considerations exposed in the previous section;
an ouput state conditional on a random measurement result is not acceptable, while an
ensemble of conditional states with adequate probability is an acceptable output
state. Therefore, integrate eq.~(\ref{eq:CharOutNotMix}) over $x'_{u}$ and $p'_{v}$
to obtain the normalized, ensemble state
\begin{align}
\chi_{out}(\xi_{B}) =&\, \mathcal{C}^{-1}\,\int dx'_{u} \,
dp'_{v}\;\chi_{out}(\xi_{B};\:x'_{u},p'_{v}) \notag\\
=&\,2\,\pi^{-2}\, \int dx'_{u}\,dp'_{v}\:\int\,
d^{2}\xi_{in}\:d^{2}\xi_{A}\;e^{-\,2\,i\,\left(w_{B}\,g_{p}\,2^{1\!/2}p'_{v}\,-\,z_{B}\,g_{x}\,2^{1\!/2}x'_{u}\right)}\notag \\
&\times\;\delta(w_{in}\,-\,w_{A})\:\delta(z_{in}\,+\,z_{A})\:e^{-\,2\,i\,\left(z_{in}\,2^{1\!/2}x'_{u}\,-\,w_{in}\,2^{1\!/2}p'_{v}\right)}\notag \\
&\times \;\chi_{AB}(\xi_{A};\:\xi_{B})\;\chi_{in}(\xi_{in}) \label{eq:CharOutMix}
\end{align}

The integration of eq.~(\ref{eq:CharOutMix}) is entirely straightforward, giving the
output state
\begin{equation}\label{eq:CharOutFin}
  \chi_{out}(\xi_{B})=\,\chi_{in}(g_{p}w_{B}\,+\,i\,g_{x}z_{B})\;\chi_{AB}(g_{p}w_{B}\,-\,ig_{x}z_{B};\:\xi_{B})
\end{equation}

which is identical to that obtained in eq.~(\ref{eq:WignerCharOutFin}), after a much
shorter and more elegant calculation.

The formalism just outlined is general for any combination of resource and input
states and gives a very simple expression for the output of Quantum Teleportation in
\textbf{CV}. It is possible to use resources that are mixed states, reflecting the
results of a conditional operation performed during the preparation of said resource.
Or to construct input states that are mixtures of the states (or likely
superpositions thereof) used to encode qudits in quantum information processing with
adequate probabilities; thus constructing the general input ensemble of a quantum
\textbf{CV} channel of teleportation, for which the fidelity can be calculated.

A basic analysis of eq.~(\ref{eq:CharOutFin}) shows that obtaining a result other
than a random, "classical" guess depends on having a high measurement gain
($g_{x},g_{p}\approx 1$) and on having a resource characteristic function
$\chi_{AB}(g_{p}w_{B}\,-\,ig_{x}z_{B};\xi_{B})$ that is nearly constant. This last
condition is fulfilled by states approximating the \textbf{EPR} state of
eq.~(\ref{eq:CharEPRSimp}). For example two-mode squeezed vacuums at high squeezing
$r$; or other states showing great similarity with a two-mode squeezed vacuum at high
squeezing.

\section{Teleportation: a projective measurement onto a mixed state}
\label{Form:TelepMix}

We have produced a simple formalism for the representation of the elementary
transformations and projective measurements performed in \textbf{CV} teleportation
and produced a compact expression for the characteristic function of a general output
state, for all resource states. It is conceivable that the first and most obvious
change in the teleportation protocol involves the \textbf{EPR} state onto which we
project to represent an homodyne measurement (see section~\ref{Back:Qtele} and
eq.~(\ref{eq:CharProjMeas})).

The first choice if we are interested in the introduction of "imperfect" homodyne
measurements would be a suitable mixture of \textbf{EPR} states. What would a mixture
imply? That we are still doing an homodyne measurement of the variables $x_{u},p_{v}$
and projecting onto an \textbf{EPR} state. We do not know \emph{which} state
precisely, even if the apparatus for homodyne detection returns a result
$x'_{u},p'_{v}$. If the apparatus is imprecise (not damaged or lacking calibration)
the distribution of outcomes will be centered on the values returned. We can define
such a projecting state as the mixture
\begin{align}
  \hat{\rho}_{mix}=&\,\pi^{-1}\,\int\,dx_{-}\,dp_{+}\;
  \mathrm{P}(x_{-},p_{+};\:x'_{u},p'_{v})\;\hat{\rho}_{x_{-}\,,\,p_{+}} \label{eq:MixDensEPR} \\
  \chi_{mix}(\xi_{in},\xi_{A})=&\,\pi^{-1}\,\int\,dx_{-}\,dp_{+}\;
  \mathrm{P}(x_{-},p_{+};\:x'_{u},p'_{v})\;\chi_{x_{-},p_{+}}(\xi_{in},\xi_{A}) \label{eq:MixCharEPR}
\end{align}
where $\chi_{x_{-},p_{+}}(\xi_{in},\xi_{A})$ is the \textbf{EPR} state of
eq.~(\ref{eq:CharEPR}) ($\Theta=\pi/4$) for eigenvalues $x_{-}$ and $p_{+}$. The
probability distribution $\mathrm{P}(x_{-},p_{+};x'_{u},p'_{v})$ is required to
fulfill the following conditions if the state in eq.~(\ref{eq:MixCharEPR}) is to
represent an imprecise measuring apparatus;
\begin{align}
  &\pi^{-1}\,\int\,dx_{-}\,dp_{+}\;
  \mathrm{P}(x_{-},p_{+};\:x'_{u},p'_{v})=1 \notag \\
    &\mathrm{P}(x_{-},p_{+};\:x'_{u},p'_{v})=\mathrm{P}(x_{-}-x'_{u}\,,\,p_{+}-p'_{v}) \notag \\
   &\bar{x}_{-}=\,x'_{u} \;\;\;\; \bar{p}_{+}=\,p'_{v}
  \label{eq:MixProbCond}
\end{align}
namely to be normalized and "centered" around $x'_{u}\,,\,p'_{v}$.

The mixture of \textbf{EPR} states of eq.~(\ref{eq:MixCharEPR}) for such a
probability distribution is given by
\begin{align}
 \chi_{mix}(\xi_{in},\xi_{A})=\;&\pi^{-1}\,\int\,dx_{-}\,dp_{+}
  \mathrm{P}(x_{-}-x'_{u},p_{+}-p'_{v})\,2\,\mathcal{C}\,e^{2i\left(z_{in}2^{1\!/2}x_{-}\,-\,w_{in}2^{1\!/2}p_{+}\right)}\notag \\
  &\times\delta(z_{in}+z_{A})\delta(w_{in}-w_{A}) \label{eq:MixCharEPR2}
\end{align}

Let us define the Fourier transform of $\mathrm{P}(x,p)$, the characteristic function
\begin{equation}\label{eq:MixProbFourier}
  \widetilde{\mathrm{P}}(w,z)= \,\pi^{-1}\:\int\,dx\,dp\;e^{2\,i\,\left(z\,x\,-\,w\,p\right)}\;\mathrm{P}(x,p)
\end{equation}

Using the definition of eq.~(\ref{eq:MixProbFourier}), we can write the
characteristic function in eq.~(\ref{eq:MixCharEPR2}) in a more elegant manner;
\begin{equation}\label{eq:MixCharEPR3}
  \chi_{mix}(\xi_{in},\xi_{A})=\,\widetilde{\mathrm{P}}(2^{1\!/2}w_{in}\,,\,2^{1\!/2}z_{in})
  \;2\,\mathcal{C}\:\delta(z_{in}\,+\,z_{A})\:\delta(w_{in}\,-\,w_{A})\;e^{2\,i\,\left(z_{in}\,2^{1\!/2}x'_{u}\,-\,w_{in}\,2^{1\!/2}p'_{v}\right)}
\end{equation}

To study the corrections to be made by \textit{Bob} we substitute the mixture state
eq.~(\ref{eq:MixCharEPR3}) into eq.~(\ref{eq:CharProjMeas}) in place of the
\textbf{EPR} state $\chi_{x'_{u},p'_{v}}$; and take the resource to be an
\textbf{EPR} state (eq.~(\ref{eq:CharEPRSimp})), yielding an output state
\begin{equation}\label{eq:MixCharMeasEPR}
\chi_{measured,mix}(\xi_{B};\:x'_{u},p'_{v})\,=\,
4\,\pi^{-2}\,\mathcal{C}^{2}\;\widetilde{\mathrm{P}}(-2^{1\!/2}w_{B}\,,\,-\,2^{1\!/2}z_{B})\;\chi_{in}(\xi_{B})\;e^{\,2\,i\,\left(w_{B}\,2^{1\!/2}p'_{v}\,-\,z_{B}\,2^{1\!/2}x'_{u}\right)}
\end{equation}

We have the product of the constant probability for a given result equal to that in
eq.~(\ref{eq:CharProjMeasSimp}); of the characteristic function
$\widetilde{\mathrm{P}}$; and of the characteristic function of the input state,
displaced in phase-space by $-(x'_{u}2^{1\!/2}\,+\,ip'_{v}2^{1\!/2})$. \textit{Bob}
will try to correct for this displacement by applying at his end the displacement
$\widehat{D}_{B}(g_{x}\,2^{1\!/2}\,x'_{u}\,+\,i\,g_{p}\,2^{1\!/2}p'_{v})$ on $B$
mode. Again, the apparatus is assumed to have non-unit gains $g_{x}$ and $g_{p}$.

Proceeding in the same manner of section~\ref{Form:TeleportIdeal}; performing the
displacement just described on the characteristic function in
eq.~(\ref{eq:CharOutNotMix}), and taking the final output state to be an ensemble of
outcomes corresponding to measurement results $x'_{u},p'_{v}$ (see
eq.~(\ref{eq:CharOutMix})) will result in the final output state
\begin{equation}\label{eq:MixCharFin}
\chi_{out,mix}(\xi_{B})=\,\widetilde{\mathrm{P}}(-2^{1\!/2}g_{p}\,w_{B}\,,\,-2^{1\!/2}g_{x}\,z_{B})\;\chi_{in}(g_{p}\,w_{B}\,+\,i\,g_{x}\,z_{B})\;\chi_{AB}(g_{p}\,w_{B}\,-\,i\,g_{x}\,z_{B};\:\xi_{B})
\end{equation}
This is the same characteristic function of the output state in
eq.~(\ref{eq:CharOutFin}) multiplied by the characteristic function
$\widetilde{\mathrm{P}}(-2^{1\!/2}g_{p}\,w_{B}\,,\,-2^{1\!/2}g_{x}\,z_{B})$.

The output state of the previous section is thus "smeared" in phase-space. This is
easily seen in the Wigner functions' language; as this product of characteristic
functions is the Fourier transform of the convolution integral between the
probability $\mathrm{P}$ and the pseudo-probability $W_{in}$. For example, a nearly
constant characteristic function $\widetilde{\mathrm{P}}$ (the Fourier transform of a
sharply peaked function $\mathrm{P}(x,p)\sim\,\delta(x)\:\delta(p)$) will yield an
output state in eq.~(\ref{eq:MixCharFin}) approximating the ideal output state of
eq.~(\ref{eq:CharOutFin}).

The natural choice for a probability distribution $\mathrm{P}(x,p)$ intended to
represent a measurement apparatus producing outcomes (as far as regards the
projection caused by the measurement) with a random deviation from the "real", mean
values $x'_{u},p'_{v}$ is the Gaussian distribution;
\begin{align}
 \mathrm{P}(x,p)=& \;e^{r+s}\,e^{-e^{2r}x^{2}\,-\,e^{2s}p^{2}}
 \label{eq:MixProbGauss}\\
 \widetilde{\mathrm{P}}(w,z)=&\; e^{-e^{-2r}z^{2}\,-\,e^{-2s}w^{2}}
 \label{eq:MixCharGauss}
\end{align}
producing a multiplying factor in eq.~(\ref{eq:MixCharFin}) of
\begin{equation}\label{eq:MixCharMult}
  e^{-\left(\,2\,e^{-2r}\,g_{x}^{2}\,z_{B}^{2}\,\right)}\;e^{-\left(\,2\,e^{-2s}\,\,g_{p}^{2}\,w_{B}^{2}\,\right)}
\end{equation}

Reducing the measurement gains $g_{x},g_{p}$ might improve the output state, for
input characteristic functions $\chi_{in}(g_{p}w_{B}\,+\,ig_{x}z_{B})$ that become
constant at a much slower rate than a Gaussian distribution when the variables
$g_{x}z_{B} \, , \, g_{p}w_{B}$ grow small, or for input characteristic functions
with a wide support, the Fourier transforms of narrow Wigner functions.

\section{Mixed teleportation resources: "superimposed" over noisy environments}
\label{Form:MixRes}

We have calculated the characteristic function for a mixture of \textbf{EPR} states
(eq.~(\ref{eq:MixCharEPR3}) in section~\ref{Form:TelepMix}. In doing so, we have
stated some properties of the mixture probability (eq.~(\ref{eq:MixProbCond})) that
ensure consistency with our hypothesis of an imperfect measurement and greatly
simplify the form of the output of teleportation when the projective measurement is
done onto this mixture (eq.~(\ref{eq:MixCharFin}).

In this section; we will derive the characteristic function of a teleportation
resource that is a mixture of pure states \emph{in} phase-space. Our first purpose is
to obtain the teleportation output when such a mixture is used as a resource. Our
second purpose is to obtain within our formalism a well-known result for normally
ordered phase-space pseudo-distributions. Namely the result of \textit{superimposing}
a state of the radiation field over another, preexisting state having a $P(\alpha)$
function~\cite{Marianthermal,Glaubercoher,Rockower}. It is generally understood that
a given (pure) resource state will propagate in a spatial volume or be prepared
ideally within the \emph{environment} given by the pure vacuum $|0\,,\,0\,\rangle$
state. In ref.~\cite{Glaubercoher} it is spoken of the \emph{switching on/off} of two
sources in order; which is equivalent to preparing one state \emph{over} another (or
letting one state propagate over the same spatial volume of the) state we have deemed
to be an \textit{initial environment}. The \textit{superimposition}~\footnote{We use
the term superimposition to avoid confusion with the fundamental Quantum Mechanics
postulate and concept of superposition.} we will consider in this section consists in
preparing (or letting propagate) a two-mode teleportation resource state when the
initial environment state is different from the vacuum and has a $P(\alpha)$
function. We will use environment states that are generally considered "noise" for
our purposes; for example, separable, two-mode thermal states.

Given the pure state $\hat{\rho}_{AB}$, let us define the phase-space mixture of
states in an similar manner to that of eq.~(\ref{eq:MixDensEPR});
\begin{equation}\label{eq:MixRes}
  \hat{\rho}^{(Mix)}_{AB}=\,\pi^{-2}\,\int\, d^{2}\alpha_{a}\,\,d^{2}\alpha_{b}\;
  \mathrm{P}_{A}(x_{a}\,,\,p_{a})\:\mathrm{P}_{B}(x_{b}\,,\,p_{b})\;\hat{\rho}_{AB,\alpha_{a};\:\alpha_{b}}
\end{equation}
where $\mathrm{P}_{A}$ and $\mathrm{P}_{B}$ are probability distributions that
fulfill  the conditions set forth in eq.~(\ref{eq:MixProbCond}). The density operator
\begin{equation}\label{eq:DispRes}
  \hat{\rho}_{AB,\alpha_{a};\:\alpha_{b}}=\,\widehat{D}^{\dag}_{B}(\alpha_{b})\;\widehat{D}^{\dag}_{A}(\alpha_{a})\;
  \hat{\rho}_{AB}\;\widehat{D}_{A}(\alpha_{a})\;\widehat{D}_{B}(\alpha_{b})
\end{equation}
is that of an ideal, pure state resource; displaced in phase-space by $\alpha_{a}$
and $\alpha_{b}$. Thus, the density operator of eq.~(\ref{eq:MixRes}) is equivalent
to that for two superimposed modes in ref.~\cite{Glaubercoher}. Note that the density
operator for any phase-space mixture; for example the mixed \textbf{EPR} state
(eq.~(\ref{eq:MixDensEPR}) is of a similar form, as phase-space displacements
transform one \textbf{EPR} state into another \textbf{EPR} state.

The characteristic function of eq.~(\ref{eq:MixRes}) is thus given by
\begin{align}
  \chi^{(Mix)}_{AB}(\xi_{A};\:\xi_{B})=\,&\pi^{-2}\,\int\, d^{2}\alpha_{a}\,d^{2}\alpha_{b}\;
  \mathrm{P}_{A}(x_{a}\,,\,p_{a})\;\mathrm{P}_{B}(x_{b}\,,\,p_{b})\notag \\
  &\times\,e^{\,2\,i\,\left(z_{A}x_{a}\,-\,w_{A}p_{a}\right)}e^{\,2\,i\,\left(z_{B}x_{b}\,-\,w_{B}p_{b}\right)}\;
  \chi_{AB}(\xi_{A};\;\xi_{B}) \label{eq:MixResChar}
\end{align}
where displacement operations correspond to phase factors in the characteristic
function. Define the characteristic functions $\widetilde{\mathrm{P}}_{A}$ and
$\widetilde{\mathrm{P}}_{B}$ as the Fourier transforms of $\mathrm{P}_{A}$ and
$\mathrm{P}_{B}$ (see eq.~(\ref{eq:MixProbFourier})). The characteristic function in
eq.~(\ref{eq:MixResChar}) will have the factorized form
\begin{equation}
  \chi^{(Mix)}_{AB}(\xi_{A};\:\xi_{B})=\, \widetilde{\mathrm{P}}_{A}(w_{A}\,,\,z_{A})\;
  \widetilde{\mathrm{P}}_{B}(w_{B}\,,\,z_{B})\;\chi_{AB}(\xi_{A};\:\xi_{B})\label{eq:MixResCharDef}
\end{equation}

The characteristic function of eq.~(\ref{eq:MixResCharDef}) can be made equivalent to
the \textit{superimposition} of the resource state $\hat{\rho}_{AB}$ over
\emph{initial} states in modes $A$ and $B$. If the Wigner functions for such initial
states can be substituted for the probability distributions $\mathrm{P}_{A}$ and
$\mathrm{P}_{B}$ in the mixture. This can be done only for Wigner functions
fulfilling certain conditions; being finite, nonnegative and normalized. In other
words, for Wigner functions that are genuine probability distributions. In
refs.~\cite{Glaubercoher,Marianthermal,Rockower}, the equivalent condition is set
forth that the initial states possess a $P(\alpha)$ function representation.

The Wigner function that is the Fourier transform of eq.~(\ref{eq:MixResCharDef}) is
given by
\begin{equation}\label{eq:MixResWigner}
W^{(Mix)}_{AB}(\alpha_{A};\:\alpha_{B})=\,\pi^{-2}\,\int\,
d^{2}\alpha_{a}\,d^{2}\alpha_{b}\;
W_{AB}(\alpha_{A}\,-\,\alpha_{a};\:\alpha_{B}\,-\,\alpha_{b})\;\mathrm{P}_{A}(x_{a},p_{a})\\;\mathrm{P}_{B}(x_{b},p_{b})
\end{equation}
which is a convolution integral of the Wigner functions of the states. This relation
is equivalent to that of the  $P(\alpha)$ functions for "superposed" excitations in
ref.~\cite{Glaubercoher}.

%%That such preparations take the form of mixtures of states ( eq.~(\ref{eq:MixResCharDef})) does
%%not mean that the end result has purity less than $1$ and is mixed. The result of superimposing a
%%pure state on a mixed state (for example a thermal state) is a mixed state. Superimposing a pure
%%state on a pure state (say, the vacuum or other starting state with positive Wigner function) will
%%produce a pure state.

The general mixed states that we have used as substitutes for the \textbf{EPR} states
in projective measurement in section~\ref{Form:TelepMix} can be thought of as
\emph{superimpositions} of a pure state on a "noisy" initial environment. The
probability distribution in eq.~(\ref{eq:MixDensEPR}) is nonlocal; therefore it must
be the nonnegative Wigner function of a nonlocal state. General, Gaussian entangled
states, having nonnegative Wigner functions would be the first choice. This kind of
superimposition can be interesting; but it falls outside of the scope of our work. We
consider entangled states "pure" resources to be prepared; not noisy environments to
"prepare over".

We have chosen separate probabilities $\mathrm{P}_{A}$ and $\mathrm{P}_{B}$ instead
of a joint probability in eq.~(\ref{eq:MixRes}; it is our purpose to represent
preparation of a resource over "noisy" separable states and study the changes wrought
by this new preparation on separability and teleportation fidelity, with respect to a
pure state prepared over a two-mode vacuum.

Thermal states on modes $A$ and $B$ constitute a \emph{noisy} initial state over
which the pure resource state can be prepared, resulting in a mixed resource state.
Thermal states~\cite{Glaubercoher} fulfill the requisite of having a positive Wigner
function and having a $P(\alpha)$ representation. They are prime candidates for the
representation of the simplest noisy environments. A single-mode thermal field of
mean photon-number $n_{Th}$ has a density matrix in the Fock basis
\begin{equation}
\rho_{th} = \,\sum_{k}\;\frac{n_{th}^{k}}{(1+n_{th})^{k+1}}\; |\,k\, \rangle\:
\langle\, k\,| \label{eq:ThermalDensity}
\end{equation}
and a Gaussian~\cite{DistFuncWigner,Glaubercoher} Wigner function that is a genuine
probability distribution. Which, along with its respective characteristic function,
reads
\begin{align}
W_{th}(\alpha)\,&=\,n_{th}^{-1}\;e^{-n_{th}^{-2}|\alpha|^{2}} \label{eq:ThermalWig} \\
\chi_{th}(\xi)\,&=\;e^{-n_{th}^{2}|\xi|^{2}} \label{eq:ThermalChar}
\end{align}

The characteristic functions of quantum states chosen to represent noise will appear
as multiplicative factors in the output of teleportation. Substituting the resource
of eq.~(\ref{eq:MixResCharDef}) in eq.~(\ref{eq:CharOutFin}) yields the outcome
\begin{equation}\label{eq:CharOutMixRes}
  \chi_{out}(\xi_{B})=\,\chi_{in}(g_{p}w_{B}\,+\,i\,g_{x}z_{B})\;\chi_{AB}(g_{p}w_{B}\,-\,i\,g_{x}z_{B};\:\xi_{B})
  \widetilde{\mathrm{P}}_{A}(g_{p}w_{B}\,,\,-g_{x}z_{B})\:
  \widetilde{\mathrm{P}}_{B}(w_{B}\,,\,z_{B})
\end{equation}

This output has in common with eq.~(\ref{eq:MixCharFin}) the presence of a
multiplicative factor; though of a different character. The mixture in that case is
that of the "homodyne" projective measurement as "induced" by an imprecise apparatus.
In this case the factor corresponds to a phase-space mixture that is the result of
preparation of the resource in a noisy environment.

\section{Losses and noise admission in the projective measurement}
\label{Form:TelepNoise}

In the previous sections we have used the characteristic functions language to
produce an expression for teleportation outcome. We have introduced modifications to
the protocol that imply, in our view, imprecisions in the homodyne measurement. We
have formulated a "noisy", mixed resource state.

A modification to the teleportation procedure is the next step. We will modify the
teleportation protocol by the introduction of \emph{fictitious} elements, with the
initial purpose of modelling a realistic, lossy and noisy homodyne measurement.
Setups introducing fictitious elements to homodyne detection were originally
developed and adapted to the study of non ideal, "realistic" homodyne detection of
single mode quadratures in the $Q(\alpha)$ function
representation~\cite{RealHomodLeonhardt}; and later independently developed for non
ideal teleportation in a coherent state representation~\cite{JanszkyNonIdTelep}.

The realistic homodyne measurement will be realized by the addition of two
beam-splitters of rather high transmittances $\cos^2(\phi)$ and $\cos^2(\varphi)$,
where $\phi\,\approx\,\varphi\,\approx\,1$; and two "external" modes that will be
mixed by these beam-splitters with the nonlocal modes (those originally to be
measure: $u$ and $v$ in fig.~(\ref{fig:telesetup})) before the actual process of
homodyne measurement takes place. These beam-splitters will also deduct a small loss
of intensity to the original nonlocal modes.

\begin{figure}[bt] \label{fig:realhomodyne}
\begin{centering}
\includegraphics[width=15cm]{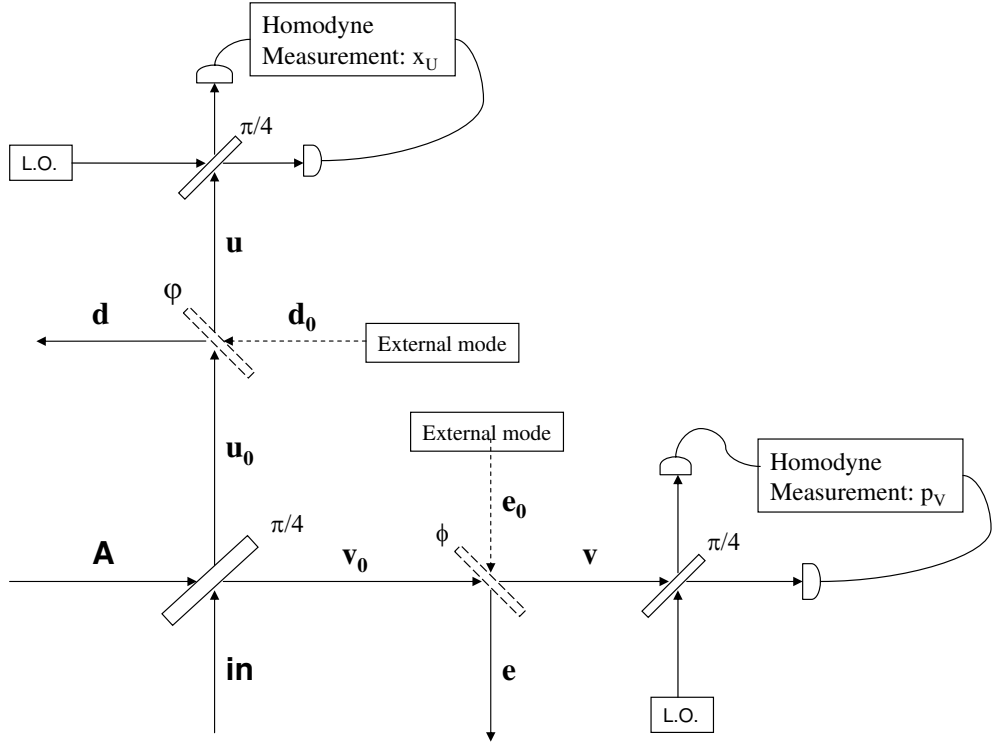}
\end{centering}
\caption[The Experimental Setup for Realistic Homodyne Measurement in \textbf{CV}
Teleportation]{The Experimental Setup for Realistic Homodyne Measurement in
\textbf{CV} Teleportation. The external modes $d_{0}$ and $e_{0}$ are mixed with the
nonlocal modes $u_{0}$ and $v_{0}$ by the "$\varphi$" and "$\phi$" beam splitters.
The resulting $u$ and $v$ modes undergo homodyne measurement and "projection" onto an
\textbf{EPR} state.}
\end{figure}

The setup for this lossy, noisy homodyne measurement is illustrated in
fig.~(\ref{fig:realhomodyne}) (see the left half of fig.~(\ref{fig:telesetup}) for a
comparison). The beam-splitters "$\phi$" and "$\varphi$" and modes $d_{0}$ and
$e_{0}$ are the new elements in the homodyne setup. For the external modes a choice
can be made of thermal states, vacuum states, or general one-mode Gaussian states
including the aforementioned as special cases.

The teleportation procedure will (as always) consist in the performance of a
projective measurement of the quadratures $x_{u},p_{v}$. These are derived from the
initial "$0$" quadratures by the beam-splitter transformations for the modes
illustrated in fig.~(\ref{fig:realhomodyne}),
\begin{align}
 \begin{pmatrix}
  \alpha_{u_{0}} \\
  \alpha_{v_{0}}
\end{pmatrix}  =& \begin{pmatrix}
  2^{-1\!/2} & -  2^{-1\!/2} \\
  2^{-1\!/2} &    2^{-1\!/2}
\end{pmatrix} \begin{pmatrix}
  \alpha_{in} \\
  \alpha_{A}
\end{pmatrix} \notag \\ \notag \\
\begin{pmatrix}
  \alpha_{u} \\
  \alpha_{d}
\end{pmatrix}  =& \begin{pmatrix}
  \cos (\varphi) & - \sin (\varphi) \\
  \sin (\varphi) & \cos (\varphi)
\end{pmatrix} \begin{pmatrix}
  \alpha_{u_{0}} \\
  \alpha_{d_{0}}
\end{pmatrix} \notag \\ \notag \\
\begin{pmatrix}
  \alpha_{v} \\
  \alpha_{e}
\end{pmatrix}  =& \begin{pmatrix}
  \cos (\phi) & - \sin (\phi) \\
  \sin (\phi) & \cos (\phi)
\end{pmatrix} \begin{pmatrix}
  \alpha_{v_{0}} \\
  \alpha_{e_{0}}
\end{pmatrix}\label{eq:NoiseUnitTrans}
\end{align}

With these transformations at hand, it is easy to write the Wigner function of the
appropriate \textbf{EPR} state for homodyne measurement (see
eq.~(\ref{eq:WignerEPR})) having eigenvalues $x'_{u}$ and $p'_{v}$;
\begin{align}
  W_{x'_{u}\,,\,p'_{v}}(
  \alpha_{u_{0}},\alpha_{d_{0}};\:\alpha_{v_{0}},\alpha_{e_{0}})\,=&\,\mathcal{C}\;
  \delta\left(\cos(\varphi)\,x_{u_{0}}\,-\,\sin(\varphi)\,x_{d_{0}}\,-\,x'_{u}\right)\notag\\
  &\times\, \delta\left(\cos(\phi)\,p_{v_{0}}\,-\,\sin(\phi)\,p_{e_{0}}\,-\,p'_{v}\right)\label{eq:WignerNoiseQuad}
\end{align}

The characteristic function for the \textbf{EPR} state is obtained in a
straightforward manner by Fourier transformation of the above expression, and is
given by
\begin{align}
  \chi_{x'_{u}\,,\,p'_{v}}(
  \xi_{u_{0}},\xi_{d_{0}};\:\xi_{v_{0}},\xi_{e_{0}})\,=&\,\pi^{2}\,\mathcal{C}\;
  \delta(w_{d_{0}})\;\delta(z_{v_{0}})\;\delta(w_{u_{0}})\delta(z_{e_{0}})\;
\delta\left(\sin(\varphi)\,z_{u_{0}}\,+\,\cos(\varphi)\,z_{d_{0}}\right)\notag \\
&\times\, \delta\left(\sin(\phi)\,w_{v_{0}}\,+\,\cos(\phi)\,w_{e_{0}}\right)
e^{-\,2\,i\,\left(z_{d_{0}}\,x'_{u}\,\csc(\varphi)\,-\,w_{e_{0}}\,p'_{v}\,\csc(\phi)\right)}\label{eq:CharNoiseQuad}
\end{align}

The state of the joint system $AB, in$ to be measured after the first, symmetric
($\pi/4$) beam-splitter reads
\begin{equation}\label{eq:NoiseSysState}
\chi_{in}(2^{-1\!/2}\xi_{u_{0}}\,+\,2^{-1\!/2}\xi_{v_{0}})\;
\chi_{AB}(2^{-1\!/2}\xi_{v_{0}}\,-\,2^{-1\!/2}\xi_{u_{0}};\:\xi_{B})\; \chi_{ext.\:
u}(\xi_{d_{0}})\;\chi_{ext.\: v}(\xi_{e_{0}})
\end{equation}
where $\chi_{ext.\: u}$ and $\chi_{ext.\: v}$ are the characteristic functions of the
two external modes $d_{0}$ and $e_{0}$, respectively.

We follow the teleportation protocol as described in section~\ref{Form:TeleportIdeal}
for the physical system in eq.~(\ref{eq:NoiseSysState}); using for the projective
measurement the \textbf{EPR} state of eq.~(\ref{eq:CharNoiseQuad}). As before, we
have Bob perform a displacement of
$\widehat{D}_{B}(x'_{u}2^{1\!/2}\,+\,ip'_{v}2^{1\!/2})$ to correct the output state;
only with the gains $g_{x}=g_{p}=1$; for the procedure we are performing already
includes losses in intensity and is intended to model a lossy measurement. The output
state will be, as before, an ensemble of all possible measurement outcomes. Thus,
\begin{align}
  \chi_{out,n}(\xi_{B})=&\,\mathcal{C}^{-1}\,\int\,dx'_{u}\,dp'_{v}\,e^{\,2\,i\,\left(z_{B}\,2^{1\!/2}\,x'_{u}\,-\,w_{B}\,2^{1\!/2}\,p'_{v}
  \right)}\notag \\
  &\times\,\pi^{-4}\,\int\,d^{2}\xi_{u_{0}}\,d^{2}\xi_{v_{0}}\,d^{2}\xi_{d_{0}}\,d^{2}\xi_{e_{0}}\;\chi_{in}(2^{-1\!/2}\xi_{u_{0}}\,+\,2^{-1\!/2}\xi_{v_{0}}) \notag \\
  &\times\,\chi_{AB}(2^{-1\!/2}\xi_{v_{0}}\,-\,2^{-1\!/2}\xi_{u_{0}};\:\xi_{B})\; \chi_{ext.\: u}(\xi_{d_{0}})\;\chi_{ext.\: v}(\xi_{e_{0}})\notag\\
  &\times\; \chi_{x'_{u}\,,\,p'_{v}}(
  \xi_{u_{0}},\xi_{d_{0}};\:\xi_{v_{0}},\xi_{e_{0}})\label{eq:NoiseCharOutIn}
\end{align}

After some (entirely straightforward) integration of the above expression, we obtain
the output state
\begin{align}
  \chi_{out,n}(\xi_{B})\,=&\,\chi_{in}(\cos(\phi)\,w_{B}\,+\,i\,\cos(\varphi)\,z_{B})\;\chi_{AB}(\cos(\phi)\,w_{B}\,-\,i\,\cos(\varphi)\,z_{B};\:\xi_{B})\notag \\
  &\times\,\chi_{ext.
  \:u}(i\,2^{1\!/2}\sin(\varphi)\,z_{B})\;\chi_{ext.\:v}(2^{1\!/2}\sin(\phi)\,w_{B}) \label{eq:NoiseCharFin}
\end{align}
The first obvious trait of this state is the multiplication by the characteristic
functions of the external modes. These have arguments that are scaled by the (small)
factors $\sin(\varphi)$ and $\sin(\phi)$, while the characteristic functions of input
and resource are scaled by factors (close to $1$) $\cos(\varphi)$ and $\cos(\phi)$
which are the transmission coefficients for the beam-splitters "$\phi$" and
"$\varphi$", respectively. For small angles $\phi$ and $\varphi$,
eq.~(\ref{eq:NoiseCharFin}) approximates eq.~(\ref{eq:CharOutFin}) for gains given as
$g_{x}=\,\cos(\varphi)$ and $g_{p}=\,\cos(\phi)$. In the limit case $\phi=\varphi=0$
eq.~(\ref{eq:NoiseCharFin}) equals eq.~(\ref{eq:CharOutFin}), as the characteristic
functions of the external modes with argument $0$ are constant.

In sections~\ref{Form:TelepMix} and~\ref{Form:MixRes} we have multiplying
characteristic functions of  \emph{probability distributions} (which can be also
Wigner functions) as multiplying factors in teleportation output states. But there
are two main differences. First, it is not required by the formalism we have used
that $\chi_{ext.\:u}$ and $\chi_{ext.\:v}$ be the Fourier transforms of nonnegative
Wigner functions. Second and most important, the multiplying factors in the arguments
of said characteristic functions are $\sin(\varphi)$ and $\sin(\phi)$: not the gains;
but rather the complementary quantities of those which may be considered  measurement
gains ($\cos(\varphi)$ and $\cos(\phi)$). Only in the case where the Wigner functions
of the external modes are positive and $\phi=\varphi=\pi/4$ (or
$g_{x}=g_{p}=2^{-1/2}$) are the two cases entirely equivalent. These two differences
illustrate the "lossy" nature of the teleportation protocol outlined here, as
compared with the merely "noisy" protocols.

There is an obvious choice of external modes' states to represent losses and added
noise in a teleportation setup~\cite{RealHomodLeonhardt,JanszkyNonIdTelep}. It is the
general one-mode Gaussian vacuum states, which may be squeezed and of non-minimum
uncertainty (thermal). Such thermal states have a characteristic function
\begin{equation} \label{eq:ThermGaussVac}
\chi_{Th(n,s)}(\xi)=e^{-\,\bar{n}^2(e^{-2s}z^{2}+e^{2s}w^{2})}
\end{equation}
for an average number of photons $n$ and a squeezing parameter $s$ (squeezing applied
on $\hat{x}$ quadrature). Given similarly squeezed thermal states with the same
average number of photons for both the external modes, the multiplying factor in
eq.~(\ref{eq:NoiseCharFin}) will be
\begin{equation} \label{eq:NoiseCharMult}
e^{\,-\,2\bar{n}^{2}\,e^{-2s}\,\sin(\varphi)^{2}\,z_{B}^{2}}\;e^{\,-\,2\,\bar{n}^2\,e^{2s}\,\sin(\phi)^{2}\,w_{B}^{2}}
\end{equation}

As explained before, the main difference with the examples of Gaussian mixtures in
the previous sections is that this "smearing factor" will tend to become constant as
the measurement gain goes to $1$.

\section{Teleportation fidelity with characteristic functions}
\label{Form:Fide}

We have adopted the definition of fidelity proposed in ref.~\cite{CriteriaCVTelep}.
In the characteristic function representation this is equal to
eq.~(\ref{eq:FideChar}). This is the scalar product, or the trace of the product of
density operators, for the input and output states. Let us take, for simplicity of
exposition, the output state we have derived in eq.~(\ref{eq:CharOutFin}) (for a
maximum gain $g_{x}=g_{p}=1$) and insert it in eq.~(\ref{eq:FideChar});
\begin{equation}\label{eq:FidCharOutFin}
\mathcal{F}=\,\pi^{-1}\,\int\,d^{2}\xi_{B}\;\chi_{in}(\xi_{B})\;\chi_{in}(-\xi_{B})\:\chi_{AB}(-\xi_{B}^{*};\:-\xi_{B})
\end{equation}

A summary analysis might be made of this expression and of its integrand. We have
inside the integrand that
$\chi_{in}(\xi_{B})\:\chi_{in}(-\xi_{B})=|\chi_{in}(\xi_{B})|^2$. This quantity is a
real, even, nonnegative characteristic function itself, corresponding to the square
of the density operator $\hat{\rho}_{in}^{2}$ (see eqs.~(\ref{eq:TraceTwoChar})
and~\ref{eq:TrSqWigChar})). Its trace integral over $\xi_{B}$ is equal to the purity
of the input state. In eq.~(\ref{eq:FidCharOutFin}) it is multiplied by the
characteristic function of the resource state $\chi_{AB}$. When the resource state
approximates an \textbf{EPR} state, its characteristic function becomes nearly
constant with a value of $1$ (which is not square integrable). The result is ideal
teleportation with maximal fidelity: $1$ for pure input states and the value of the
purity for mixed input states.

Given that the fidelity is real, positive and bounded from above, we remark that
$\chi_{AB}(\pm\xi_{B}^{*};\pm\xi_{B})$ has to be bounded and must have an imaginary
part that is odd or $0$. Its real part must be even and positive (in the sense that
an application is positive).

The fidelity in eq.~(\ref{eq:FidCharOutFin}) is quadratic in the input state and
linear in the resource state. The fidelities corresponding to the output states
derived in sections~\ref{Form:TeleportIdeal},~\ref{Form:TelepMix}
and~\ref{Form:TelepNoise} share these two features; even though there are additional
multiplying factors.

One very simple conclusion to be drawn from the linearity of the teleportation output
and of the fidelity with respect to the resource state is that we need only calculate
the fidelity of teleportation for \emph{pure} resource states, as the fidelity for a
simple mixture of resource states comes about trivially.  Given a mixture of pure
resource states $|\,\psi_{i}\,\rangle_{AB}$, and given the characteristic function's
linearity with respect to the density operator (see eq.~(\ref{eq:CharTracDisp})), we
have that
\begin{align}
\hat{\rho}_{AB}=&\,\sum_{i}\: \mathrm{P}_{i} \;|\,\psi_{i}\,\rangle_{AB} \;\langle\,\psi_{i}\,|_{AB}\notag \\
\chi_{AB}(\xi_{A};\:\xi_{B})=&\, \sum_{i}\: \mathrm{P}_{i}
\;\chi_{\psi_{i}\,,\,AB}(\xi_{A};\:\xi_{B}) \label{eq:SumMixtCharRes}
\end{align}
where $\chi_{\psi_{i}\,,\,AB}$ are the characteristic functions of the pure states
$|\,\psi_{i}\,\rangle_{AB}$.

Given the linearity of the fidelity (eq.~(\ref{eq:FidCharOutFin})) with respect to
the above characteristic function; it follows that the fidelity of teleportation
using this mixed resource state is simply the weighted average of the fidelities
$\mathcal{F}_{i}$ resulting from the use of the pure resource states singly as
teleportation resources;
\begin{equation}\label{eq:FidMixRes}
  \mathcal{F}=\,\sum_{i}\: \mathrm{P}_{i} \; \mathcal{F}_{i}
\end{equation}

Therefore, we need only concern ourselves with pure resource states, at least as
regards discrete mixtures of resources. These discrete mixtures appear whenever one
considers conditional preparation procedures for resources; such as photon
subtraction and addition for two-mode squeezed Gaussian states, which involve
measurements.

The fidelity of teleportation is quadratic with respect to the input state. For an
input state that is a mixture of other (pure) input states with characteristic
functions $\chi_{in,j}(\xi_{in})$, we have
\begin{equation}\label{eq:SumMixtCharInp}
\chi_{in}(\xi_{in})= \sum_{j} \mathrm{P}_{j} \;\chi_{in,j}(\xi_{in})
\end{equation}

The fidelity of teleportation in eq.~(\ref{eq:FidCharOutFin} for the above mixture is
given by
\begin{equation}\label{eq:FidMixInp}
  \mathcal{F}=\,\sum_{j,k}\: \mathrm{P}_{j}\: \mathrm{P}_{k}\;\pi^{-1}\,\int\,d^{2}\xi_{B}\;\chi_{in,j}(\xi_{B})\;\chi_{in,k}(-\xi_{B})\;\chi_{AB}(-\xi_{B}^{*};\:-\xi_{B})
\end{equation}
which is rather complicated. The integrals summed in eq.~(\ref{eq:FidMixInp}) are
traces of the input pure state "$i$" with the output state resulting from the
teleportation "$j$". The integrals are not (individually) \textit{bona fide}
teleportation fidelities; save for the case where $i=j$, but must sum to a
teleportation fidelity that is bounded. The summation over the integrals is not
equivalent to the average of fidelities in eq.~(\ref{eq:SumMixtCharInp}. Even in the
case of a near-ideal resource; eq.~(\ref{eq:FidMixInp}) includes terms that are the
casual overlap of the components of the input mixture, and the (nearly) perfect
transcriptions resulting from teleportation. This can be seen as a limitation in the
concept of fidelity we have adopted.

If we consider the mixture above described as the ensemble of likely (one qudit)
inputs of a quantum teleportation channel for a \textbf{CV}~\cite{VanLoockBraunstein}
or hybrid quantum computation scheme~\cite{Hybrid}; we can count on the likely inputs
for the channel to be far away from each other as regards scalar product: nearly
orthogonal or orthogonal, so as to be easily distinguishable by measurement apparatus
even under noisy conditions. In that case, and for a good enough teleportation
resource, we can consider eq.~(\ref{eq:FidMixInp}) to include mainly the contribution
from the $i=j$ traces (bona fide teleportation fidelities) that sum to a weighted
average fidelity.

\chapter{Teleportation with degaussified, squeezed Fock and squeezed Bell-like resources}
\label{NGau}

In the previous chapter we introduced a formalism for the study of teleportation in
\textbf{CV} based on the Wigner characteristic function representations of quantum
states. The expression for the output state derived for the ideal \textbf{CV}
protocol (section~\ref{Form:TeleportIdeal}) is general for any combination of
teleportation resource and input having Wigner and Wigner characteristic functions.

The objective of the investigation of teleportation resources is to produce feasible
teleportation schemes that improve the teleportation fidelity for the input states
most likely to be used in \textbf{CV} quantum
computation~\cite{VanLoockBraunstein,LloydBraunstein} or in hybrid schemes that use
\textbf{CV} states for communication~\cite{Hybrid,Communication}. The original
protocol~\cite{BraunsteinKimble} was developed for Gaussian two-mode squeezed states,
and these will be used as the benchmark from which it is desirable to improve output
state fidelity, for a fixed squeezing parameter $r$. The parameter $r$ assumes the
meaning of indicator of technological capability in the preparation of resources.

An initial analysis of non-Gaussian resources can begin with the class of
"degaussified" two-mode squeezed states that have been shown to produce an
improvement in the teleportation fidelity with respect to Gaussian states of both a
comparable covariance matrix and similar squeezing parameter
$r$~\cite{Opatrny,Cochrane,Olivares,KitagawaPhotsub}.

"Degaussification"~\footnote{Not to be confused with the demagnetization procedure}
is the production of a non-Gaussian state of the radiation field by
\emph{conditional} photon subtraction (addition) on an initially Gaussian state. This
procedure is both performed and verified by appropriate single-photon measurements.
The "simulation" of the conditional measurement and the use of the subsequent
mixtures of degaussified and Gaussian resources has been done~\footnote{In
representations different from the characteristic functions'.} in
refs.~\cite{Opatrny,Olivares,KitagawaPhotsub}. Within the characteristic function
representation, we have shown (see section~\ref{Form:Fide}) that the use of any such
mixture as a resource gives results for fidelity that can be trivially derived from
the results for pure state resources.

We have found useful for the search of improvements in teleportation fidelity to
formulate classes of resource states that encompass the largest possible categories
of Gaussian, degaussified or simply non-Gaussian states. These generalizing classes
of resources are by definition non-Gaussian. They are necessarily superpositions of
the "special case" resources or involve additional unitary transformations on such
states; moreover, they reduce to the special case states for particular choices of
the superposition and unitary transformation parameters. These same parameters can be
arbitrarily chosen, allowing us to sculpt teleportation resources; even to optimize
the resource for maximal fidelity of teleportation of a given input state. A further
and related advantage lies in the ability to compare optimal resources with special
case resource states for which preparation strategies have been devised or
experimentally tested; having been proposed in previous work for \textbf{CV} quantum
information resources.

In this chapter, we will first introduce the non-Gaussian resources and inputs used
throughout the chapter, in the characteristic function representation. We will then
study the fidelity in the teleportation of selected \emph{pure} input states; using
photon-subtracted and photon-added squeezed states, two-mode squeezed Gaussian
states, two-mode squeezed Fock states; and as implied before, a class of states that
includes the aforementioned resource states as special cases. These we have named the
\textit{squeezed Bell-like states} in previous work~\cite{TelepNonGauss}; general
superpositions of the two-mode, Gaussian squeezed vacuum state and the two-mode
squeezed Fock state of number $1,1$, which is non-Gaussian. For the squeezed
Bell-like resource states, we have calculated an optimal fidelity (with respect to
the superposition parameters) for each of the selected input
states~\cite{TelepNonGauss} and compared this optimal fidelity with the (smaller)
fidelities obtained for the resources mentioned above.

The optimal squeezed Bell-like states and the "special case" resources are also
studied and compared as to entanglement, non-Gaussian character and affinity with
Gaussian states; with the purpose of establishing the characteristics of a
non-Gaussian state that better correlate with teleportation fidelity. The parameters
for comparison are the von Neumann entropy~\cite{BennettEntropy}; a non-Gaussianity
measure~\cite{GenoniNonGaussy}, and a two-mode squeezed vacuum "affinity" measure
devised for non-Gaussian teleportation resources~\cite{TelepNonGauss}.

\section{Resources and inputs: state vectors and characteristic functions}
\label{NGau:Chzation}

In this section, we will introduce the two-mode non-Gaussian states that are the
object of our study as resources for \textbf{CV} teleportation in this chapter.

\subsection{Squeezed Fock states}
\label{NGau:Chzation:SqFock}

The two-mode squeezed Fock state is prepared by the performance of a two-mode
squeezing operation on a \textbf{CV} system that is in a (separable) two-mode Fock
state. The state vector for this wholly non-Gaussian state is given by,

\begin{equation}\label{eq:SqFocket}
  |\zeta\,;\: m_{A}\,,\, m_{B}\,\rangle \, = \, \widehat{S}_{AB}(\zeta) \; |\,m_{A}\,,\,m_{B}
\,\rangle_{AB}
\end{equation}

where $\widehat{S}_{AB}(\zeta)$ is the two-mode squeezing operator of
eq.~(\ref{eq:TwoModSqOp}); with the separable state vector $|\,m_{A}\,,\,m_{B}\,
\rangle_{12} \equiv |\,m_{A}\,\rangle_{A} \otimes |\,m_{B}\,\rangle_{B}$ being the
two-mode Fock state.

The two-mode squeezed vacuum state $\widehat{S}_{AB}(\zeta)
\;|\,\zeta\,;\,0,0\,\rangle$ and the two-mode squeezed Fock state
$\widehat{S}_{AB}(\zeta) |\,\zeta\,;\,1\,1\,\rangle$ are special instances of this
class of states. This last state will be referred to as two-mode squeezed Fock state
throughout the chapter.

The calculation of the characteristic function for the two-mode squeezed Fock state
is straightforward, given that it is pure state; hence it's characteristic function
is given by eq.~(\ref{eq:CharPurDisp}). Thus
\begin{equation}\label{eq:CharTrSqFock}
  \chi_{SN}^{(1,1)}(\xi_{A} ; \xi_{B})=\;\langle
  1,1|\,\widehat{S}^{\dag}_{AB}(\zeta)\:\widehat{D}_{A}(\xi_{A})\:\widehat{D}_{B}(\xi_{B})\:{S}_{AB}(\zeta)|\,1,1\rangle
\end{equation}

Recall the Bogoliubov transformation effected by the \emph{unitary} two-mode
squeezing operator on the displacement operators (see eq.~(\ref{eq:TwoModBogDisp}))
and the expression for the matrix element of the displacement operator in the Fock
basis~\cite{CahillGlauber1};
\begin{equation}
\langle \,m\,|\: \widehat{D}(\xi) |\,n\, \rangle \,=\,
\left(\frac{n!}{m!}\right)^{1/2}\xi^{m-n}\:e^{-\frac{1}{2}|\,\xi\,|^{2}}
\:L_{n}^{(m-n)}(|\,\xi\,|^{2}) \label{eq:DispFockElem}
\end{equation}
where $L_{n}^{(m-n)}$ is the associated Laguerre polynomial~\cite{GRADRIZ}.

Given the above relations, the characteristic function for the state $|\,\zeta\,;\:
1\,,1\rangle$ is easily seen to be
\begin{equation}
\chi_{SN}^{(1,1)}(\,\xi_{A} ;\: \xi_{B}\,) =\, e^{-1\
,/\,2\:\left(|\,\xi'_{A}\,|^{2}\,+\,|\,\xi'_{B}\,|^{2}\right)}\;\left(1-|\,\xi'_{A}\,|^{2})(1-|\,\xi'_{B}\,|^{2}\right)
\label{eq:CharSqFockTwo}
\end{equation}
where the variables $\xi'_{A}$ and $\xi'_{B}$ are related to the variables $\xi_{A}$
and $\xi_{B}$ via the Bogoliubov transformation described in
eq.~(\ref{eq:TwoModBogChar}) for the displacement operator arguments.

\subsection{Degaussified resource states}
\label{NGau:Chzation:Degauss}

The degaussified resource states are generated by photon subtraction (or addition) to
each of the modes $A$ and $B$ of a two-mode squeezed Gaussian state. Although the
photon subtraction (addition) procedure is conditional in most experimental setups,
it will be assumed here that it has been successfully performed and verified,
obtaining a pure resource state. Thus, the photon subtracted (added) states have the
state vectors
\begin{align}
|m_{A}^{(-)}\,,\, m_{B}^{(-)} \,;\: \zeta\, \rangle \, =& \,
\mathcal{N}_{AB}^{\,(-)}\; a_{A}^{m_{A}}\; a_{B}^{m_{B}} \;\widehat{S}_{AB}(\zeta)
\; |\,0\,,\,0\, \rangle_{AB} \label{eq:PhotSubket} \\
|m_{A}^{(+)}\,, m_{B}^{(+)} \,; \zeta \rangle \, =& \, \mathcal{N}_{AB}^{\,(+)}
\;a_{A}^{\dag \; m_{A}}\; a_{B}^{\dag \; m_{B}} \;\widehat{S}_{AB}(\zeta) \;
|\,0\,,\,0\, \rangle_{AB} \label{eq:PhotAddket}
\end{align}
where $\mathcal{N}_{AB}^{\,\pm}$ are normalization constants specific to each state
vector. Henceforth, and in keeping with the proposals for the experimental generation
of these states~\cite{DeGauss0,DeGauss1,DeGauss2,DeGauss3,DeGauss4}, we will restrict
our analysis to the case where $m_{A}=m_{B}=1$; to the single-photon subtracted
(added) states.

To easily calculate the normalization constants and the respective characteristic
functions, and to put the degaussified states in a proper perspective with respect to
the other resources studied in this chapter; we write the states described in
eqs.~(\ref{eq:PhotSubket}) and~(\ref{eq:PhotAddket}), taking into account the
transformation that results from the application of the two-mode squeezing operator
on the annihilation and creation operators (see eq.~(\ref{eq:TwoModBogCom})),
\begin{align}
|1^{(-)} \,,\, 1^{(-)} ;\: \zeta\, \rangle  = & \,\mathcal{N}\; e^{i\phi}
\widehat{S}_{AB}(\zeta)\; \left(-|\,0\, ,\, 0\, \rangle_{AB} \:+\: e^{i\phi}
\,\tanh(r)\, |\,1 \,,\, 1\, \rangle_{AB} \right) \label{eq:PhotSubketSup}\\
|1^{(+)}\, , \,1^{(+)} ;\: \zeta\, \rangle  = &\,  \mathcal{N}\; e^{-i\phi}
\widehat{S}_{AB}(\zeta)\; \left( -\tanh(r)\, |\,0 \,,\, 0 \,\rangle_{AB} \:+\:
e^{i\phi}\, |\,1\, ,\, 1 \,\rangle_{AB} \right) \label{eq:PhotAddketSup}
\end{align}
where $\mathcal{N} =\, [\,1 \,+\, \tanh^{2}(r)\,]^{-1/2}$ is the normalization
constant for both desgaussified states. Thus, degaussified resources are shown to be
superpositions of the two-mode squeezed vacuum and the two-mode squeezed Fock state.

However, eqs.~(\ref{eq:PhotSubketSup}) and~(\ref{eq:PhotAddketSup}) substantially
differ in the exchange of the hyperbolic coefficients for the superposition, and in
the character of the state for vanishing squeezing. Even if both states become
separable for $r=0$, the photon-added squeezed state reduces to the non-Gaussian
two-mode Fock state; while the photon-subtracted squeezed state becomes the Gaussian
the two-mode vacuum.

The calculation of the characteristic functions for these degaussified resources
makes use of the same relations
(eqs.~(\ref{eq:TwoModBogDisp}),~(\ref{eq:TwoModBogChar}),~(\ref{eq:CharTrSqFock})
and~(\ref{eq:DispFockElem})) used in the derivation of the characteristic function of
the two-mode squeezed Fock state (eq.~(\ref{eq:CharSqFockTwo})). Recall that the
degaussified resources are superpositions of two-mode squeezed Fock states
(eqs.~(\ref{eq:PhotSubketSup}) and~(\ref{eq:PhotAddketSup})). The density matrices
have the form
\begin{align}
\hat{\rho}&=\,|\,K_{0}\,|^{2}|\,\zeta\,,\,0\,,\,0\,\rangle\:\langle\,\zeta\,,\,0\,,\,0\,|\:+\:|\,K_{1}\,|^{2}|\,\zeta\,,\,1\,,\,1\,\rangle\:\langle\,\zeta\,,\,1\,,\,1|\notag
\\
&+\,K_{0}^{*}K_{1}\,|\,\zeta\,,\,1\,,\,1\,\rangle\:\langle\,\zeta\,,\,0\,,\,0\,|\:+\:K_{1}^{*}K_{0}\,|\,\zeta\,,\,0\,,\,0\,\rangle\:\langle\,\zeta\,,\,1\,,\,1\,|\label{eq:DensNonGau}
\end{align}
with the factors $K_{0,1}$ taking the appropriate values for photon-subtracted or
added states. Terms in eq.~(\ref{eq:DensNonGau}) correspond to (Fock basis) matrix
elements of the displacement operator (see eq.~(\ref{eq:DispFockElem})) in the
calculation of the characteristic function. The characteristic functions of the
photon-subtracted and photon-added two-mode squeezed states are thus given by
\begin{align}
\chi_{PSS}^{\,(1,1)}\,(\xi_{A} ;\: \xi_{B})  = &\,
{\mathcal{N}}^{\,2}\;e^{-1/2\:\left(|\,\xi'_{A}\,|^{2}\,+\,|\,\xi'_{B}\,|^{2}\right)}
\;\{ 1 \,-\, 2 \tanh(r) \: \mathrm{Re}[e^{-\,i\, \phi}
\xi'_{A}\, \xi'_{B}] \notag \\
 & + \, \tanh^{2}(r)\: (1-|\,\xi'_{A}\,|^{2})\: (1-|\,\xi'_{B}\,|^{2}) \}
 \label{eq:CharPSS} \\
\notag \\
\chi_{PAS}^{\,(1,1)}(\xi_{A} ;\: \xi_{B})  = &\,
{\mathcal{N}}^{\,2}\;e^{-1/2\:\left(|\,\xi'_{A}\,|^{2}\,+\,|\,\xi'_{B}\,|^{2}\right)}
\;\{ \tanh^{2}(r) \,-\, 2 \tanh(r) \:\mathrm{Re}[e^{-\,i\, \phi} \xi'_{A}\, \xi'_{B}] \notag \\
& +\,  (1-|\,\xi'_{A}\,|^{2})\: (1-|\,\xi'_{B}\,|^{2}) \} \label{eq:CharPAS}
\end{align}
where the relation of the variables $\xi'_{A}$ and $\xi'_{B}$ to the variables
$\xi_{A}$ and $\xi_{B}$ is described by the Bogoliubov transformation of
eq.~(\ref{eq:TwoModBogChar}) for the displacement operator arguments.

Comparing eq.~(\ref{eq:CharGaussVac}) with
eqs.~(\ref{eq:CharSqFockTwo}),~(\ref{eq:CharPSS}), and~(\ref{eq:CharPAS}); we see
that the polynomial terms that define the non-Gaussian character of the state are
always multiplied by a Gaussian factor equal to the two-mode squeezed state
characteristic function of eq.~(\ref{eq:CharGaussVac}).

Lastly; it is worth remarking that the (non-Gaussian) two-mode photon-subtracted
squeezed state can be defined as the first-order truncation of the (Gaussian)
two-mode squeezed state. Let us consider the two-mode squeezed vacuum given by
$|-\,2\,r\,\rangle \,=\, \widehat{S}_{AB}(-\,2\,r)\:|\,0\,,\,0\,\rangle_{AB}$.
Recalling eq.~(\ref{eq:TwoModSqVacRFock}); such a state can be written as
\begin{align}
 |-2\,r\,\rangle_{AB} \,=&\,
\widehat{S}_{AB}(-r)\;\widehat{S}_{AB}(-r)\;|\,0\,,\,0\,\rangle_{AB}\notag \\
=&\,\widehat{S}_{AB}(-r)\;
\cosh^{-1}(r)\;\sum_{n=0}^{\infty}\:\left(\,\tanh(-r)\,\right)^{n}\:|\,n\,,\,n\,\rangle_{AB}
\label{eq:TruncatGauss}
\end{align}

Truncating the series of eq.~(\ref{eq:TruncatGauss}) beyond $n=1$ , we recover the
photon-subtracted resource state (eq.~(\ref{eq:PhotSubketSup})), with $\phi=\pi$;
that is $|1^{(-)}\,,\, 1^{(-)} \,; -r \,\rangle$. Moreover, this state coincides with
that of the photon-subtracted state introduced in ref.~\cite{KitagawaPhotsub} for the
ideal case of a beam splitter with unity transmittance.

\subsection{General non-Gaussian resources: Squeezed Bell-like states}
\label{NGau:Chzation:Bellike}

In order to unify the study of the properties, of the teleportation performance, and
of the experimental methods for the generation of the above mentioned resource
states, we take into consideration a class of states that have been named the
\textit{squeezed Bell-like states} in previous work~\cite{TelepNonGauss}. The state
vector for the squeezed Bell-like state is given by
\begin{equation}
|\,\Psi \,\rangle_{SBell} \, = \, \widehat{S}_{AB}(\zeta) \: \left( \,\cos(\delta)
\:|\,0\,,\,0\,\rangle_{AB} \; + \; e^{i \theta}\, \sin(\delta)\:
|\,1\,,\,1\,\rangle_{AB} \right) \label{eq:SqueBellket}
\end{equation}
which is a superposition of the two-mode squeezed vacuum and the two-mode squeezed
Fock state. The superposition coefficients are parameterized as $\cos(\delta)$ and
$e^{i\,\theta}\sin(\delta)$, which is convenient as the free parameters $\delta$ and
$\theta$ involved can be chosen arbitrarily; always having a normalized state vector.

The state in eq.~(\ref{eq:SqueBellket}) is two-mode squeezed. That it is Bell-like is
apparent on inspection of eq.~(\ref{eq:BellStatesDV}). For $r=0$ and $\delta=\pi/4$
the squeezed Bell-like state reduces to
$|\,0\,,\,0\,\rangle_{AB}\,+\,e^{i\theta}|\,1\,,\,1\,\rangle_{AB}$; the Bell state
for the two-dimensional Hilbert space spanned by Fock states $|\,0\,\rangle$ and
$|\,1\,\rangle$; and the ideal resource for discrete variables quantum
teleportation~\cite{Bennett} of states belonging to that two-dimensional Hilbert
space.

The squeezed Bell-like state will have as special cases all of the two-mode states
that we consider as teleportation resources in this chapter; for appropriate values
of the parameters $\delta$ and $\theta$, and will interpolate between these
resources. Taking $\zeta=r$, we obtain a Gaussian resource for $\delta=0$, a squeezed
Fock state for $\delta=\pi/2$, a photon-subtracted squeezed state for
$\delta=\cos^{-1}(\mathcal{N}),\theta=\pi$ and a photon-added squeezed state for
$\delta=\cos^{-1}(\mathcal{N}\tanh(r)),\theta=\pi$. The squeezed Bell-like state is
inseparable for $r=0$, except for the aforementioned trivial choices of $\delta$
corresponding to squeezed vacuum and squeezed Fock state.

The advantage of interpolating between known resource states allows for an unifying
study of their characteristics and of their performance as entangled resources for
quantum information in \textbf{CV}. This advantage is compounded by the possibility
of choosing the parameters $\delta$ and $\theta$ in an arbitrary manner, thus
\emph{sculpting} the entangled resource. For instance, given an input state, and an
analytical expression for teleportation fidelity; this fidelity can be optimized with
respect to the superposition parameters, for a fixed squeezing $r$. Even if squeezed
Bell-like states are considered nothing more than theoretical constructs for the
study of the teleportation fidelity for a wide variety of resources; it will always
be possible to pick a suitable "special case" to sculpt in an experimentally feasible
manner so as to better approximate the optimal squeezed Bell-like state resource.

The squeezed Bell-like state is, like the degaussified states, a superposition of
two-mode squeezed states. The density matrix for the squeezed Bell-like state is of
the general form described in eq.~(\ref{eq:DensNonGau}). Calculating the
characteristic function for this state involves repeating the procedure outlined
above for the degaussified states; for a superposition with coefficients
$\cos(\delta)$ and $e^{i \theta} \sin(\delta)$. The characteristic function of the
squeezed Bell-like state reads
\begin{align}
\chi_{SBell}(\xi_{A} ;\: \xi_{B}) = & \,
e^{-1/2\,\left(|\,\xi'_{A}\,|^{2}\,+\,|\,\xi'_{B}\,|^{2}\right)}\: \{
\cos^{2}(\delta) \:+\: 2 \cos(\delta)\,\sin(\delta)\, \mathrm{Re}[e^{\,i\, \theta}
\xi'_{A}\, \xi'_{B}]
\notag \\
 & \,+ \, \sin^{2}(\delta)\,(1\,-\,|\,\xi'_{A}\,|^{2})\,(1\,-\,|\,\xi'_{B}\,|^{2}) \} \label{eq:CharSqueBell}
\end{align}
where, as before; $\xi'_{A}$ and $\xi'_{B}$ are related to $\xi_{A}$ and $\xi_{B}$ by
the Bogoliubov transformation of eq.~(\ref{eq:TwoModBogChar}) for the displacement
operator arguments.

\subsection{Selected input states}
\label{NGau:Chzation:Input}

The following single-mode input states have been considered for the study of
teleportation with non-Gaussian resources, among a variety of Gaussian and
non-Gaussian states that are likely inputs both in \textbf{CV} quantum computation
and hybrid quantum computation: Coherent states; squeezed vacuum states;
single-photon Fock states; squeezed single-photon Fock states and photon-added
coherent states.

The characteristic functions for most of these one-mode states are straightforward to
derive, and can be found in the literature, for example in
refs.~\cite{Barnett,QOptWalls}.

For the coherent state $|\,\beta\,\rangle$ the characteristic function reads
\begin{equation}
\chi_{coh}(\xi_{in}) \,=\, e^{-\frac{1}{2}|\,\xi_{in}\,|^{2}\,+\,2\,i\,
\mathrm{Im}[\xi_{in}\beta^{*}]} \label{eq:CharCohIn}
\end{equation}

The one-mode squeezed vacuum state $|\,\varepsilon\,\rangle
\,=\,\widehat{S}(\varepsilon)\:|\,0\,\rangle$ (where $(\varepsilon = e^{i
\varphi}s)$) has a characteristic function given by
\begin{equation}
\chi_{sq}(\xi_{in}) \,=\, e^{-\frac{1}{2}|\,\xi'_{in}\,|^{2}} \label{eq:CharSqVacIn}
\end{equation}
where $\xi'_{in} \,=\, \xi_{in} \cosh(s) \,+\, \xi_{in}^{*} e^{i \varphi}\sinh(s)$, a
Bogoliubov transformation of variables.

The characteristic section for the squeezed Fock state
$\widehat{S}(\varepsilon)\:|\,1\,\rangle$ can be calculated easily given
eq.~(\ref{eq:DispFockElem}) and reads
\begin{equation}\label{eq:CharSqFockIn}
  \chi_{sqF}(\xi_{in}) =\,
e^{\,-\,\frac{1}{2}|\,\xi'_{in}\,|^{2}}\:(1\,-\,|\,\xi'_{in}\,|^{2})
\end{equation}
with $\xi'_{in}$ given by the above Bogoliubov transformation.

The characteristic function for the Fock state $|\,1\,\rangle_{in}$ is given by the
trivial limit $s=0$ of eq.~(\ref{eq:CharSqFockIn}).

The (non-Gaussian) photon-added coherent state is prepared by the application of the
creation operator $\hat{a}^{\dag}$ to a coherent state;
\begin{equation}\label{eq:CohPAddket}
  (1\,+\,|\,\beta\,|^{2})^{-1/2}\:\hat{a}^{\dag}\:|\,\beta\,\rangle
\end{equation}
where, for $\beta=0$ eq.~(\ref{eq:CohPAddket}) reduces to the single-photon Fock
state $|\,1\,\rangle$. The derivation of the characteristic function of the
photon-added coherent state is; however, straightforward
\begin{equation}\label{eq:CharCohPAdd}
 \chi_{pac}(\xi_{in}) \, =\,  (1\,+\,|\,\beta\,|^{2})^{-1}
\;e^{-\frac{1}{2}|\,\xi_{in}\,|^{2}\,+\,3\,i\,
\mathrm{Im}[\xi_{in}\,\beta^{*}]}\;(1\,+\,
|\,\beta\,|^{2}\,-\,|\,\xi_{in}\,|^{2}\,+\,2\,i\, \mathrm{Im}[\xi_{in}\,\beta^{*}])
\end{equation}

\section{Teleportation with degaussified and squeezed Fock resources}
\label{NGau:Degauss}

In this section, we compare the behavior of the teleportation fidelity for the input
states described above; using as resources the two-mode squeezed Fock state
(eq.~(\ref{eq:SqFocket})), and both the degaussified states
(eqs.~(\ref{eq:PhotSubketSup}) and~(\ref{eq:PhotAddketSup})).

We will assume ideal \textbf{CV} teleportation. Therefore, the output state
characteristic function is given by eq.~(\ref{eq:CharOutFin}) (for $g_{x}=g_{p}=1$),
and the teleportation fidelity is given by eq.~(\ref{eq:FidCharOutFin}). The fidelity
is analytically computable for the the input states and entangled resources
considered here; the integral in eq.~(\ref{eq:FidCharOutFin}) can be exactly
calculated in terms of finite sums of averages over Gaussian distributions.

The fidelities of teleportation will be analyzed and compared for fixed and equal
squeezing parameter $r$ for all resources. The performance of the Gaussian two-mode
squeezed state will also be displayed for reference purposes, as it is generally
considered a benchmark~\cite{CriteriaCVTelep,CriteriaCVTelep2}. This choice is made
to compare the performance of Gaussian and chosen non-Gaussian resources, given the
possession of the same technological means for their generation.

\subsection{Teleportation of Gaussian input states}
\label{NGau:Degauss:Gaussinp}

\begin{figure}[p] \label{fig:FidGaussIn}
\begin{centering}
\includegraphics[width=12cm]{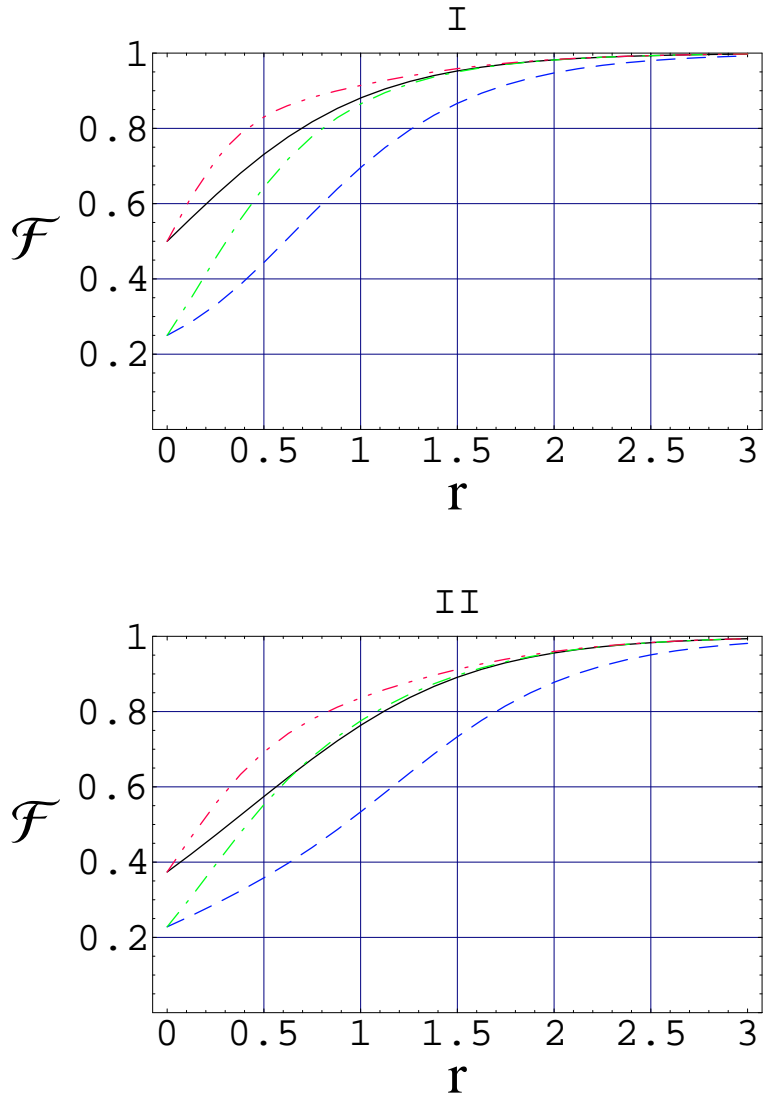}
\end{centering}
\caption[Teleportation fidelity using degaussified and two-mode squeezed Fock
resources, for the coherent and the squeezed vacuum input states]{Fidelity of
teleportation $\mathcal{F}$, as a function of the squeezing parameter $r$, with
$\phi=\pi$, for input coherent state $|\beta\rangle$ (Panel I) and input squeezed
vacuum state $|\,\varepsilon\,\rangle$ (Panel II). Comparison is given for different
two-mode entangled resources: \textit{(a)}~squeezed state (full line); \textit{(b)}
~squeezed Fock state (dashed line); \textit{(c)}~photon-subtracted squeezed state
(double-dotted, dashed line); \textit{(d)}~photon-added squeezed state (dot-dashed
line). In plot I the value of $\beta$ is arbitrary. In plot II the squeezing
parameter of the input state is fixed at $s=0.8$.}
\end{figure}

We begin our analysis with the teleportation of Gaussian input states. In
fig.~(\ref{fig:FidGaussIn}) we plot the fidelity of teleportation (see
eq.~(\ref{eq:FideChar})) having as inputs the coherent state $|\,\beta\,\rangle_{in}$
(Panel I), and the input squeezed vacuum state $|\,\beta\,\rangle_{in}$ (Panel II),
for the resource states introduced above, with the exception of the squeezed
Bell-like state. The phase of squeezing will be fixed as $\phi=\pi$ from now on; in
keeping with the conventions established in the \textbf{CV} protocol outlined in
chapter~\ref{Form}.

For both inputs in fig.~(\ref{fig:FidGaussIn}); the choice of the photon-subtracted
squeezed state (eq.~(\ref{eq:CharPSS})) as entangled resource is the most convenient
one. It corresponds to the highest value of the fidelity $\mathcal{F}$ for any fixed
value of the squeezing $r$ in the realistic range $[0,1]$. On the contrary, the
choice of the squeezed Fock state (eq.~(\ref{eq:CharSqFockTwo})) as entangled
resource is the least convenient, yielding the poorest performance; even when
compared to the Gaussian squeezed resource. Finally, regarding the use of the
photon-added squeezed state (eq.~(\ref{eq:CharPAS})) as entangled resource, it allows
for a very modest improvement in the fidelity when compared to the Gaussian resource;
and this only for a small interval of values around $r=1$. For smaller squeezing, its
performance is poorer than that of the Gaussian resource.

The poor performance of the photon-added and two-mode squeezed Fock resources is not
surprising for small squeezing parameters. The input in both cases is a Gaussian
state; and both these resource states reduce to a two-mode Fock state for $r=0$. In
Panel I of fig.~(\ref{fig:FidGaussIn}) we see that the overlap of a coherent input
state with a mixture of displaced~\footnote{Randomly displaced, given the
separability of the "resource"} coherent states for classical teleportation is $0.5$,
as expected from previous work on "classical"
teleportation~\cite{CriteriaCVTelep,CriteriaCVTelep2}. The overlap of the coherent
input state with a mixture of randomly displaced one-photon Fock states is $0.25$.
The probability for a given displaced element of the mixture is given by the
probability of a result equal to the displacement; when the joint measurement of
non-local quadratures of the $A$ and $in$ modes is performed (see
subsection~\ref{Back:Qtele:Fide}).

We must, finally, remark that the fidelity of teleportation is significantly higher
for the coherent state input (Panel I,fig.~(\ref{fig:FidGaussIn})) than for the
squeezed vacuum input (Panel II,fig.~(\ref{fig:FidGaussIn})), for \emph{all}
resources, for small $r$. This is due to the fact that the one-mode squeezing
operation has an asymptotic limit $s\,\rightarrow\infty$ in which all squeezed states
will approximate a quadrature eigenstate. In the case of quasi-classical
teleportation, for small $r$, the randomly displaced, nearly uniform mixture in mode
$B$ has little overlap with a squeezed input state; even for the moderate squeezing
($s=0.8$) chosen.

\subsection{Teleportation of non-Gaussian input states}
\label{NGau:Degauss:NGaussinp}

\begin{figure}[p] \label{fig:FidNonGaussIn1}
\begin{centering}
\includegraphics[width=12cm]{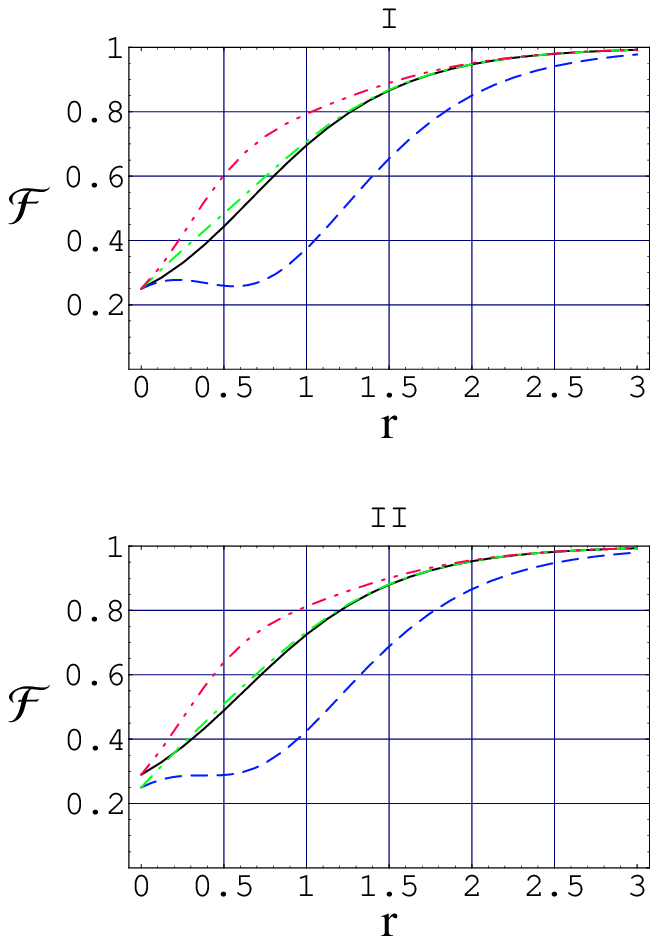}
\end{centering}
\caption[Teleportation fidelity using degaussified and two-mode squeezed Fock
resources, for the single-photon Fock and the photon-added coherent input
states]{Behavior of the fidelity of teleportation $\mathcal{F}$ as a function of the
squeezing parameter $r$, with $\phi=\pi$, for two different non-Gaussian input
states: The Fock state $|1\rangle$ (Panel I), and the photon-added coherent state
$(1+|\beta|^{2})^{-1/2}\hat{a}^{\dag}|\beta\rangle$ (Panel II). We compare the
performances obtained by using different two-mode entangled Gaussian and non-Gaussian
resources:
 \textit{(a)}~squeezed state (full line); \textit{(b)}~squeezed Fock state (dashed
line); \textit{(c)}~photon-subtracted squeezed state (double-dotted, dashed line);
 \textit{(d)}~photon-added squeezed state (dot-dashed line). In Panel II the value of
the coherent amplitude of the input photon-added coherent state is fixed at
$\beta=0.3$.}
\end{figure}

Let us now consider the teleportation performance for non-Gaussian input states; the
single-photon Fock state $|\,1\,\rangle_{in}$, the squeezed single-photon Fock state
$\widehat{S}_{in}(\varepsilon)\:|\,1\,\rangle_{in}$, and the photon-added coherent
state $(1\,+\,|\,\beta\,|^{2})^{-1/2}\:\hat{a}^{\dag}\:|\,\beta\,\rangle$. In
fig.~(\ref{fig:FidNonGaussIn1}), we plot the fidelity of teleportation for two
non-Gaussian input states: the single-photon Fock state (Panel I), and the
photon-added coherent state (Panel II).

In Panel I, we observe that both the degaussified resources lead to an improvement of
the fidelity with respect to the Gaussian resource. The photon-subtracted squeezed
resource again performs better than the photon-added resource; while the squeezed
Fock state yields the poorest performance when compared to all the resources. From
Panel II we see that once more the photon-subtracted resource yields the best
performance for any fixed squeezing parameter; that the photon-added squeezed
resource allows for a very modest improvement in the fidelity with respect to the
squeezed Gaussian reference for small squeezing.

Differently from the teleportation of Gaussian inputs (see
fig.~(\ref{fig:FidGaussIn})), the photon-added resource is no worse than the Gaussian
state for small squeezing $r$. That the two-mode squeezed Fock state is a poor
resource even at a small squeezing seems unusual. Let us recall that the overlap in
eq.~(\ref{eq:FidCharOutFin}) for classical or nearly classical teleportation
($r\,\rightarrow\,0$) in this case is that of a non-Gaussian, pure input state with a
mixture of randomly displaced Fock states; not with a single-photon Fock state.

It is interesting to note that teleportation fidelities (for the same squeezing value
$r$) are higher for all the resource states when the input is a photon-added coherent
input (compare Panels I and II of fig.~(\ref{fig:FidNonGaussIn1})); higher than when
the input is the Fock state that is the zero amplitude limit of the photon-added
coherent state. The improvement is small, however, for the input and resource states
we have considered. This is not unsurprising, given that the photon-added coherent
state, at amplitude $|\beta|^{2}\gg 1$ approximates a coherent state. The protocol
for \textbf{CV} teleportation seems more suitable for the teleportation of coherent
states than for the teleportation of Fock states, at least as far as our chosen
definition of fidelity is concerned.

\begin{figure}[ht] \label{fig:FidNonGaussIn2}
\begin{centering}
\includegraphics[width=12cm]{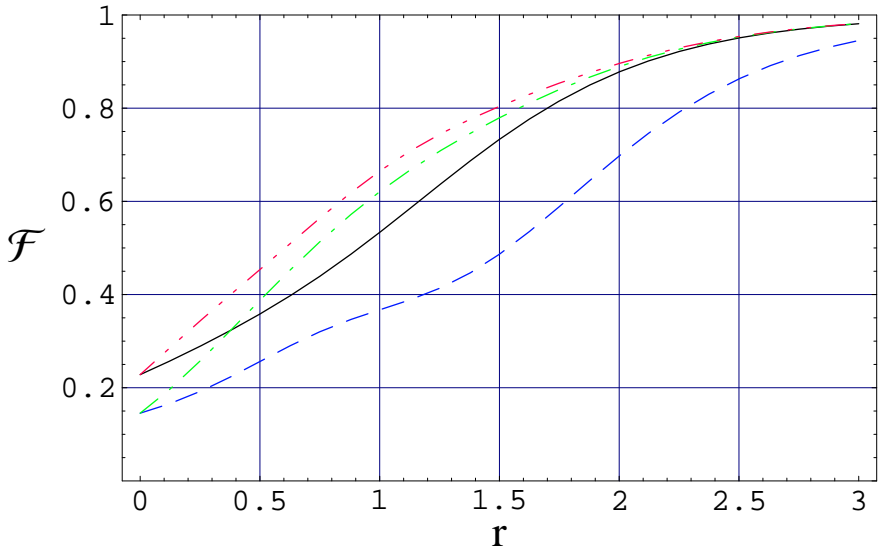}
\end{centering}
\caption[Teleportation fidelity using degaussified and two-mode squeezed Fock
resources, for the squeezed single-photon Fock input state]{Behavior of the fidelity
of teleportation $\mathcal{F}$ as a function of the squeezing parameter $r$, with
$\phi=\pi$, for the squeezed Fock input state $S(s)\;|\,1\,\rangle$, using different
two-mode Gaussian and non-Gaussian entangled resources:  \textit{(a)}~squeezed state
(full line); \textit{(b)}~squeezed Fock state (dashed line); \textit{(c)}~
photon-subtracted squeezed state (double-dotted, dashed line);
\textit{(d)}~photon-added squeezed state (dot-dashed line). The value of squeezing is
for the input state is fixed at $s=0.8$.}
\end{figure}

In fig.~(\ref{fig:FidNonGaussIn2}) we compare the fidelity of teleportation
$\mathcal{F}$ for the case of a squeezed Fock input state and different Gaussian and
non-Gaussian resources. Comparing with Panels I and II of
fig.~(\ref{fig:FidNonGaussIn1}), we see that the qualitative behavior of
teleportation is different from both previous examples, and more reminiscent of that
for the Gaussian inputs (see fig.~(\ref{fig:FidGaussIn})). This we explain by noting
that the one-mode squeezed Fock state is (at $s=0.8$) of a more Gaussian character
than the previous examples of Fock state and photon-added coherent state at small
amplitudes.

From all of the above investigations, we conclude that the photon-subtracted squeezed
state (eq.~(\ref{eq:PhotSubketSup})) is always to be preferred as entangled resource
compared either to the Gaussian ones or to non-Gaussian states that are obtained by
combining squeezing and photon pumping. The reason explaining this result will become
clear in the next sections when we will discuss the general class of states that
include as particular cases all the resources introduced so far; and attempt to
single out some properties of resource states that are related to improved
performance in quantum teleportation.

\section{Squeezed Bell-like
resources for teleportation} \label{NGau:SqueBell}

In this section we study the fidelity of \textbf{CV} teleportation of the squeezed
Bell-like state of eq.~(\ref{eq:SqueBellket}). The class of squeezed Bell-like states
includes as special cases all the resources studied so far, as well as the Bell
states of the two-dimensional Hilbert space spanned by the vacuum and single-photon
Fock states. Part of the analysis will consist in the optimization of the
teleportation fidelity with respect to the free superposition parameters of squeezed
Bell-like states. The optimal squeezed Bell-like state (for a given input state) can
always point the way to an adequate (if not optimal) choice of technically feasible
resources for teleportation; and to the further sculpturing of such resources. Even
if some states in the class of Bell-like states remain out of the reach of
experimental realization.  Moreover, the study of the properties of a wide class of
resources can be unified and interpolated by the use of squeezed Bell-like states,
not only regarding teleportation fidelity, but other uses as well.

The teleportation fidelity (eq.~(\ref{eq:FidCharOutFin})) is computable in an
analytical manner for all the input states considered, using the squeezed Bell-like
resource. It is an explicit function $\mathcal{F}(r,\phi,\delta, \theta)$ of the
independent parameters $r$, $\phi$, $\delta$ and $\theta$ that describe the squeezed
Bell-like resource.

We do not report here the explicit analytic expressions of the fidelities associated
to the squeezed Bell-like resource and to each input state, because they are rather
long and cumbersome. Some of them are reported in Appendix~\ref{APPA}.

An explicit analysis has established that all the fidelities are monotonically
increasing functions of the squeezing parameter $r$ at maximally fixed phase
$\phi=\pi$. In the following we assume that $\phi=\pi$ and $\theta=0$. It can be
checked that non vanishing values of $\theta$ do not lead to an improvement of the
fidelity in \textbf{CV} teleportation.

\subsection{Teleportation with a squeezed Bell state}
\label{NGau:SqueBell:SymBell}

At finite squeezing $r$ and for $\delta=\frac{\pi}{4}$ and $\theta=0$, state
(eq.~(\ref{eq:SqueBellket})) reduces to a squeezed Bell state. This resource is, as
we have remarked in subsection~\ref{NGau:Chzation:Bellike} intrinsically
nonclassical; as well as having as a $r=0$ limit the Bell state for the
two-dimensional Hilbert space spanned by the vacuum and the single-photon Fock state.

\begin{figure}[ht] \label{fig:Fidsqsuperpos1}
\begin{centering}
\includegraphics[width=12cm]{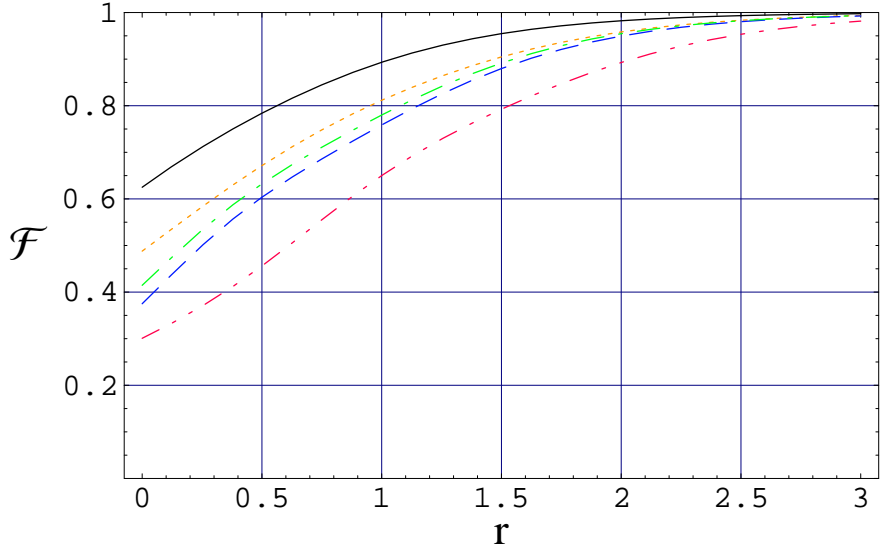}
\end{centering}
\caption[Teleportation fidelity using the squeezed Bell resource ($\phi=\pi$,
$\delta=\frac{\pi}{4}$, $\theta=0$) for all the selected input states]{Behavior of
the fidelity of teleportation $\mathcal{F}(r,\phi,\delta, \theta)$ associated to the
squeezed Bell resource with $\phi=\pi$, $\delta=\frac{\pi}{4}$, $\theta=0$, plotted
as a function of the squeezing parameter $r$ for the following input states:
\textit{(a)}~coherent state (full line); \textit{(b)}~squeezed state $|s\rangle =
S(s)|0\rangle$, with $s=0.8$ (dotted line); \textit{(c)}~Fock state $|1\rangle$
(dashed line); \textit{(d)}~photon-added coherent state
$(1+|\beta|^{2})^{-1/2}\hat{a}^{\dag}|\beta\rangle$, with $\beta=0.3$ (dot-dashed
line); \textit{(e)}~squeezed Fock state $|s\rangle = S(s)|1\rangle$, with $s=0.8$
(double-dotted, dashed line).}
\end{figure}

We may assess analytically the performance of such an entangled resource as far as
teleportation is concerned. In fig.~(\ref{fig:Fidsqsuperpos1}) we show the behavior
of the fidelity as a function of the squeezing parameter $r$, with $\phi=\pi$,
$\delta=\frac{\pi}{4}$ and $\theta=0$ for the five different input states considered
in the previous Section. It is straightforward to observe that the squeezed Bell
state; used as an entangled resource, leads to a relevant improvement of the
performance, when compared to all the other Gaussian and non-Gaussian resources that
we have investigated in the previous section.

The squeezed Bell state will always have a better performance for Gaussian inputs
than non-Gaussian inputs. The "best" non-Gaussian input, the small amplitude
($|\beta|^{2}=0.09$) photon-added coherent state, shows a noticeable improvement in
fidelity with respect to both Fock and squeezed Fock states. This suggests that the
\textbf{CV} teleportation protocol will always have a better performance for Gaussian
(or approximatively Gaussian) inputs.

\subsection{Optimized squeezed Bell-like resources for teleportation}
\label{NGau:SqueBell:OptBell}

We proceed to maximize, for every input state, the fidelity
$\mathcal{F}(r,\pi,\delta, 0)$ over the Bell-superposition angle $\delta$. At a fixed
squeezing $r=\tilde{r}$, we define the optimized fidelity as
\begin{equation}
\mathcal{F}_{opt}(\tilde{r}) \,=\, \max_{\delta} \;
\mathcal{F}(\,\tilde{r},\,\pi,\,\delta,\, 0\,) \; . \label{eq:FidOptim}
\end{equation}

For a coherent state input the maximization of $\mathcal{F}(\,r,\pi,\delta, 0\,)$, at
fixed $r$, leads to the following determination for the optimal Bell-superposition
angle $\delta_{max}^{(c)}$;
\begin{equation}
\delta_{\,max}^{(c)} \,=\, \frac{1}{2}\,\arctan(1\,+\,e^{-2\,r}\,)
\label{eq:deltaoptC}
\end{equation}

For a single-photon Fock state input, the optimal angle is given by
\begin{equation}
\delta_{\,max}^{(F)} \,=\, \frac{1}{2}\,\arctan\left(
\frac{e^{-2\,r}(1-e^{2\,r}+e^{4\,r}+3e^{6\,r})}{3(e^{2\,r}-1)^{2}} \right)
\label{eq:deltaoptF}
\end{equation}

\begin{figure}[ht] \label{fig:OptimalFid}
\begin{centering}
\includegraphics[width=12cm]{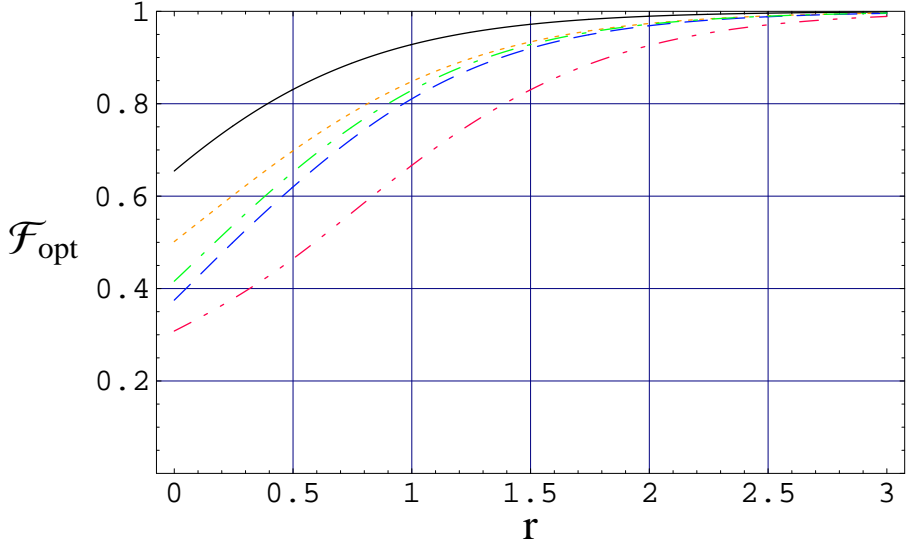}
\end{centering}
\caption[Teleportation fidelity $\mathcal{F}_{opt}(r)$; using optimal squeezed
Bell-like resources for each of the selected input states]{Plot of the fidelity
$\mathcal{F}_{opt}(r)$, using optimal squeezed Bell-like resources, as a function of
$r$ for the following input states: \textit{(a)}~coherent state (full line);
\textit{(b)}~squeezed vacuum $|\,s\,\rangle$, with $s=0.8$ (dotted line);
 \textit{(c)}~single-photon Fock state $|\,1\,\rangle$ (dashed line); \textit{(d)}~photon-added
coherent state $(1+|\beta|^{2})^{-1/2}\hat{a}^{\dag}|\beta\rangle$, with $\beta=0.3$
(dot-dashed line); \textit{(e)}~squeezed Fock state $|1,\,s\,\rangle$, with $s=0.8$
(double-dotted, dashed line).}
\end{figure}

We report, in fig.~(\ref{fig:OptimalFid}) the behavior of the optimized fidelities
$\mathcal{F}_{opt}(r)$ as functions of $r$ for all the input states considered in
this chapter. In all: coherent state, squeezed vacuum, single-photon Fock state,
photon-added coherent state and squeezed single-photon Fock state.

A relevant improvement of the fidelity is observed in all cases, even at vanishing
squeezing, due to the persistent nonclassical character of the optimal, squeezed
Bell-like entangled resource; even in the limit $r \rightarrow 0$.

The Gaussian states, and the photon-added coherent state show higher teleportation
fidelities than the wholly non-Gaussian states for all $r$ values. This result;
together with similar results in subsections~\ref{NGau:Degauss:NGaussinp}
and~\ref{NGau:SqueBell:SymBell} strongly suggest that some local operations; a
suitable displacement, for instance on a non-Gaussian input state \emph{previous} to
\textbf{CV} teleportation might improve the fidelity of teleportation, for any
resource used.

To quantify the increase in the probability of success for teleportation when we use
the optimal squeezed Bell-like state for a resource; we define a relative fidelity
coefficient $\Delta\,\mathcal{F} (r)$; relative with respect to the fidelity
$\mathcal{F}_{ref}(r,\pi)$ achieved using with one of the "special case" resources as
a benchmark . A natural definition of such a relative quantity is
\begin{equation}
\Delta\,\mathcal{F} (r) \,=\, \frac{\mathcal{F}_{\,opt\,}(r) -
\mathcal{F}_{\,ref\,}(r,\pi)}{\mathcal{F}_{\,ref\,}(r,\pi)} \label{eq:Deltafidelity}
\end{equation}

\begin{figure}[p] \label{fig:DeltaFid}
\begin{centering}
\includegraphics[width=12cm]{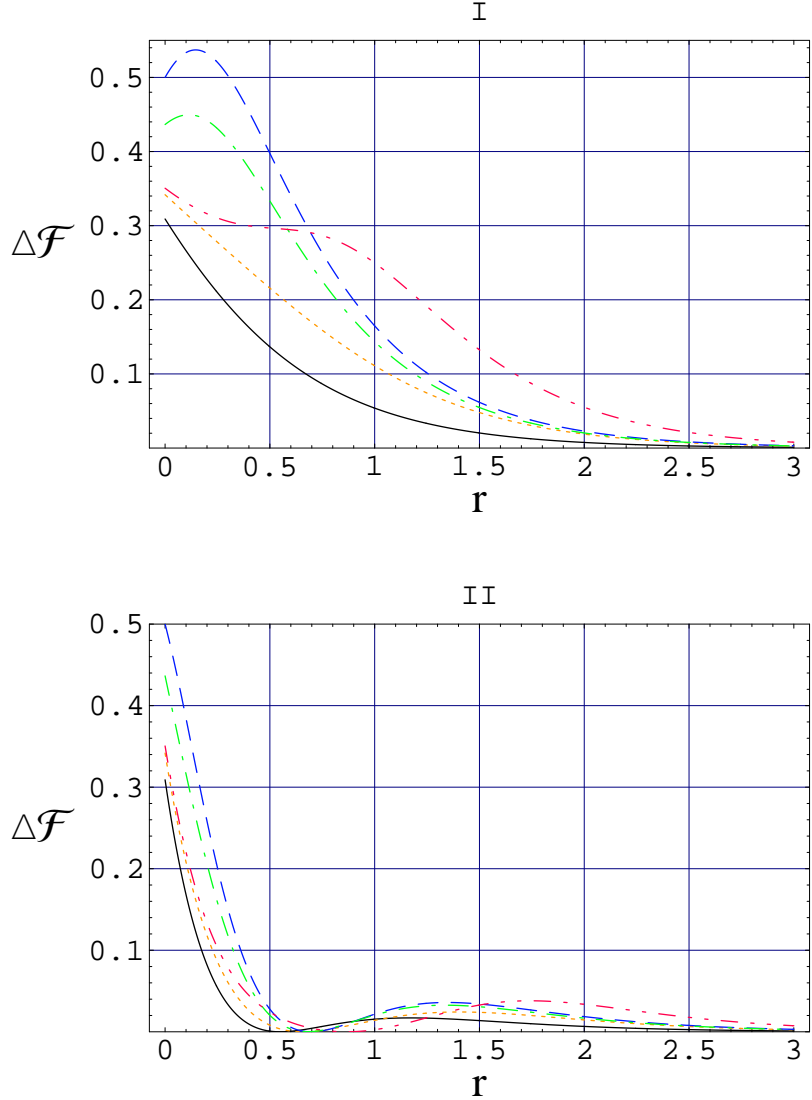}
\end{centering}
\caption[Relative teleportation fidelities using optimized squeezed Bell-like
resources for each of the selected input states; with respect to the two-mode
squeezed vacuum and the two-mode photon subtracted squeezed state]{Behavior of the
relative fidelity $\Delta\mathcal{F}(r)$ using an optimal squeezed Bell-like
resource, as a function of $r$ for the following input states: \textit{(a)}~coherent
state (full line); \textit{(b)}~squeezed state $|s\rangle$, with $s=0.8$ (dotted
line); \textit{(c)}~single-photon Fock state $|1\rangle$ (dashed line);
\textit{(d)}~photon-added coherent state
$(1+|\beta|^{2})^{-1/2}\hat{a}^{\dag}|\beta\rangle$, with $\beta=0.3$ (dot-dashed
line); \textit{(e)}~squeezed Fock state $|1,s\rangle$, with $s=0.8$ (double-dotted,
dashed line). In Panel I the benchmark resource is the two-mode squeezed vacuum; in
Panel II the benchmark resource is the two-mode photon-subtracted squeezed state.}
\end{figure}

In fig.~(\ref{fig:DeltaFid}), we plot the relative fidelity $\Delta\,\mathcal{F}(r)$
as a function of the squeezing parameter $r$ for two different choices of benchmark
resources that we have deemed the most interesting, given the high values of the
fidelities obtained when using them. In Panel I, the benchmark resource is the
Gaussian two-mode squeezed vacuum; in Panel II the benchmark resource is the
non-Gaussian two-mode photon-subtracted squeezed state.

In Panel I of fig.~(\ref{fig:DeltaFid}) we see that at fixed squeezing, the use of
optimized squeezed Bell-like resources leads to a strong percent enhancement of the
teleportation fidelity (up to more than $50\%$) with respect to that attainable
exploiting the standard two-mode squeezed vacuum. The greatest relative improvements,
and the \emph{maxima} in relative fidelity, occur in the teleportation of the
non-Gaussian inputs, particularly with the Fock state and the photon-added coherent
state (for which the relative improvement is not as great). Obviously, in the
asymptotic limit of very large squeezing, the two resources converge to unity
teleportation efficiency, as the resources themselves converge to \textbf{EPR}
states.

Panel II shows that the use of the optimized squeezed Bell-like resource leads to a
significant advantage with respect to the use of the photon-subtracted squeezed state
resource for low values (up to $r \simeq 0.5$) of the squeezing. The different curves
corresponding to the different input states, exhibit the same qualitative behavior;
however, the greater relative improvements are those for the fidelity of
teleportation of non-Gaussian inputs, particularly the Fock state and the
photon-added coherent state (again, with a lesser relative improvement than the Fock
state). Starting from large values, $\Delta\mathcal{F} (r)$ decreases monotonically
and vanishes at different points in the interval $[0.5 \leq r \leq 0.9]$. After
vanishing, the relative fidelity exhibits peaks at different intermediate values of
the squeezing, before vanishing asymptotically for large values of $r$.
Unsurprisingly; for these \emph{minima} values of $r=\bar{r}$ such that
$\Delta\mathcal{F} (\bar{r}) = 0$, the optimized Bell-like state and the
photon-subtracted squeezed state are identical.

\section{A comparison of entanglement, non-Gaussian character and Gaussian affinity of resource states}
\label{NGau:Compaprop}

In this section, we analyze the characteristics of the resource states studied in
this chapter, to determine which properties of the resources have an influence on
teleportation fidelity. Particularly, we study the entanglement and non-Gaussian
character of optimized Bell-like states and compare them with those of
photon-subtracted and photon-added two mode squeezed states. This comparison is to be
particularly illuminating, given the difference in teleportation fidelity obtained
using these degaussified states as resources; and given the very different
characteristics (obvious for $r\rightarrow 0$) of these two types of resources that
seem otherwise similar.

First, we study entanglement, as given for pure states by the von Neumann Entropy.
Then, we study non-Gaussian character as given by a measure of the distance between
the resource under study and a reference Gaussian state having the same covariance
matrix and first order averages as the resource. Thirdly, we define and study
Gaussian affinity as given by the overlap of the resource and a two-mode squeezed
vacuum, maximized over the squeezing parameter of the latter.

We analyze these measures in conjunction for the resources above mentioned, paying
special attention to the behavior of these measures for optimized Bell-like states.

\subsection{The von Neumann entropy of non-Gaussian resources}
\label{NGau:Compaprop:EntropyAll}

The bipartite entanglement; the \textit{Schmidt rank} of the pure non-Gaussian states
chosen for study in this chapter can be quantified in an unique manner using the
partial Von Neumann entropy (entropy of entanglement) $E_{vN}$. Given the density
operator for the two-mode state $\hat{\rho}$, the von Neumann entropy is equal to the
Shannon~\cite{BennettEntropy} entropy of the partial traces:
\begin{equation}\label{eq:VonNeumannEnt}
  E_{vN}=\,-\mathrm{Tr}\left(\,\hat{\rho}^{\,Tr(A)}\:\log_{2}(\hat{\rho}^{\,Tr(A)})\right)\;=\;-\,\mathrm{Tr}\left(\,\hat{\rho}^{\,Tr(B)}\:\log_{2}(\hat{\rho}^{\,Tr(B)})\right)
\end{equation}
where the partial traces are indicated by
$\hat{\rho}^{\,Tr(A)}=\,\mathrm{Tr}_{A}(\hat{\rho})$, and by
$\hat{\rho}^{\,Tr(B)}=\,\mathrm{Tr}_{B}(\hat{\rho})$

To simplify calculations involving powers of the density operator of the
aforementioned resource states (see section~\ref{NGau:Chzation}), the series
expansion for the logarithmic function in eq.~(\ref{eq:VonNeumannEnt}) has been
truncated to the first-order.

\begin{figure}[ht] \label{fig:vonNeumann}
\begin{centering}
\includegraphics[width=12cm]{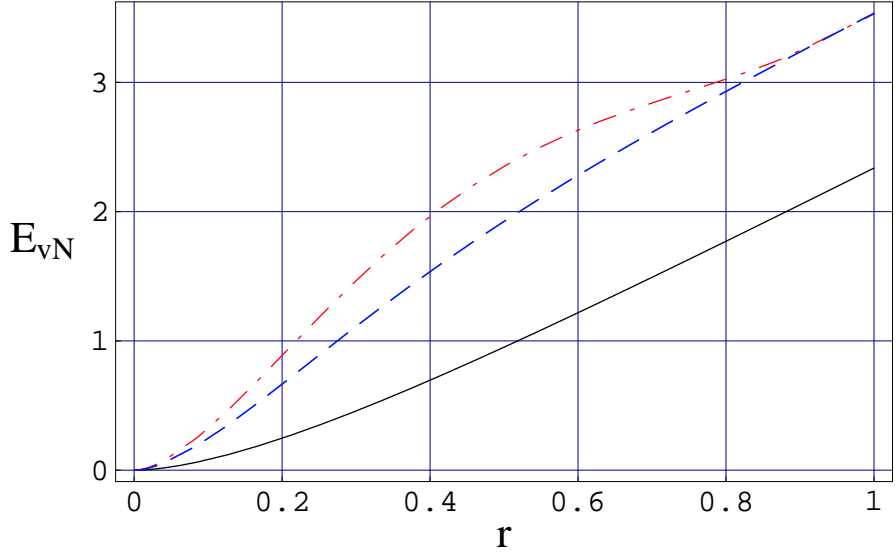}
\end{centering}
\caption[The von Neumann entropy for the two-mode squeezed Fock and degaussified
resource states]{Behavior of the von Neumann entropy $E_{vN}$ for two-mode squeezed
Fock, two-mode photon-added and two-mode photon subtracted squeezed states, as a
function of the modulus squeezed parameter $r$. The upper curve (dot-dashed line)
corresponds to the squeezed Fock state $|r\,; 1\,, 1\,\rangle$; the intermediate
curve (dashed line) corresponds equivalently to the photon-subtracted squeezed state
$|1^{(-)}\,, 1^{(-)} \,;\: r \,\rangle$ and to the photon-added $|1^{(+)}\,,\,
1^{(+)} \,;\: r \rangle$ squeezed state . The lower curve corresponds to the Gaussian
two-mode squeezed state $|\,r\,\rangle$.}
\end{figure}

For the two-mode squeezed Fock and the degaussified states, the von Neumann entropy
depends only on the modulus $r$ of the squeezing parameter $\zeta$. It is plotted in
fig.~(\ref{fig:vonNeumann}) and compared to that of the Gaussian two-mode squeezed
state.

At a given squeezing, all the non-Gaussian states show an entanglement larger than
that of the Gaussian two-mode squeezed vacuum. In the range of experimentally
realistic values $0<r<1$ of the squeezing, the two-mode squeezed Fock state is the
most entangled state. Moreover, the photon-added and the photon-subtracted two-mode
squeezed states exhibit exactly the same amount of entanglement at any $r$.

However, the results in subsections~\ref{NGau:Degauss:Gaussinp}
and~\ref{NGau:Degauss:NGaussinp} indicate that, for a given squeezing $r$, a higher
von Neumann Entropy (and greater entanglement) does not correspond to a higher
teleportation fidelity. The two-mode squeezed Fock state obtains the smallest
teleportation fidelities of any resource while having the greatest von Neumann
entropies. And the photon-added two-mode squeezed state obtains lower teleportation
fidelities than the photon-subtracted state, even if they are indistinguishable by
their von Neumann entropy.

\begin{figure}[p] \label{fig:vonNeumSBell}
\begin{centering}
\includegraphics[width=12cm]{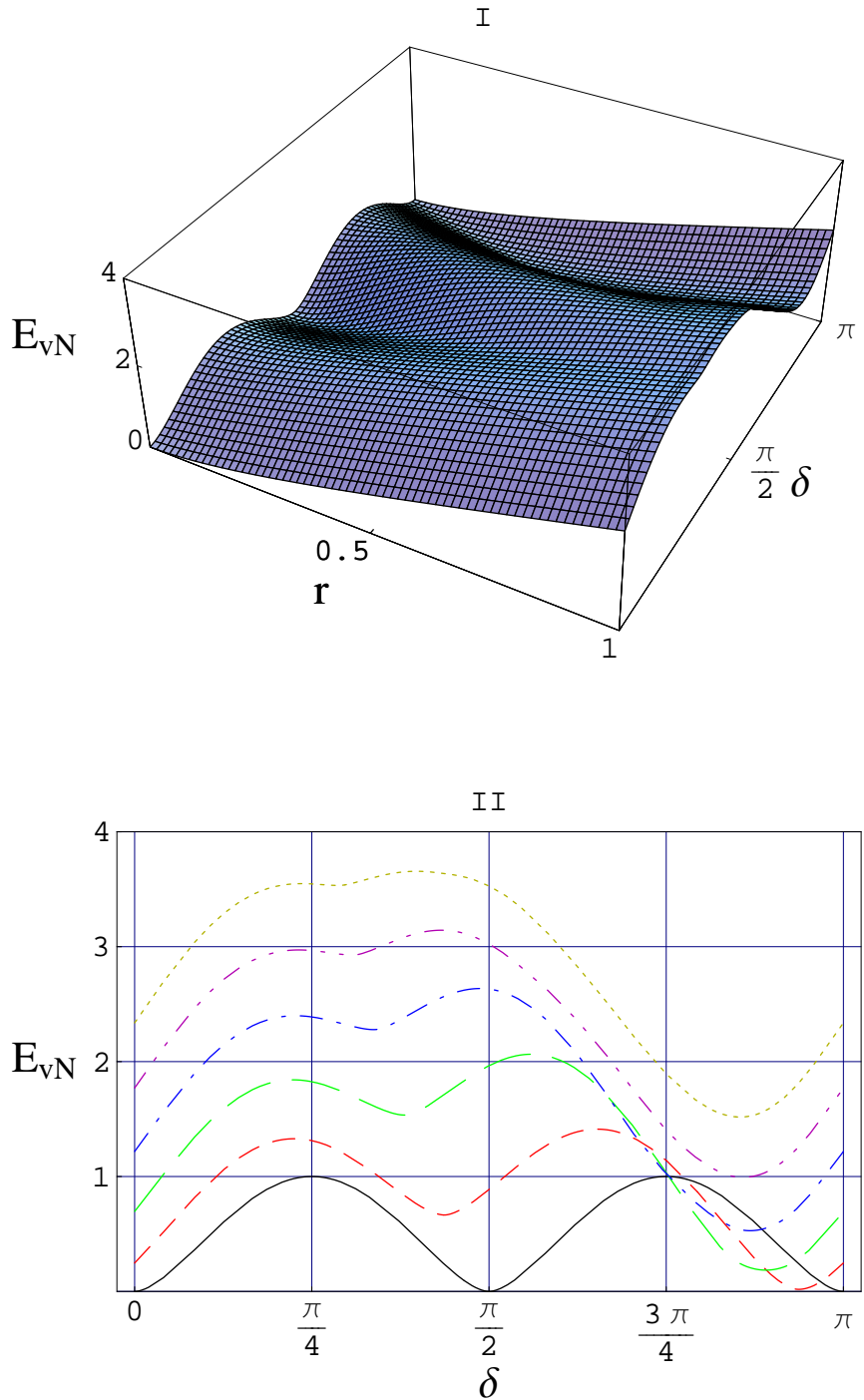}
\end{centering}
\caption[The von Neumann entropy for the squeezed Bell-like state as a function of
$r$ and $\delta$.]{The von Neumann entropy $E_{vN}$ for the squeezed Bell-like state
as a function of its defining parameters $r$ and $\delta$. Panel I displays the
three-dimensional plot of $E_{vN}$. Panel II displays two-dimensional projections at
fixed squeezing strength $r$. Curves from bottom to top correspond to the different
sections of $E_{vN}$ as functions of $\delta$ for $r=0,\,0.2,\,0.4,\,0.6,\,0.8,\,1$.}
\end{figure}

The von Neumann entropy of the general squeezed Bell-like states is plotted in
fig.~(\ref{fig:vonNeumSBell}). In panel I we plot $E_{vN}$ as a function of $r$ and
$\delta$. In panel II, we can observe how the regular, oscillating behavior of the
entropy for the Bell-like state $(r=0)$ becomes gradually deformed by the optical
pumping $(r\neq 0)$, leading to a peculiar pattern of correlation properties for the
squeezed Bell-like state.

Note that for $\delta=0$ and $\delta=\pi/2$, we have the von Neumann entropy of the
Gaussian state and the squeezed Fock state, respectively in Panel II of
fig.~(\ref{fig:vonNeumSBell}).

The squeezed Bell-like states have maxima of von Neumann Entropy for $r \rightarrow
0$ near $\pi/4$ and $3\pi/4$. However, it suffices to evaluate
eqs.~(\ref{eq:deltaoptC}) and~(\ref{eq:deltaoptF}) for $r=0$ (recall that the
Bell-like state is entangled even if it is not squeezed) to realize that, while the
optimal fidelity for the teleportation of Fock state inputs is obtained for
$\delta^{(F)}_{max}(r=0)=\pi/4$ (the Bell state of discrete variables settings), this
is not the case for the coherent state inputs, with $\delta^{(C)}_{max}(r=0)=0.554$.
Even in the case of Bell state resources for \textbf{CV} teleportation, the greatest
von Neumann entropy does not maximum determine fidelities for certain classes of
states.

\subsection{Optimized squeezed Bell-like resources: A comparison of von Neumann entropies}
\label{NGau:Compaprop:EntropyOpt}

\begin{figure}[ht] \label{fig:vonNeumSBelloptCF}
\begin{centering}
\includegraphics[width=12cm]{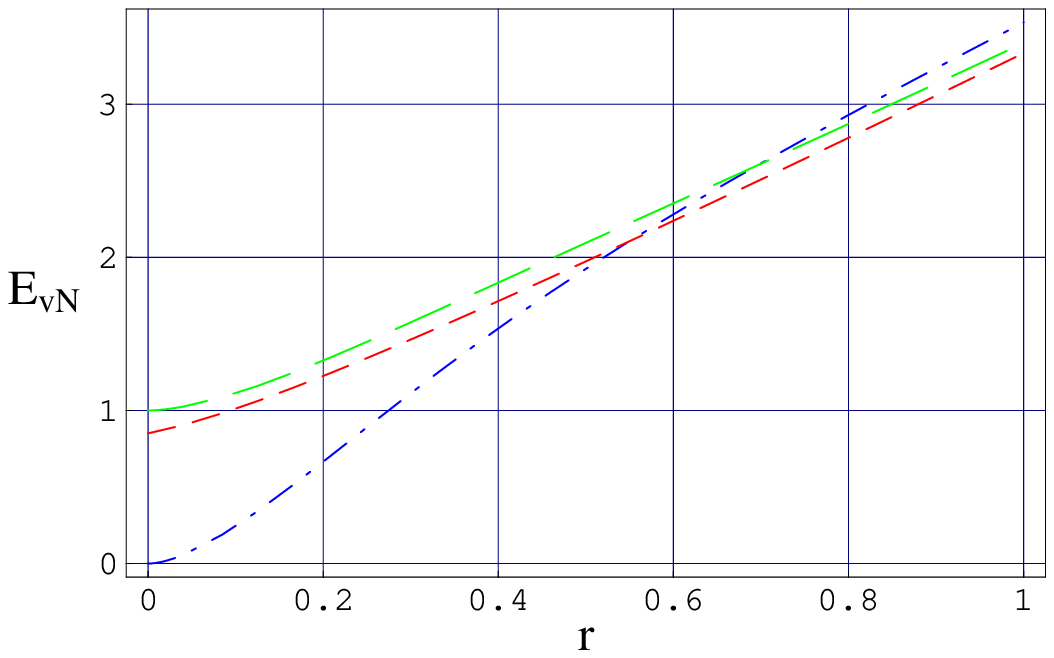}
\end{centering}
\caption[The von Neumann entropy of squeezed Bell-like states optimized for coherent
and Fock state inputs: a comparison with the degaussified states]{The von Neumann
entropy $E_{vN}$ for the optimized squeezed Bell-like state, as a function of $r$.
Dashed line: Optimized resource for a coherent state input, $\delta =
\delta_{max}^{(c)}$; Long dashed line: Optimized resource for a Fock state input,
$\delta = \delta_{max}^{(F)}$. The von Neumann entropy of the degaussified states is
reported for comparison purposes (dot-dashed line).}
\end{figure}

In fig.~(\ref{fig:vonNeumSBelloptCF}), we show the behavior of the von Neumann
entropy $E_{vN}$ for two different squeezed Bell-like resources; the first one
optimized for the teleportation of an input coherent state, with
$\delta_{max}^{(c)}(r)$ (see eq.~(\ref{eq:deltaoptC})); the second one optimized for
the teleportation of a single-photon Fock state, with $\delta_{max}^{(F)}(r)$ (see
eq.~(\ref{eq:deltaoptF})). This behavior is compared with that of the von Neumann
entropy of both the degaussified states. The intersections between the curves in the
figure correspond to the values os squeezing $\bar{r}$ for which the squeezed
Bell-like state reduces to a photon-subtracted or to a photon-added squeezed state.

It can be noticed in the range $0 < r < \bar{r}$; in which the fidelity of
teleportation using optimized Bell-like resources is substantially higher than that
for the photon-subtracted squeezed resource (see fig.~(\ref{fig:DeltaFid}), Panel
II), that the entanglement (quantified by the von Neumann Entropy) of the squeezed
Bell-like state is always larger than that of the photon-subtracted (and
photon-added) squeezed states. Therefore, a partial explanation of the better
performance of squeezed Bell-like resources lies in their higher degree of
entanglement compared to other non-Gaussian resources at lower levels of squeezing.
However, from the graphs it can be seen that for $r
> \bar{r}$ the entanglement of photon-subtracted and/or
added resources is larger; nevertheless, the fidelity of teleportation is below that
associated to optimized squeezed Bell-like resources. Much lower, in the case of the
photon-added resource, we remark.

Therefore, entanglement is not the only parameter necessary to determine the
teleportation performance of different non-Gaussian resources. The example chosen
here validates this conclusion specially well; nonetheless because the squeezed
Bell-like state is a generalization of degaussified resources, that reduces to a
degaussified resource for certain values of the superposition angle $\delta$.

\subsection{The non-Gaussianity: A character measure}
\label{NGau:Compaprop:NonGaussianity}

We have concluded that entanglement alone does not suffice to determine the
teleportation fidelity for a non-Gaussian resource. We have seen that the two-mode
squeezed Fock state, the wholly non-Gaussian resource having the highest entanglement
of the resources studied in this work, performs poorly as a teleportation resource.
We have seen that the degaussified states, resources of a different character; but
equal entanglement, perform the teleportation of the same input with very different
fidelities. We have seen that the optimal choices of squeezed Bell-like resources for
teleportation (at a fixed squeezing $r$) do not always correspond to maximal
entanglement. We have seen notable differences in the teleportation fidelities of
Gaussian and non-Gaussian inputs, using the same resource.

It is natural, as we study non-Gaussian resources and observe improvements in
teleportation fidelity over Gaussian resources, to look for a quantification of the
non-Gaussian character of the different resources we have considered, and in general
of the optimized squeezed Bell-like resources in order to compare their performance
in teleportation with respect to both this quantity and the entanglement in
conjunction. The task is to define a reasonable measure of \textit{non-Gaussianity}
that is endowed with some nontrivial operative meaning.

Recently, inspired by work~\cite{ExtremalGaussian} on the extremal nature of Gaussian
states at fixed covariance matrix; a measure of non-Gaussianity has been introduced
in terms of the Hilbert-Schmidt distance between a given non-Gaussian state and a
reference Gaussian state with the same covariance matrix~\cite{GenoniNonGaussy}.
Given a generic state with density operator $\hat{\rho}$; its non-Gaussian character
can be quantified through the distance $d_{nG}$ between $\hat{\rho}$ and the
reference Gaussian state $\hat{\rho}_{G}$, defined according to the following
relation:
\begin{equation}
d_{nG} =
\frac{\mathrm{Tr}((\,\hat{\rho}\,-\,\hat{\rho}_{G}\,)^{2})}{2\:\mathrm{Tr}(\,\hat{\rho}^{2}\,)}=
\frac{\mathrm{Tr}(\,\hat{\rho}^{2}\,)\:+\:Tr(\,\hat{\rho}_{G}^{2}\,)\:-\:2\,\mathrm{Tr}(\,\hat{\rho}\,
\hat{\rho}_{G})\,}{2\: \mathrm{Tr}(\,\hat{\rho}^{2}\,)} , \label{eq:nonGaussianity}
\end{equation}
where the Gaussian state $\rho_{G}$ is completely determined by the same covariance
matrix and the same first order average values of the quadrature operators associated
to state $\rho$.

\begin{figure}[p] \label{fig:nonGaussMeas}
\begin{centering}
\includegraphics[width=12cm]{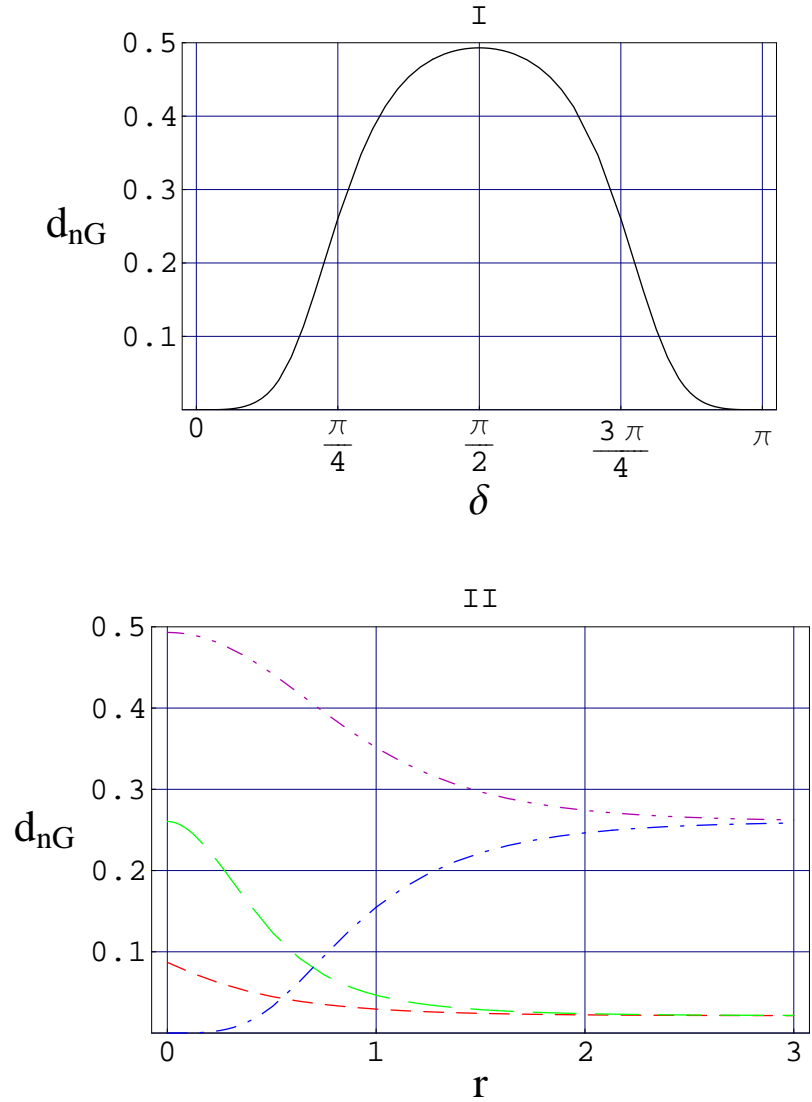}
\end{centering}
\caption[Non-Gaussianity measure for squeezed Bell-like resources: As a function of
$\delta$ for arbitrary $r$; and optimized for teleportation of coherent and Fock
state inputs]{Non-Gaussianity measure $d_{nG}$ for the squeezed Bell-like state.
Panel I shows $d_{nG}$ for the squeezed Bell-like state as a function of $\delta$,
for arbitrary $r$. Panel II shows $d_{nG}$ for the squeezed Bell-like state as a
function of $r$, for $\delta$ fixed at the optimized values for coherent state inputs
$\delta = \delta_{max}^{(C)}$ (dashed line); and $\delta = \delta_{max}^{(F)}$ for
Fock state inputs (long dashed line). For comparison, the values of $d_{nG}$ for the
photon-subtracted resource (dot-dashed line) and photon-added resource (double
dotted, dashed line) are also reported.}
\end{figure}

Using this definition, in fig.~(\ref{fig:nonGaussMeas}) we report the behavior of the
{\it non-Gaussianity} $d_{nG}$ for the squeezed Bell-like state.

The quantity $d_{nG}$ depends only on the superposition angle $\delta$ of the
squeezed Bell-like state (see Panel I), as the non-Gaussianity of the state cannot
change under squeezing operations; the squeezing induced on the state measured would
translate into squeezing of the reference Gaussian state. For $\delta$ in the
interval $[0,\pi]$, $d_{nG}$ attains its maximum at $\delta=\frac{\pi}{2}$: At that
point, the squeezed Bell-like state reduces to a two-mode squeezed Fock state. It is
expected for a (two-mode squeezed) Fock state to be more non-Gaussian than a
(two-mode squeezed) superposition of the vacuum and of a Fock state.

In Panel II, we plot the behavior of $d_{nG}$ for the squeezed Bell-like resources
optimized for the teleportation of a coherent state input and a single-photon Fock
state input, respectively, with $\delta = \delta_{max}^{(C)}$ and $\delta =
\delta_{max}^{(F)}$. For comparison, we plot as well the non-Gaussianity $d_{nG}$ for
the photon-added and the photon-subtracted squeezed states. The intersection points
occur once again at the points $\bar{r}$ where the optimized squeezed Bell-like
states reduce to the photon-subtracted squeezed states. For $r$ in the range
$[0,\bar{r}]$, the optimized squeezed Bell-like resources are not only more
entangled; they are more non-Gaussian than the photon-subtracted squeezed states. We
note that for $\lim_{r\rightarrow \infty} \delta_{max}^{(C)} \,=\,\lim_{r\rightarrow
\infty} \delta_{max}^{(F)} \,=\, 1$. Thus, for very large squeezing the two optimized
squeezed Bell-like resources tend to the state $\widehat{S}_{AB}(-r)\;\{\cos
\frac{\pi}{8}|\,0\,,\,0\,\rangle_{AB}\, +\, \sin
\frac{\pi}{8}\,|\,1\,,\,1\,\rangle_{AB}\}$, exhibiting a dominating Gaussian
component. On the other hand, for large $r$, the squeezed photon-added and
photon-subtracted squeezed states asymptotically tend to a squeezed Bell state
(corresponding to $\delta_{max}=\frac{\pi}{4}$), which has balanced Gaussian and
non-Gaussian contributions. While for smaller values of $r$, they have very different
values of non-Gaussianity; which is not surprising if we recall the enormous
difference in character of these states for $r\rightarrow\,0$.

It is also remarkable that, for $r$ in the range $[0,\bar{r}]$, that the optimal
squeezed Bell-like resources for teleportation of the non-Gaussian Fock state
($\delta_{max}^{(F)}$) have a much higher non-Gaussianity than those optimal squeezed
Bell-like resources for teleportation of the Gaussian coherent state
($\delta_{max}^{(C)}$). This might be understandable for completely "classical"
teleportation with an squeezed resource of the type studied so far, where fidelity is
just the overlap of the input state $in$ with a mixture of states constituted of
different instances of random displacements of mode $B$. But this is not the case
here; the squeezed Bell-like resources are always entangled even for $r=0$, with the
exception of some trivial choices for $\delta$.

\subsection{The Gaussian resource affinity}
\label{NGau:Compaprop:Affinity}

We have developed a non-Gaussianity measure for entangled resources that is in effect
a distance between the resource state we are studying and a Gaussian state of the
same covariance matrix. But this Gaussian reference is different for every resource
studied, and the interest of such a measure lies in making comparisons of the
non-Gaussianity for different non-Gaussian resources. Moreover, we do not know that
the Gaussian state constructed for every particular measure is a resource for
teleportation, for it is not described as such.

Observing that the squeezed Bell-like states and the photon-added and
photon-subtracted squeezed states are all obtained through a degaussification
protocol from a pure squeezed state, one could modify the definition of
eq.~(\ref{eq:nonGaussianity}) by taking the two-mode squeezed vacuum
$|\,\zeta'\,\rangle_{AB}$ $(\zeta'=r'\, e^{i\phi'})$ as the universal reference
Gaussian state $\hat{rho}_{G}$. This choice makes it possible to compare our
non-Gaussian resources to an unique reference that is also a well studied
teleportation resource. This will be particularly important if we are to analyze
measures of non-Gaussianity, teleportation fidelities and von Neumann entropies
together.

Adopting this modified definition, and observing that the non-Gaussian states to be
compared and the reference Gaussian state are all pure, eq.~(\ref{eq:nonGaussianity})
reduces to $d_{nG} \,=\, \min_{r',\,\phi'} \{1-Tr[\hat{\rho} \, \hat{\rho}_{G}]\}$,
where the minimization is constrained to run over the squeezing parameters $\zeta'$
of the reference twin-beam. However, it turns out that this modified definition
provides results and information qualitatively analogous to those obtained by
applying the original definition.

There is still one, and apparently opposite property that plays a crucial role in the
sculpturing of an optimized non-Gaussian entangled resource. From
figs.~(\ref{fig:vonNeumSBelloptCF}) and~(\ref{fig:nonGaussMeas}) we see that at
sufficiently large squeezing the photon-added and photon-subtracted squeezed
resources have entanglement comparable to that of the optimized squeezed Bell-like
states and, moreover, possess stronger non-Gaussianity. Yet they are not able to
perform better than the optimized Bell-like resources. This fact can be understood as
follows, leading to a definition of {\em squeezed vacuum affinity}: It is well known
that the Gaussian two-mode squeezed state in the limit of infinite squeezing realizes
the \textbf{EPR} state which is the equivalent of the Bell
(eq.~(\ref{eq:BellStatesDV})) state for two-dimensional systems. These two ideal
resources, respectively in the \textbf{CV} and discrete variables case, allow perfect
quantum teleportation with maximal fidelity. Therefore, we argue that, even when
exhibiting enhanced properties of non-Gaussianity and entanglement, any efficient
resource for \textbf{CV} quantum information tasks should enjoy a further property,
i.e. to resemble the form of a two-mode squeezed vacuum as much as is possible, in
the large $r$ limit. The {\em squeezed vacuum affinity} can be quantified by the
following maximized overlap:
\begin{equation}
\mathcal{G} \,=\, \max_{s} |\,_{AB}\langle\,
-s\,|\,\psi_{res}(r)\,\rangle_{AB}\,|^{2} \label{eq:OverlapTWB}
\end{equation}
where $|\,-s\,\rangle_{AB}$ is a two-mode squeezed vacuum with real squeezing
parameter $-s$, and $|\,\psi_{res}(r)\,\rangle_{AB}$ is any entangled two-mode
resource that depends on the squeezing parameter $r$. This definition applies
straightforwardly to the photon-added and photon-subtracted squeezed resources, and
as well to the squeezed Bell-like resources whenever they are optimized with respect
to a particular input state. The maximization over $s$ is imposed in order to
determine, at fixed $r$, the two-mode squeezed vacuum that is most affine to the
non-Gaussian resource being considered.

\begin{figure}[ht] \label{fig:OverlapTWBRes}
\begin{centering}
\includegraphics[width=12cm]{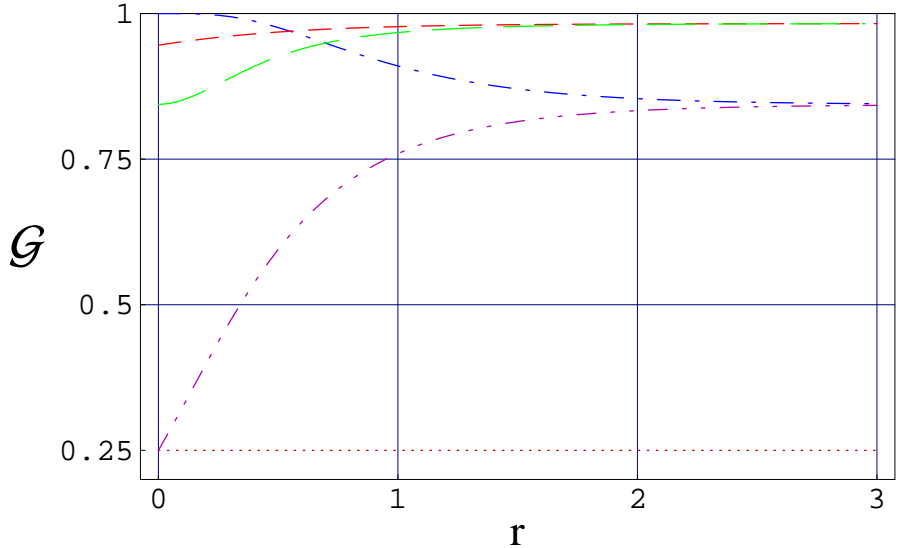}
\end{centering}
\caption[Maximized overlap $\mathcal{G}$ between a two-mode squeezed vacuum and
different non-Gaussian entangled resources as a function of $r$]{Maximized overlap
$\mathcal{G}$ between a two-mode squeezed vacuum and different non-Gaussian entangled
resources as a function of $r$: squeezed Bell-like state with $delta$ fixed at the
optimized value $\delta = \delta_{max}^{(c)}$ for coherent state inputs (dashed
line); the same state with $delta$ fixed at the optimized value $\delta =
\delta_{max}^{(F)}$ for Fock state inputs (long dashed line). For comparison purposes
the maximized overlap with the photon-subtracted squeezed state (dot-dashed line);
photon-added squeezed state (double-dotted, dashed line); and the single-photon
squeezed Fock state (dotted line) are displayed as well.}
\end{figure}

In fig.~(\ref{fig:OverlapTWBRes}) we study the behavior of the overlap $\mathcal{G}$
as a function of the squeezing $r$ for different non-Gaussian entangled resources.

Comparing fig.~(\ref{fig:nonGaussMeas}) and fig.~(\ref{fig:OverlapTWBRes}) we can see
that the behaviors of the non-Gaussianity and squeezed vacuum affinity seem to be
complementary. Take the curve for any one state and observe growth and asymptotic
behaviors. For example the photon added resource: non-Gaussianity approaches its
asymptotic value from above; it is obviously a convex function; two-mode vacuum
affinity approaches its asymptotic value from below, and it is a concave function.

Observe (in fig.~(\ref{fig:OverlapTWBRes})) that the photon-subtracted and
photon-added resources that have equal entanglement for every $r$ are perfectly
distinguishable by squeezed vacuum affinity.

In fig.~(\ref{fig:OverlapTWBRes}), the greatest affinity $\mathcal{G}$ is always
achieved at large values of the squeezing parameter by the optimized squeezed
Bell-like resources, while the lowest, constant affinity is always exhibited by the
squeezed Fock states. Observe additionally that the Squeezed Bell-like resource
optimized for coherent states ($\delta_{max}^{(c)}$) has a higher affinity than the
resource optimized for Fock states (($\delta_{max}^{(F)}$)), specially for the
smaller values of $r$.

In conclusion, optimized squeezed Bell-like resources are the ones that in all
squeezing regimes are closest to the simultaneous maximization of entanglement,
non-Gaussianity, and affinity to the two-mode squeezed vacuum. The optimized
interplay of these three properties explains the ability of squeezed Bell-like states
to yield better performances, when used as resources for CV quantum teleportation, in
comparison both to Gaussian resources at finite squeezing and to the standard
degaussified resources such as the photon-added and the photon-subtracted squeezed
states.

\section{Experimental generation of degaussified and squeezed Bell-like states}
\label{NGau:ExpGene}

While two-mode Gaussian squeezed states are currently produced in the laboratory, the
experimental generation of non-Gaussian states in quantum optics is still a complex
task, as it requires the availability of large nonlinearities and/or the arrangement
of proper apparatus for conditional measurements. Nevertheless, some truly remarkable
realizations of single-mode non-Gaussian states have been recently carried out
through the use of parametric amplification plus
postselection~\cite{ZavattaScience,ExpdeGauss1,ExpdeGauss2}. Recently, by a
generalization of the experimental setup used in Ref.~\cite{ExpdeGauss2} to a
two-mode configuration, a method has been proposed~\cite{KitagawaPhotsub} for the
generation of a certain class of two-mode photon-subtracted states.

\begin{figure}[ht] \label{fig:NonGaussGen}
\begin{centering}
\includegraphics[width=14cm]{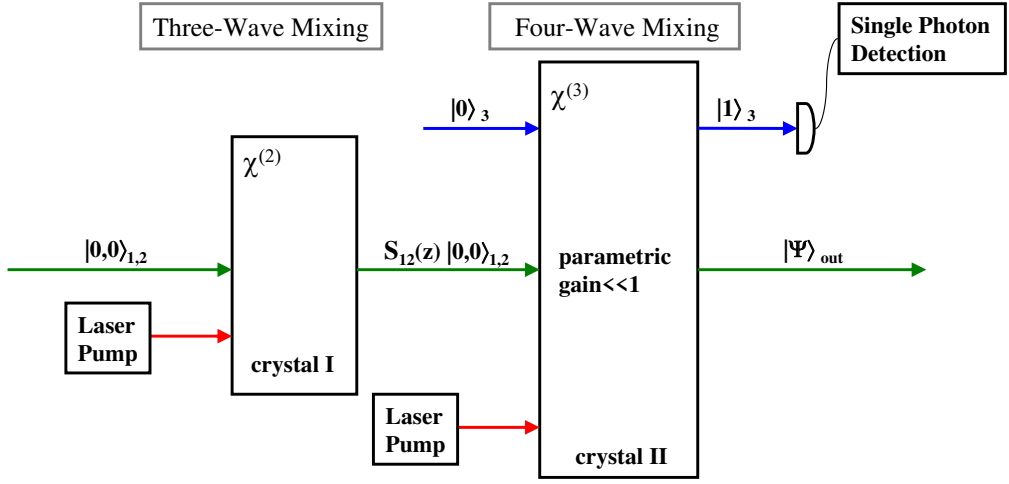}
\end{centering}
\caption[Scheme for the generation of degaussified and squeezed Bell-like resource
states]{Scheme for the generation of the photon-subtracted squeezed state and of the
photon-added squeezed state. Two nonlinear crystals are used in a cascaded
configuration. The first $\chi^{(2)}$-crystal is part of a three-wave mixer, acting
as a parametric amplifier for the production of a two-mode squeezed state. The
squeezed state seeds the successive nonlinear process, a four-wave mixing interaction
occurring in a $\chi^{(3)}$-crystal. A final conditional measurement reduces the
multiphoton state to a photon-subtracted (or added) squeezed state
$|\,\Psi\,\rangle_{out}$.}
\end{figure}

Here, in an analogous manner to that of ref.~\cite{ZavattaScience}, we propose a
possible experimental setup for the generation of the degaussified states
(eqs.~(\ref{eq:PhotAddketSup}) and~(\ref{eq:PhotSubketSup})), and of the squeezed
Bell-like states (eq.~(\ref{eq:SqueBellket})). The scheme, based on a configuration
of cascaded crystals, is depicted in fig.~(\ref{fig:NonGaussGen}).

In the first stage, by means of a three-wave mixer, functioning as a parametric
amplifier, a two-mode squeezed state $|\zeta \rangle_{12} = \widehat{S}_{12}(\zeta)
|0,0 \rangle_{12}$ is produced. In the second stage, a four-wave mixing process takes
place in a crystal with third order nonlinear susceptibility $\chi^{(3)}$. We
consider two possible multiphoton interactions, in the travelling wave configuration,
described by the following Hamiltonian operators;
\begin{align}
\widehat{H}_{I}^{(A)} \,=&\, \kappa_{A} \hat{a}_{1}\hat{a}_{2}\hat{a}_{3}^{\dag}
\,+\, \kappa_{A}^{*} \hat{a}_{1}^{\dag}\hat{a}_{2}^{\dag}\hat{a}_{3} \;,
\label{eq:Hint4wmA}\\
\widehat{H}_{I}^{(B)} \,=&\, \kappa_{B}
\hat{a}_{1}^{\dag}\hat{a}_{2}^{\dag}\hat{a}_{3}^{\dag} \,+\, \kappa_{B}^{*}
\hat{a}_{1}\hat{a}_{2}\hat{a}_{3} \;, \label{eq:Hint4wmB}
\end{align}

where $\hat{a}_{i}$ $(i=1,2,3)$ denotes three quantized modes of the radiation field.
The complex parameters $\kappa_{A}$ and $\kappa_{B}$ are proportional to the third
order nonlinearity and to the amplitude of an intense coherent pump field, treated
classically in the regime of parametric approximation. The two-mode squeezed state
seeds modes $1$ and $2$; mode $3$ is initially in the vacuum state
$|\,0\,\rangle_{3}$; mode $4$ is the classical pump. Energy conservation and phase
matching are assumed throughout. Let us remark that, due to the typical orders of
magnitudes of the third order susceptibilities, the parametric gains are very small
$|\,\kappa_{A}\,|\,, \;\;|\,\kappa_{B}\,|\ll 1$. The propagation (time evolution) in
the crystal yields $|\,\Psi_{I}^{(L)}\,\rangle \,=\, \exp\{-\,i\, t \,
\widehat{H}_{I}^{(L)}\}|\,\zeta\,\rangle_{12} \; |0\rangle_{3}$ $(L=A,B)$. Truncating
the series expansion of the evolution operator at the first order in
$\tilde{\kappa}_{L}=\,-\,i\, t\, \kappa_{L}$ we get
\begin{align}
|\,\Psi_{I}^{(A)}\,\rangle \,\approx &\, \{1 \,+\, \tilde{\kappa}_{A}\,
\hat{a}_{1}\hat{a}_{2}\hat{a}_{3}^{\dag} \} \, |\,\zeta\,\rangle_{12} \,
|\,0\,\rangle_{3}
\label{eq:Hint4wmATev}\\
|\,\Psi_{I}^{(B)}\,\rangle \,\approx &\, \{1 \,+ \,\tilde{\kappa}_{B}\,
\hat{a}_{1}^{\dag}\hat{a}_{2}^{\dag}\hat{a}_{3}^{\dag} \} \, |\,\zeta\,\rangle_{12}
\, |\,0\,\rangle_{3} \label{eq:Hint4wmBTev}
\end{align}

Finally, a conditional measurement is performed on mode $3$, consisting in a
single-photon detection; a projection onto the state $|1\rangle_{3}$. The
postselection reduces the states of eq.~(\ref{eq:Hint4wmATev}) and
eq.~(\ref{eq:Hint4wmBTev}) to the states of eq.~(\ref{eq:PhotSubketSup}) and
eq.~(\ref{eq:PhotAddketSup}), respectively. It is worth noting that the low values of
the parametric gains do not affect the implementation of the process. In fact, it is
analogous to require low reflectivity of a beam splitter to generate
photon-subtraction (addition) by using linear optics.

Regarding the production of the squeezed Fock state (eq.~(\ref{eq:SqFocket})), it can
be generated, in principle, by seeding a parametric amplifier with a single-photon
states in the two modes.

Let us now turn to the experimental generation of the squeezed Bell-like states
(eq.~(\ref{eq:SqueBellket})). They can be engineered by using the same setup
illustrated in fig.~(\ref{fig:NonGaussGen}), and by simultaneously realizing inside
the nonlinear crystal the processes corresponding to the interactions in
eq.~(\ref{eq:Hint4wmA}) and eq.~(\ref{eq:Hint4wmB}). In this case the fundamental
requirements are that of energy conservation and phase-matching for each multiphoton
interaction must hold simultaneously at each stage. This condition can be satisfied
by suitably exploiting the phenomenon of birefringence in a negative uniaxial
crystal~\cite{MidwinterWarner}. In particular, the following set of equations must
hold:
\begin{align}
\Omega_{1} =  \omega_{1}+\omega_{2}+\omega_{3}  \nonumber \\
K_{1}^{ext} =  k_{1}^{ord}+k_{2}^{ord}+k_{3}^{ext} \label{eq:cond1proc} \\
\Omega_{2}+\omega_{1}+\omega_{2} = \omega_{3}\nonumber \\
K_{2}^{ord} + k_{1}^{ord}+k_{2}^{ord} = k_{3}^{ext} \label{eq:cond2proc}
\end{align}
where $\omega_{j}$ and $k_{j}^{\lambda}$ $(j=1,2,3)$ represent the frequencies and
the wave vectors of the quantized modes with polarization $\lambda$; $\Omega_{j}$ and
$K_{j}^{\lambda}$ $(j=1,2)$ represent the frequencies and the wave vectors of the
classical pump fields; the superscript $ord$ and $ext$ denote, respectively, the
ordinary and extraordinary polarizations for the propagating waves. A collinear
configuration is assumed for the geometry of propagation inside the crystal. Then, at
fixed $\omega_{1}$ and $\omega_{2}$, the energy conservation relations, the type-II
phase matching condition in eq.~(\ref{eq:cond1proc}), and the type-I phase matching
condition in eq.~(\ref{eq:cond2proc}) can be, in principle, satisfied by a suitable
choice of $\omega_{3}$, $\Omega_{1}$, $\Omega_{2}$, and of the phase-matching angle
between the direction of propagation and the optical axis. Various examples of such
simultaneous multiphoton processes have been demonstrated both theoretically and
experimentally~\cite{PhysRep,SimultProc1,SimultProc2,SimultProc3}. The final
conditional measurement on mode $3$ yields the superposition state
\begin{equation}
|\Psi_{I}\rangle \, \approx \, \tilde{\kappa}_{A} \,\hat{a}_{1}\,\hat{a}_{2}
\,\widehat{S}_{12}(\zeta)\: |\,0\,,\,0\,\rangle_{12}\; + \; \tilde{\kappa}_{B}\,
\hat{a}_{1}^{\dag}\,\hat{a}_{2}^{\dag} \,\widehat{S}_{12}(\zeta)\:
|\,0\,,\,0\,\rangle_{12} \label{eq:psiSimuint}
\end{equation}

By applying a standard Bogoliubov transformation and after a little algebra, it is
straightforward to show that the superposition state (eq.~(\ref{eq:psiSimuint}))
reduces to the squeezed Bell-like state (eq.~(\ref{eq:SqueBellket})) if
\begin{align}
 c_{1} = & \, -(e^{-i
\phi}\,\tilde{\kappa}_{B}\,\tanh(r)\: +\:
e^{i\, \phi}\,\tilde{\kappa}_{A}) \nonumber \\
c_{2}  = & \, \tilde{\kappa}_{B} \:+\: e^{2\,i\, \phi}\,\tilde{\kappa}_{A}\, \tanh(r)
\label{eq:GenSqBellCond}
\end{align}

The latter conditions can be successfully implemented by observing that the complex
parameters $\tilde{\kappa}_{A}$ and $\tilde{\kappa}_{B}$ can be controlled to a very
high degree by means of the amplitudes of the external classical pumps.

\chapter{Teleportation with "truncated" Gaussians and squeezed cat-like resources}
\label{SSFCat}

We introduce, in section~\ref{SSFCat:Trunk}, the class of squeezed symmetric
superpositions of Fock states~\cite{TelNoisyCats} as a further generalization of the
squeezed Bell-like states (see eq.~(\ref{eq:SqueBellket})), and apply to them the
optimization procedure defined in ref.~\cite{TelepNonGauss} and used in
chapter~\ref{NGau}; obtaining a maximal fidelity of teleportation of a given input,
over the superposition parameters defining the character of the resource. We find
that the optimal resources for teleportation in this class of states are necessarily
constrained to be second-order "truncations" of the two-mode squeezed Gaussian states
(see eq.~(\ref{eq:TruncatGauss})).

In section~\ref{SSFCat:Cat} we introduce the class of squeezed cat-like states;
optimize this class of states for maximal fidelity of teleportation over the
available parameters, and compare results with those obtained in the previous
section.

\section{Truncated squeezed Gaussians: Symmetric Superpositions of Fock States and Bell-like states}
\label{SSFCat:Trunk}

In chapter~\ref{NGau}, the squeezed Bell-like states have been exploited as
non-Gaussian entangled resources that generalize the search for an optimal
teleportation fidelity among various classes of input states, which these states
interpolate. The squeezed Bell-like states (eq.~(\ref{eq:SqueBellket})) can be
formulated as the application of the two-mode squeezing operator
(eq.~(\ref{eq:TwoModSqOp})) to a general superposition of the two-mode vacuum and
two-mode single-photon Fock state (having two photons one in each mode), similar to a
Bell state of discrete variables (eq.~(\ref{eq:BellStatesDV})).

The optimization of teleportation fidelity over the superposition parameters yields a
remarkable enhancement in the success probability of teleportation for various input
states~\cite{TelepNonGauss} using squeezed Bell-like states. In this section, we
produce a further generalization on the squeezed Bell-like states deemed the Squeezed
Symmetric Superposition of Fock states~\cite{TelNoisyCats}; and show that the optimal
fidelity choice for this new class of states (and all the squeezed Bell-like states)
can be regarded as truncations (on the photon number) of two-mode squeezed vacuums.

\subsection{Definition: truncated Gaussians}
\label{SSFCat:Trunk:Char}

The squeezed Bell-like states, having only two superposition components and the
constraint of normalization, can be parameterized in alternative ways. A
parameterization based on the first order ($n=1$) truncation (see
eqs.~(\ref{eq:TwoModSqVacRFock}) and~(\ref{eq:TruncatGauss})) of the two-mode
squeezed vacuum state is possible. Take a trivial generalization of the
aforementioned expressions,
\begin{align}
|-(r+s)\,\rangle_{AB} \,=\,&
\widehat{S}_{AB}(-r)\;\widehat{S}_{AB}(-s)\;|\,0\,,\,0\,\rangle_{AB} \notag\\
&\,=\,\widehat{S}_{AB}(-r)\;
\left(\cosh(s)\right)^{-1}\:\sum_{n=0}^{\infty}\:\left(\,\tanh(-s)\,\right)^{n}\:|\,n\,,\,n\,\rangle_{AB}\label{eq:SqVacTruncat}
\end{align}
and delete all the terms of order $n>1$. The normalization of the state holds for all
$s$; \emph{genuine squeezing} is performed with parameter $-r$, however. The only
additional complication with respect to the former parameterization is the relation
between both; equations involving trigonometric functions of $\delta$ and exponential
functions of $s$. Thus, the squeezed Bell-like state can necessarily be regarded as
the first-order truncation of some two-mode squeezed vacuum.

A further and obvious generalization of the squeezed Bell-like states lies in
considering squeezed, \emph{symmetrical} Fock states of higher number in the
superposition. Let us consider the squeezed superposition
\begin{align}
|\,\Psi\,\rangle_{SSF}
&=\,\widehat{S}_{AB}(\zeta)\;\left(\,c_{0}^{2}\,+\,c_{1}^{2}\,+\,c_{2}^{2}\,\right)^{-1/2}\notag\\
&\times\,\left\{c_{0} |\,0\,,\,0\, \rangle_{AB}
 \:+\; e^{\,i\, \theta_{1}}\, c_{1}
|\,1\,,\,1\, \rangle_{AB} \,+\, e^{\,i\, \theta_{2}}\, c_{2} |\,2\,,\,2\,
\rangle_{AB}\right\} \label{eq:squeezSSF}
\end{align}
where $c_{i}$ and $\theta_{i}$ are real constants and phases, respectively. This
class of states, termed the squeezed symmetric superposition of Fock
states~\cite{TelNoisyCats} is not necessarily a second-order ($n=2$) truncation of
eq.~(\ref{eq:SqVacTruncat}); given that there are three components of the
superposition and only one normalization constraint. The second-order truncations of
two-mode squeezed vacuums are special cases of the squeezed symmetric superposition
of Fock states (\textbf{SSSF}). However, the choice of superposition parameters for
the \textbf{SSSF} state that interests us most corresponds to an optimal fidelity of
teleportation for a given input state.

A convenient parameterization of the $c_{i}$ coefficients in eq.~(\ref{eq:squeezSSF})
is provided by the hyper-spherical coordinates in three dimensions:
$c_{0}=\,\cos(\delta_{1})$, $c_{1}=\,\sin(\delta_{1}) \cos(\delta_{2})$,
$c_{2}=\,\sin(\delta_{1})\, \sin(\delta_{2})$. It is expected that having an
additional free superposition parameter will allow us to find an optimal
teleportation fidelity greater than the optimal fidelity obtained using the squeezed
Bell-like state; for the latter is the special case of the \textbf{SSSF}, for
$\delta_{2}=0$.

The two-mode characteristic (eq.~(\ref{eq:CharTracDisp}))function $\chi_{SSF}$ of the
\textbf{SSSF} state (\ref{eq:squeezSSF}) can be calculated easily. Let us recall
eqs.~(\ref{eq:CharPurDisp}) for the pure state characteristic function;
eq.~(\ref{eq:TwoModBogChar}) for the Bogoliubov transformation effected by two-mode
squeezing on a displacement operator;  eq.~(\ref{eq:DispFockElem}) for the matrix
element of the displacement operator. Following a procedure similar to the
calculation of the characteristic function for the degaussified resource states (and
for the squeezed Bell-like state, \emph{see} subsection~\ref{NGau:Chzation:Degauss})
we obtain, for $\theta_{1}=\theta_{2}=0$;
\begin{align}
\chi_{SSF}(\xi_{A};\:\xi_{B})=&\,\frac{e^{\,-1/2\,(|\,\xi_{A}'|^{2}+|\,\xi_{B}'|^{2})}}{(\,c_{0}^{\,2}\,+\,c_{1}^{\,2}\,+\,c_{2}^{\,2}\,)}\notag\\
&\times\,\sum_{m=0}^{2}\sum_{n=0}^{2}\:c_{n}^{*}\,c_{m}\;\frac{m!}{n!}\;(\xi_{A}'\,\xi_{B}')^{n-m}\;L_{\,m}^{\,(n-m)}(|\,\xi_{A}'|^{2})\:L_{\,m}^{\,(n-m)}(|\,\xi_{B}'|^{2})
\label{eq:SSFChar}
\end{align}
where $\xi_{A}'$ and $\xi_{B}'$ are related to $\xi_{A}$ and $\xi_{B}$ by the
Bogoliubov transformation for the displacement operator arguments in
eq.~(\ref{eq:TwoModBogChar}); and $L_{m}^{(n-m)}$ is the associated Laguerre
polynomial.

\subsection{Optimized teleportation fidelities with truncated Gaussians}
\label{SSFCat:Trunk:Fid}

We proceed now to calculate the teleportation fidelities for ideal \textbf{CV}
teleportation (see eq.~(\ref{eq:FidCharOutFin})) for coherent state and Fock state
inputs, using the non-Gaussian resource states outlined above as resources (for the
description and characteristic functions of the input states see
subsection~(\ref{NGau:Chzation:Input}).

For the \textbf{SSSF} state, and the squeezed Bell-like state, we have calculated the
teleportation fidelities and performed the optimization of said fidelity over the
superposition parameters regarding each state. The squeezed Bell-like state is
optimized over its sole free parameter $\delta$ (see
subsection~\ref{NGau:SqueBell:OptBell}). The \textbf{SSSF} teleportation fidelity is
in principle a function of parameters $r$, $\phi$, $\delta_{1}$, $\theta_{1}$,
$\delta_{2}$, $\theta_{2}$; but $r$ is fixed as a technical capability indicator and
the phase $\phi=\pi$ is fixed as required by the conventions in the \textbf{CV}
protocol of chapter~\ref{Form}. Fidelity is thus to be optimized over the set of
superposition parameters
$\mathcal{P}=\{\delta_{1},\theta_{1},\delta_{2},\theta_{2}\}$;
\begin{equation}
\mathcal{F}_{opt}=\,\max_{\mathcal{P}} \;
\mathcal{F}(\,r\,,\,\pi\,,\,\delta_{1}\,,\,\theta_{1}\,,\,\delta_{2}\,,\,\theta_{2}\,)
\label{eq:OptimalFid}
\end{equation}
for a given class of input states. The optimization is performed numerically on the
analytically calculated fidelities (reported in appendix~\ref{APPA}). The results of
optimization are independent of the phases $\theta_{1}$ and $\theta_{2}$, which can
be set to zero in consequence. In this manner the \textbf{SSSF} resource is
\emph{sculpted} for the teleportation task at hand.

The procedure leads to a \emph{very remarkable} result: Optimization over the set
$\mathcal{P}$ yields maximal parameters for an \textbf{SSSF} state that is a
\emph{second-order truncation} of the two-mode squeezed vacuum state (that of
eq.~(\ref{eq:SqVacTruncat})). The optimal resources for \textbf{CV} teleportation
form a subset of the class \textbf{SSSF} having the form
\begin{align}
|\,\psi'\,\rangle_{SSF} &=\, [1\,+\,(\tanh(s))^{2}\,+\,(\tanh(s))^{4}]^{-1/2}\notag\\
&\times\,\widehat{S}_{AB}(-r)\;\left\{\,|\,0\,,\,0\, \rangle_{AB} \:+\: \tanh(s)\,
|\,1\,,\,1\, \rangle_{AB} \:+\: (\tanh(s))^{2} \,|\,2\,,\,2\, \rangle_{AB}\,\right\}
\label{eq:squeezSSFTrun}
\end{align}
where, in accordance with the optimization performed in eq.~(\ref{eq:OptimalFid}),
the real parameter $s$ is fixed to the optimal value $s=\tilde{s}$. We have performed
 a change of parameterization that simplifies the description of the optimal resource;
nevertheless $\mathcal{F}_{opt}(\tilde{s})=\max_{s}\mathcal{F}(r,s)$.

\begin{figure}[ht] \label{fig:FidSculpTelep}
\begin{centering}
\includegraphics[width=13.5cm]{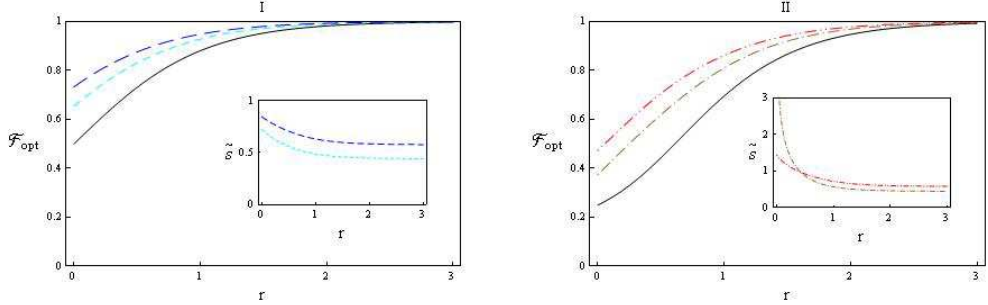}
\end{centering}
\caption[Optimal fidelity of teleportation using truncated-Gaussian
resources]{Optimal fidelity of teleportation $\mathcal{F}_{opt}$, as a function of
the squeezing parameter $r$, with $\phi=\pi$ for truncated Gaussian resources. In
Panel I, we report the fidelity for the teleportation of input coherent states
$|\,\beta\,\rangle$ using different two-mode entangled resources: \textit(a)~Gaussian
two-mode squeezed state (full line); \textit(b)~squeezed Bell-like state (dashed
line); \textit(c)~squeezed symmetric superposition of Fock states (long-dashed line).
In Panel II, we report the fidelity for the teleportation of input single-photon Fock
state $|\,1\,\rangle$ using different two-mode entangled resources:
\textit(a)~Gaussian two-mode squeezed state (full line); \textit(b)~squeezed
Bell-like state (dot-dashed line); \textit(c)~squeezed symmetric superposition of
Fock states (double-dot-dashed line). In plot I the value of $\beta$ is arbitrary.
The insets in both Panels give the maximal values of the parameter $s=\tilde{s}$ as a
function of $r$ for a given entangled resource and fixed input state. The plot styles
are chosen as specified above.}
\end{figure}

In fig.~\ref{fig:FidSculpTelep}, we report the behavior of $\mathcal{F}_{opt}$ as a
function of $r$ for coherent state inputs (Panel I) and the single-photon Fock state
input (Panel II) with a \textbf{SSSF} state as the entangled resource. For
comparison, the curve for optimal fidelity corresponding to a squeezed Bell-like
resource and the curve for fidelity corresponding to a two-mode squeezed vacuum are
shown. The parameterization chosen for the squeezed Bell-like resource is that of the
Gaussian truncation illustrated before (see eq.~(\ref{eq:SqVacTruncat}) and related
discussion); thus, an optimal "superposition" squeezing $\tilde{s}$ is given for both
non-Gaussian resources. We observe that, as expected, a further enhancement of the
fidelity is obtained for the \textbf{SSSF} resource compared to the squeezed
Bell-like resource. Even if the number of "free" parameters (one, $s$) appears to be
the same for both non-Gaussian resources, the extra dimension in the Hilbert space
spanned by the superposition allows an increase in fidelity.

\subsection{Entanglement, non-Gaussian character and Gaussian affinity for truncated Gaussians}
\label{SSFCat:Trunk:Compaprop}

In section~\ref{NGau:Compaprop}, to understand the properties of the optimized
teleportation resources, we have investigated three quantities: the von Neumann
entropy $E_{vN}$ (eq.~(\ref{eq:VonNeumannEnt})) as a measure of the amount of
entanglement in the resource; the non-Gaussianity $d_{nG}$
(eq.~(\ref{eq:nonGaussianity})), to provide a measure of the non-Gaussian character
of the resource; and the squeezed vacuum affinity $\mathcal{G}$
(eq.~(\ref{eq:OverlapTWB})), in order to determine the degree of resemblance of the
resource to the closest two-mode squeezed vacuum.

\begin{figure}[ht] \label{fig:vonNeumSTruncTwB}
\begin{centering}
\includegraphics[width=12cm]{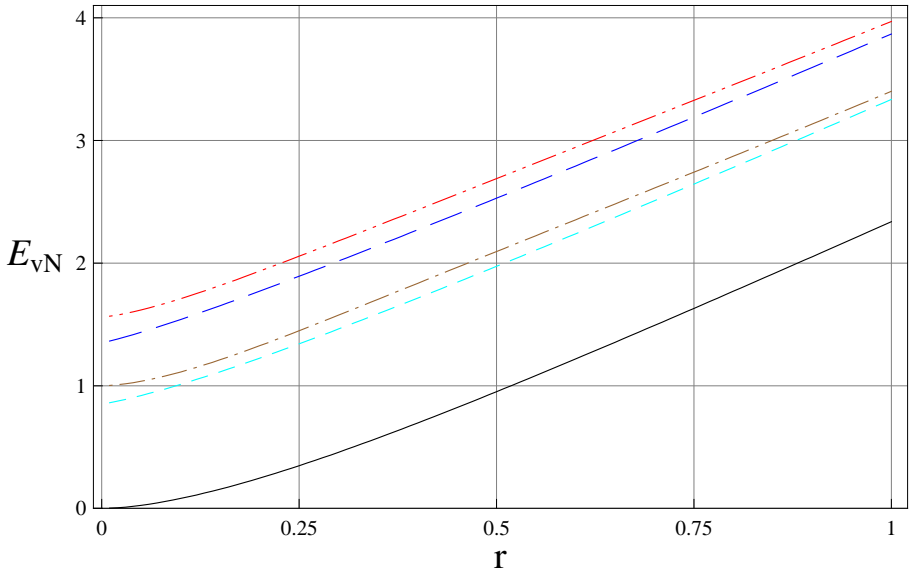}
\end{centering}
\caption[The von Neumann entropy for the truncated-Gaussian resources]{Plot of the
von Neumann entropy $E_{vN}$ as a function of $r$, with $s=\tilde{s}$ (see the insets
in fig.~\ref{fig:FidSculpTelep}), corresponding to the following entangled resources:
squeezed symmetric superposition of Fock states, squeezed Bell-like states, and
two-mode squeezed vacuum states, optimized for the teleportation of coherent state
inputs, and the single-photon Fock state input. The plot styles are chosen as
specified in fig.~(\ref{fig:FidSculpTelep}).}
\end{figure}

In fig.~(\ref{fig:vonNeumSTruncTwB}), we plot the entanglement measure for pure
states $E_{vN}$ (eq.~(\ref{eq:VonNeumannEnt})); for the optimized non-Gaussian
entangled resources above mentioned and used as teleportation resources, and for the
two-mode squeezed vacuum. All the curves exhibit very similar behaviors. At fixed
$r$, and for a given input state, the optimized \textbf{SSSF} state is the most
entangled; the optimal resource for the teleportation of a single-photon Fock state
having more entanglement than the resource for the optimal teleportation of a
coherent state; this can lead to the conclusion that the non-Gaussian input requires
more entanglement for optimal teleportation.

The behavior of the von Neumann entropies across different classes of resources is in
agreement with the behavior of the optimal fidelities. In fact, for a given input
state, a non-Gaussian resource with a higher teleportation fidelity (in
fig.~(\ref{fig:FidSculpTelep})) is associated with a higher amount of entanglement
content for any $r$.

It is not remarkable that the Gaussian truncation (see eq.~(\ref{eq:SqVacTruncat}))
to higher order ($n=2$, optimal \textbf{SSSF} state)) shows a higher von Neumann
entropy that the Gaussian truncation to lower order ($n=1$, optimal Bell-like
resource). We can only speculate that a higher-order ($n>2$) truncation will have a
higher von Neumann entropy for the same squeezing $r$. What seems remarkable, is that
for the same squeezing parameter $r$, the un-truncated Gaussian has the lowest von
Neumann entropy. Before beginning to speculate as to the truncation order in which
the von Neumann entropy begins to diminish, let us think again about what is meant by
Gaussian "truncation". Inspecting eq.~(\ref{eq:SqVacTruncat}); and on seeing the
squeezing $-(r+s)$ performed on the resource to be "truncated", we can only conclude
that to be "fair" to the Gaussian resource, the optimal non-Gaussian resources used
here (defined by their parameters $-r$ and $-\tilde{s}$, see
eq.~(\ref{eq:squeezSSFTrun})) should be compared to a two-mode squeezed vacuum of
squeezing $-(r+\tilde{s})$ instead. We rephrase our claim: "truncating" the Gaussian
resource~\footnote{We do not necessarily think that "truncation" is a physically
feasible operation on a Gaussian resource.} \emph{raises the entanglement} of the
resource: from that associated to a squeezing $r$ for a two-mode squeezed vacuum, to
that associated to a squeezing $r+s'$ for the same state. This \textit{added
squeezing} $s'$ is a free superposition parameter to be chosen. From inspection of
figs.~(\ref{fig:FidSculpTelep}) and~(\ref{fig:vonNeumSTruncTwB}): for the optimal
resources we have considered the added squeezing $\tilde{s}$ (and its effect on
entanglement) is of the order of magnitude of the squeezing $r$ (and its associated
levels of entanglement).

\begin{figure}[ht] \label{fig:NonGaussAffSTruncTwB}
\begin{centering}
\includegraphics[width=14cm]{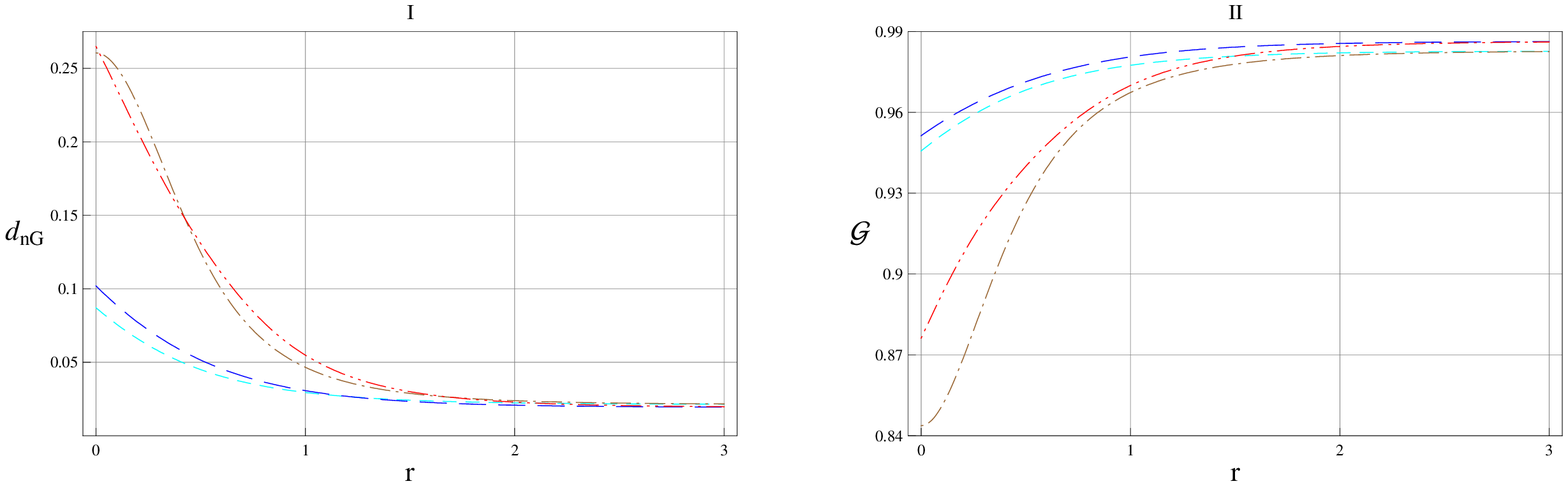}
\end{centering}
\caption[Plot of non-Gaussianity and squeezed vacuum affinity for truncated-Gaussian
resources]{Plot of the non-Gaussianity $d_{nG}$ (panel I) and of the squeezed vacuum
affinity $\mathcal{G}$ (panel II) as a function of $r$, with $s=\tilde{s}$ (see the
insets in fig.~(\ref{fig:FidSculpTelep})), corresponding to the following entangled
resources: squeezed symmetric superposition of Fock states, and squeezed Bell-like
states, optimized for the teleportation of input coherent states and the
single-photon Fock input state. The plot styles are chosen as specified in
fig.~(\ref{fig:FidSculpTelep}). }
\end{figure}

Next, we examine the behavior of the non-Gaussianity $d_{nG}$, a measure of the
difference between the resource and a reference Gaussian state with the same first
and second-order moments. In fig.~(\ref{fig:NonGaussAffSTruncTwB}), panel I: we plot
$d_{nG}$ (eq.~(\ref{eq:nonGaussianity})) for the \textbf{SSSF} resources and the
squeezed Bell-like resources, optimized both for the coherent inputs and for the
single-photon Fock input. The difference in non-Gaussianity between the coherent
input and Fock input cases for $r=0$ is remarkable. While the resources are entangled
even with no squeezing; it seems that optimization for very low $r$ is just an
attempt at (nearly) classical teleportation. The behavior of the pairs of $d_{nG}$
curves corresponding to the teleportation of the same input states (\emph{either}
coherent \emph{or} Fock) for different resources (\textbf{SSSF} \emph{and} Bell-like)
is very similar; a pair of curves (the Fock state input's) crosses at a certain value
of $r$. All the curves tend to the same small asymptotic value of $d_{nG}$ for very
large $r$, as expected. Therefore, according to this measure of non-Gaussianity: the
non-Gaussian resource with a higher teleportation fidelity (the \textbf{SSSF} state)
does not exhibit a significantly higher non-Gaussian character for any $r$.

Lastly, we turn our attention to the Gaussian resource affinity $\mathcal{G}$
(eq.~(\ref{eq:OverlapTWB})), given as a measure of the maximum similarity of the
resource to an ideal two-mode squeezed vacuum. In
fig.~(\ref{fig:NonGaussAffSTruncTwB}) panel II: we plot $\mathcal{G}$ for the
\textbf{SSSF} and squeezed Bell-like resources, as always optimized for the
teleportation of given input states. We can see a mirror of the behaviors of $d_{nG}$
for $r\rightarrow\,0$. All the curves have a high asymptotic value of $\mathcal{G}$
for large $r$, a fact consistent with the results for non-Gaussianity $d_{nG}$. Most
importantly: For the same class of input state (\emph{either} coherent \emph{or}
Fock), the optimized \textbf{SSSF} state possesses a greater squeezed vacuum affinity
than that of the squeezed Bell-like state, for any $r$. This result is in complete
agreement with the behaviors of the corresponding optimal teleportation fidelities
and von Neumann entropies studied before.

Comparing the behaviors associated with resources optimized for the efficient
teleportation of the same input states, we conclude that the following hierarchy can
be established: The enhancement of the teleportation fidelity corresponds to an
enhancement of the entanglement content, and to an enhancement of squeezed vacuum
affinity. The measure of non-Gaussianity $d_{nG}$ seems not to have a place in this
hierarchy. This result shows that such a measure must be used and interpreted with
attention, case by case.

Further conclusions can be drawn from the above discussion: The squeezed truncated
Gaussian resource, with truncation at $n=2$ (\textbf{SSSF} state, with a four-photon
term) is an entangled non-Gaussian resource suitable for \emph{sculpturing} and
optimization as regards \textbf{CV} quantum teleportation. We argue that this result
can be generalized to higher orders ($n>2$) in expansions of the two-mode squeezed
vacuum.

Therefore, we conjecture that in general, efficient entangled non-Gaussian resources
for \textbf{CV} quantum teleportation should be characterized by a suitable balance
between the entanglement content and the degree of affinity to squeezed vacuum
states.

\section{Squeezed cat-like states}
\label{SSFCat:Cat}

In this section, we define and use as teleportation resources a special class of
two-mode squeezed superposition. This time, instead of using (solely) Fock states for
the superposition we use coherent states. Namely the vacuum (which is a coherent
state) and an arbitrary separable coherent state with the same displacement on both
it's modes. These can be thought of as more feasible and more amenable to
manipulation than the superpositions made of the highly nonclassical Fock states. We
remark that the engineering of states of such a form is made possible, in principle,
by the recent successful experimental realization of their single-mode
equivalents~\cite{GrangierCats}.

\subsection{Definition: squeezed cat-like states}
\label{SSFCat:Cat:Char}

The squeezed cat-like state is given by the two-mode squeezed superposition of
coherent states
\begin{equation}
|\,\psi\,\rangle_{SC} =\, \mathcal{N}\: \widehat{S}_{AB}(\zeta)\;
\left\{\cos(\delta)\, |\,0\,,\,0\, \rangle_{AB} \:+\: e^{\,i\, \theta}\, \sin(\delta)
\,|\,\gamma\,,\,\gamma\, \rangle_{AB}\right\} \label{eq:squeezCat}
\end{equation}
where the normalization factor is $\mathcal{N}=\left(1\,+\,
e^{\,-|\,\gamma\,|^{2}}\:\sin(2\,\delta)\:\cos(\theta)\right)^{-1/2}$. Without the
two-mode squeezing, eq.~\ref{eq:squeezCat} becomes an entangled Schr\"{o}dinger Cat
state. The one and two-mode cat states have been proposed as qubits and as entangled
resources (respectively) for quantum information processing; together with an
universal set of operations for this purpose~\cite{Ralph}.

The state in eq.~(\ref{eq:squeezCat}) can be regarded as the substitution of a
coherent state $|\,\gamma\,,\,\gamma\, \rangle_{AB}$ for the two-photon
$|\,1\,,\,1\,\rangle_{AB}$, in the squeezed Bell-like state of
eq.~(\ref{eq:SqueBellket}). Thus producing a more classical two-mode squeezed
superposition that is non-Gaussian and entangled even for zero squeezing $r$.

Following the same reasoning used to calculate the characteristic function of the
\textbf{SSSF} state and using mostly the same identities, plus the composition law
for displacement operators (eq.~(\ref{eq:DispComp})), the characteristic function for
squeezed cat-like resource of eq.~(\ref{eq:squeezCat}) can be easily calculated. For
$\theta=0$ it is given by
\begin{align}
\chi_{SC}(\xi_{A};\:\xi_{B})=&\,\mathcal{N}^{\,2}\:e^{\,-\frac{1}{2}(|\,\xi_{A}'\,|^{2}\,+\,|\,\xi_{B}'\,|^{2})}\notag\\
&\times\,((\cos(\delta))^{2}\,+\,\frac{\sin(2\,\delta)}{2}\,e^{\,-|\,\gamma\,|^{2}}
\,\left(e^{\,\gamma^{*}(\xi_{A}'\,+\,\xi_{B}')}\,+\,e^{\,-\gamma\left(\xi_{A}'\,+\,\xi_{B}'\right)^{\,*}}\right)\notag
\\
&+\,(\sin(\delta))^{2}\,e^{\,2\,i\,\mathrm{Im}[\gamma^{*}(\xi_{A}'\,+\,\xi_{B}')]})
\label{eq:CatChar}
\end{align}
where $\xi_{A}'$ and $\xi_{B}'$ are related to $\xi_{A}$ and $\xi_{B}$ by the
Bogoliubov transformation for the displacement operator arguments in
eq.~(\ref{eq:TwoModBogChar}).

\subsection{Optimized teleportation fidelities with cat-like states}
\label{SSFCat:Cat:Fide}

We can compute analytically the fidelity of teleportation using squeezed cat-like
states as entangled resources, limiting the analysis, for simplicity's sake, to the
teleportation of coherent state inputs (eq.~(\ref{eq:CharCohIn})). The maximization
procedure is discussed in appendix~\ref{APPA}, including the analytical expression
for the fidelity. At fixed squeezing $r$, the fidelity in
eq.~(\ref{eq:FidelitySqueezKat2}) may be optimized over the real amplitude
$|\,\gamma\,|$ (the only free parameter remaining). Therefore;
\begin{equation} \label{eq:FOptCat}
\mathcal{F}_{opt}(r) \,=\,
\max_{\,|\,\gamma\,|\,}\:\mathcal{F}_{SC}\left(\,r,\,|\,\gamma\,|\,\right)
\end{equation}

\begin{figure}[ht] \label{fig:FidSculpTelep2}
\begin{centering}
\includegraphics[width=12cm]{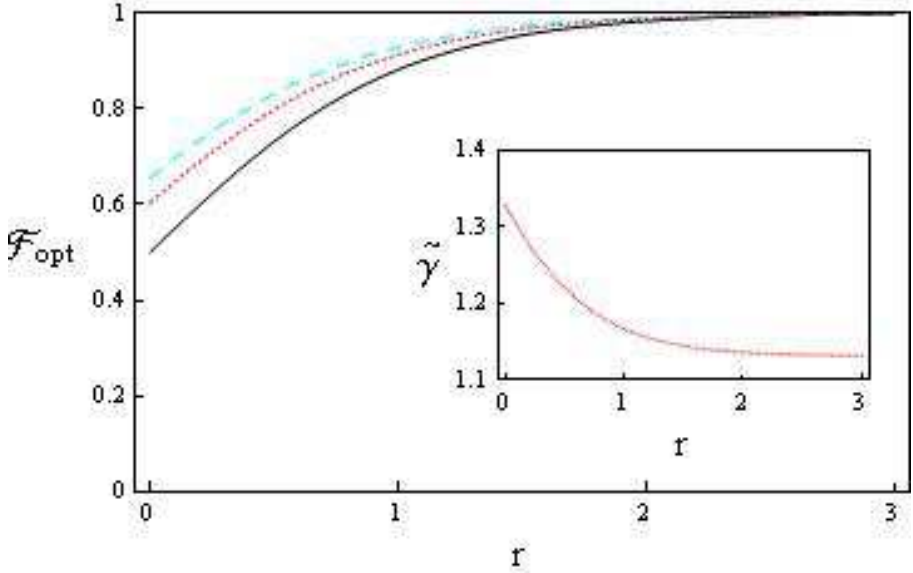}
\end{centering}
\caption[Optimal fidelity of teleportation using squeezed cat-like resources]{Optimal
fidelity of teleportation $\mathcal{F}_{opt}$, as a function of the squeezing
parameter $r$, with $\phi=\pi$, using squeezed cat-like resources. The fidelity
corresponds to the teleportation of input coherent states with a squeezed cat-like
state as entangled resource (dotted line). For comparison, we plot the fidelities
obtained with a squeezed vacuum resource (full line) and with a squeezed Bell-like
resource (dashed line). The inset gives the optimal value of the parameter
$|\,\gamma\,|=|\,\tilde{\gamma}\,|$ as a function of $r$. }
\end{figure}

In fig.~(\ref{fig:FidSculpTelep2}), we show the behavior of $\mathcal{F}_{opt}$ as a
function of $r$ for coherent state inputs, using an optimized squeezed cat-like state
as a teleportation resource. For comparison, the optimal fidelities corresponding to
the teleportation with Gaussian squeezed state resources and with optimized, squeezed
Bell-like states are plotted. The optimized squeezed cat-like resources, for all $r$,
yield a significant improvement of the fidelity with respect to the two-mode squeezed
vacuum, but are less efficient than squeezed Bell-like resources; all three curves
show the same behavior, including the same asymptotic behavior. The high value of
$\tilde{\gamma}$ for $r=0$ is a result of the optimization trying to increase the
usable entanglement that is present in a two-mode Schr\"{o}dinger cat state. For much
larger squeezing, the optimal amplitude $\tilde{\gamma}$ tends to a smaller,
asymptotic value; this behavior is consistent with the behavior of the resource,
which eventually becomes an \textbf{EPR} state, for $r\rightarrow\infty$.

\subsection{Entanglement, non-Gaussian character and Gaussian affinity for cat-like states}
\label{SSFCat:Cat:Compaprop}

\begin{figure}[ht] \label{fig:vonNeumSqCat}
\begin{centering}
\includegraphics[width=12cm]{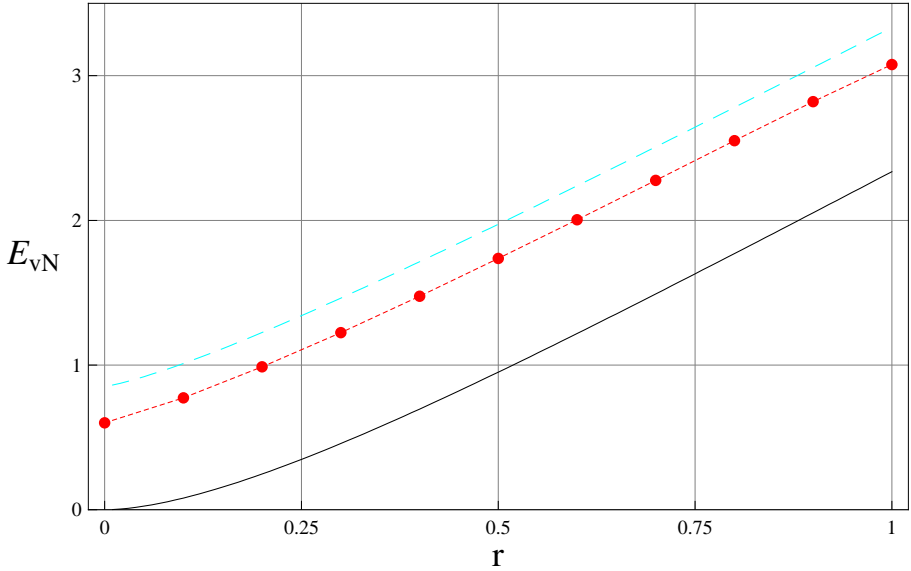}
\end{centering}
\caption[The von Neumann entropy for squeezed cat-like resources]{Behavior of the von
Neumann entropy $E_{vN}$ as a function of $r$, with $|\,\gamma\,|=\tilde{\gamma}$
(see the inset in fig.~(\ref{fig:FidSculpTelep2})) for the optimized squeezed
cat-like state (dotted line). For comparison, we also plot the quantities $E_{vN}$
corresponding to a two-mode squeezed vacuum (full line), and to a squeezed Bell-like
state (dashed line).}
\end{figure}

\begin{figure}[ht] \label{fig:dnGAffSqCat}
\begin{centering}
\includegraphics[width=14cm]{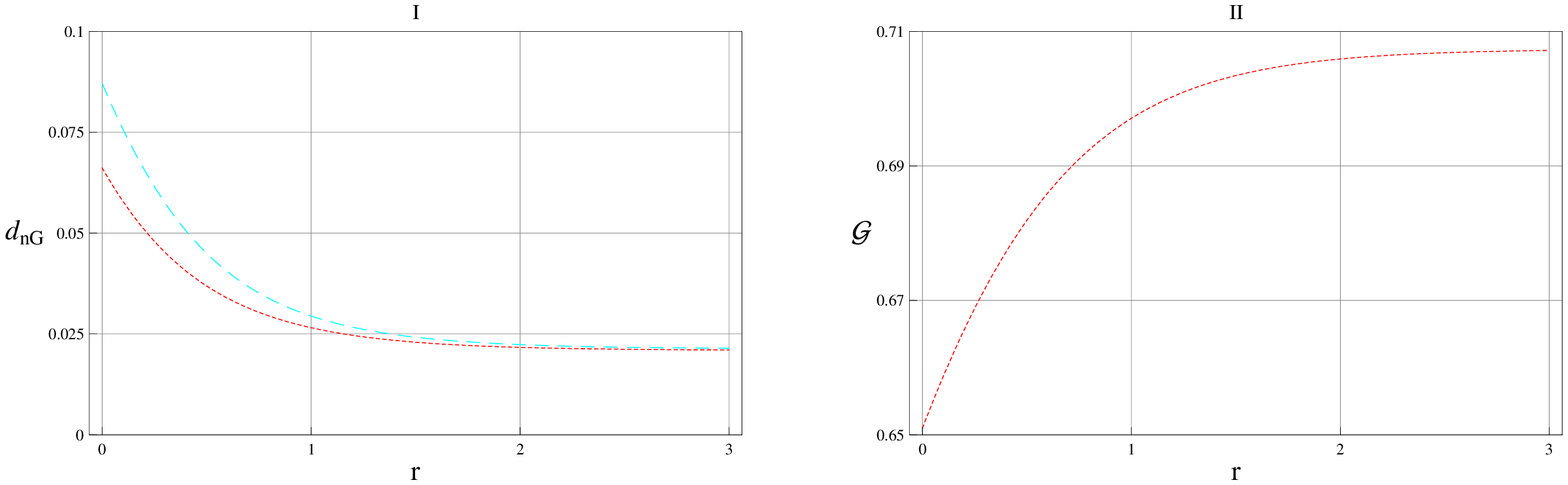}
\end{centering}
\caption[The non-Gaussianity and squeezed vacuum affinity for squeezed cat-like
resources]{Behavior of the non-Gaussianity $d_{nG}$ (panel I) and of the squeezed
vacuum affinity $\mathcal{G}$ (panel II) as a function of $r$, with
$|\,\gamma\,|=\tilde{\gamma}$ (see the inset in fig.~(\ref{fig:FidSculpTelep2})) for
the optimized squeezed cat-like state (dotted line). For comparison, in panel I we
plot $d_{nG}$ for the optimized squeezed Bell-like state (dashed line).}
\end{figure}

The behavior of the teleportation fidelity for the optimized, squeezed cat-like
resource in fig.~(\ref{fig:FidSculpTelep2}) is in agreement with the behaviors of its
von Neumann entropy (eq.~(\ref{eq:VonNeumannEnt})), non-Gaussianity
(eq.~(\ref{eq:nonGaussianity})), and squeezed vacuum affinity
(eq.~(\ref{eq:OverlapTWB})). We report these in fig.~(\ref{fig:vonNeumSqCat}) and
fig.~(\ref{fig:dnGAffSqCat}), panel I and panel II, respectively.

The optimized squeezed cat-like resource is less entangled, less non-Gaussian
(according to the $d_{nG}$ measure), and less affine to the squeezed vacuum than the
optimized squeezed Bell-like state. The entanglement as measured by the von Neumann
entropy is higher for the cat-like resource than for a Gaussian, but lower than that
for a Bell-like state for all values of $r$. The optimized squeezed cat-like resource
exhibits a markedly lower value of $\mathcal{G}$ in comparison with the truncated
Gaussian resources. This can be seen by comparing fig.~(\ref{fig:dnGAffSqCat}), panel
II with the corresponding plots shown in fig.~(\ref{fig:NonGaussAffSTruncTwB}), panel
II; in fact, $\mathcal{G}$ seems to have an asymptote at value $71\%$ (approached
from below).

\chapter{Teleportation with noisy non-Gaussian resources}
\label{Noisy}

In chapter~\ref{NGau}, we defined the squeezed Bell-like states
(eq.~(\ref{eq:SqueBellket})); a class of states that include all the other
non-Gaussian resources studied in that chapter, as well as the Gaussian two-mode
squeezed vacuum. We optimized the Squeezed Bell-like state for the teleportation of
some inputs and compared them for their properties of entanglement, non-Gaussianity
and Gaussian resource affinity with the other resource states studied in that
chapter. In chapter~\ref{SSFCat}, we introduced the squeezed cat-like
states~\cite{TelNoisyCats} (eq.~(\ref{eq:squeezCat})) by substituting a coherent
state term for the Fock state term in the superposition making up a Bell-like state,
with the intention of creating a more feasible~\cite{GrangierCats}, more
resilient~\cite{SerafiniCats} (to decoherence) resource.

In this chapter, we extend the analysis performed in chapters~\ref{NGau}
and~\ref{SSFCat} to the \emph{realistic} case of optimized, squeezed Bell-like and
squeezed cat-like resources prepared or propagated in the presence of thermal noise
(see section~\ref{Form:MixRes}), for the teleportation of coherent state
inputs~\cite{TelepNoise,TelNoisyCats}. For comparison purposes we perform the same
analysis for Gaussian two-mode squeezed vacuum states, also prepared in the presence
of thermal noise.

We analyze the behavior of inseparability parameters developed for general
\textbf{CV} contexts~\cite{InsepShchukin,InsepDellAnno}, for Bell-like and Gaussian
resource states. For a practical measure of the disappearance (or appearance) of
entanglement of teleportation, we consider the level of noise at which the resource
passes the \textit{classical teleportation
threshold}~\cite{CriteriaCVTelep,CriteriaCVTelep2} of $1/2$, for a fixed squeezing.
This last analysis is performed for both the squeezed Bell-like resource and the
squeezed cat-like resource, with the squeezed vacuum put in for comparison purposes.

\section{Fidelity of teleportation with mixed, noisy non-Gaussian resources}
\label{Noisy:Telep}

The presence of thermal noise in the preparation of the resource states can be
modelled by \emph{superimposing} a pure resource state on thermal states (see
section~\ref{Form:MixRes}) in modes $A$ and $B$ of the resource. The characteristic
function of such a preparation (see eq.~(\ref{eq:MixResCharDef})) is given by

\begin{equation}
\chi_{AB}^{(th)}(\xi_{A},\xi_{B}) \,=\,
e^{-n_{th,A}|\xi_{A}|^{2}-n_{th,B}|\xi_{B}|^{2}} \,
\chi_{AB}(\xi_{A},\xi_{B})\label{eq:MixResCharNoise}
\end{equation}
where $n_{th,A},n_{th,B}$ are the mean photon numbers; thermal parameters associated
with the modes $A$ and $B$.

The state of eq.~(\ref{eq:MixResCharNoise}) is mixed for $n_{th,A},n_{th,B}\neq\,0$.
$\chi_{AB}$ is the characteristic function for the pure resource state thus
superimposed over thermal states. We substitute in eq.~(\ref{eq:MixResCharNoise}) the
characteristic functions of either the squeezed Bell-like state
(eq.~(\ref{eq:CharSqueBell})); the squeezed cat-like resource
(eq.~(\ref{eq:CatChar})) or the two-mode squeezed vacuum
(eq.~(\ref{eq:CharGaussVac})) to produce the respective "noisy" teleportation
resources.

Substituting the resource of eq.~(\ref{eq:MixResCharNoise}) in
eq.~(\ref{eq:CharOutFin}) with a coherent state input (eq.~(\ref{eq:CharCohIn})) and
$g_{x}=g_{p}=1$; we obtain the output for \textbf{CV} teleportation with a noisy
resource. From the output state, it is straightforward to calculate the analytic
expression for teleportation fidelity as a function of the adequate superposition
parameters: $\delta$ for the squeezed Bell-like state and $|\,\gamma\,|$ for the
squeezed cat-like states. The squeezing parameter $r$ is fixed and $\phi=\pi$ in
keeping with convention. Likewise are fixed the thermal parameters $n_{th,A}$ and
$n_{th,B}$. The analytic expressions of the fidelities for the noisy non-Gaussian
resources in the teleportation of coherent states; squeezed Bell-like
($\mathcal{F}_{SB}^{\,(th)\,}(\,r\,,\,n_{th,A}\,,\,n_{th,B}\,,\,\delta\,)$) and
squeezed cat-like
($\mathcal{F}_{SC}^{(th)}(\,r\,,\,n_{th,A}\,,\,n_{th,B}\,,\,|\,\gamma\,|\,)$) and an
initial analysis of the optimization procedure performed on the aforementioned
fidelities are reported in appendix~\ref{APPA}.

For example, for the mixed squeezed Bell-like state teleporting a coherent state, the
fidelity is optimized over $\delta$, leaving other parameters fixed;
\begin{equation}
\mathcal{F}_{opt}(\,r\,,\,n_{th,A}\,,\,n_{th,A}\,) \,=\, \max_{\delta} \;
\mathcal{F}_{SB}^{\,(th)\,}(\,r\,,\,n_{th,A}\,,\,n_{th,B}\,,\,\delta) \; ,
\label{eq:FidthSBOptim}
\end{equation}

The optimization yields an optimal angle $\delta_{max}^{(c,th)}$ that has a form not
too dissimilar to that for ideal teleportation (eq.~(\ref{eq:deltaoptC}));
\begin{equation}
\delta_{max}^{(\,c\,,\,th\,)} \,=\,
\frac{1}{2}\;\arctan\left(\,1\,+\,\frac{e^{\,-\,2\,r}}{1\,+\,n_{th,A}\,+\,n_{th,B}}\right)
\label{eq:deltaoptCNoisyRes}
\end{equation}
which reduces to the pure state case if the thermal parameters $n_{th,A}=n_{th,B}=0$.
When this angle is substituted in the expression for the fidelity
(eq.~(\ref{eq:FidMixedSBell})) gives the optimal fidelity for a mixed Bell-like
resource.

In the case of mixed squeezed cat-like states, at given thermal numbers $n_{th,A}$,
$n_{th,B}$ and fixed squeezing parameter $r$, the optimal fidelity is defined as
\begin{equation}
\mathcal{F}_{opt}(\,r\,,\,n_{th,A}\,,\,n_{th,B}\,) \,=\, \max_{|\,\gamma\,|} \;
\mathcal{F}_{SC}^{\,(th)}(\,r\,,\,n_{th,A}\,,\,n_{th,B},|\,\gamma\,|) \;
\label{eq:FidthSCOptim}
\end{equation}
where the maximization is performed numerically for fixed values of $r$ and
$n_{th}=n_{th,A}=n_{th,B}$.

\begin{figure}[htp] \label{fig:FidTelepMixedNG}
\begin{centering}
\includegraphics[width=14cm]{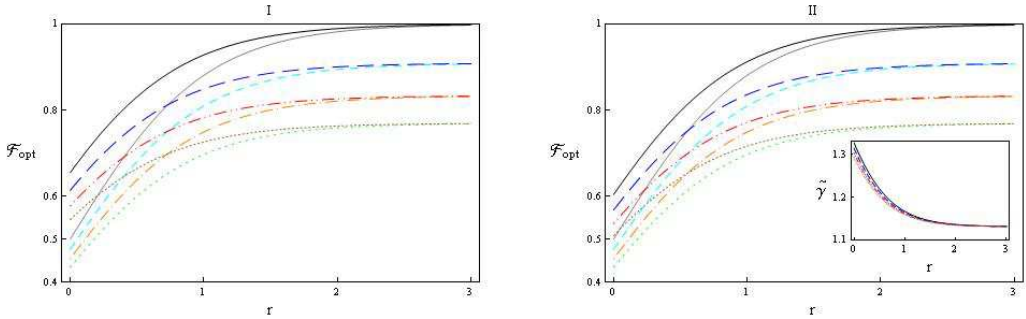}
\end{centering}
\caption[Optimal fidelity for teleportation of input coherent states using mixed
squeezed Bell-like resources and mixed squeezed cat-like resources]{Optimal fidelity
$\mathcal{F}_{opt}$, associated with the teleportation of input coherent states
 with mixed squeezed Bell-like resources (panel I) and mixed squeezed
cat-like resources (panel II), as a function of the squeezing $r$, for several
choices of the thermal parameters $n_{th,A}=n_{th,B}=n_{th}$. The curves representing
the fidelities associated with mixed non-Gaussian entangled resources are plotted
with the following plot style: $n_{th}=0$ (full black line), $n_{th}=0.05$
(long-dashed line), $n_{th}=0.10$ (double-dot dashed line), and $n_{th}=0.15$ (dotted
line). For comparison, in both panels, we also plot the fidelities associated to the
mixed squeezed vacuum entangled resources, with $n_{th}=0$ (full gray line),
$n_{th}=0.05$ (dashed line), $n_{th}=0.10$ (dot-dashed line), and $n_{th}=0.15$
(long-dotted line). The inset in panel II gives the maximal value of the parameter
$|\,\gamma\,|=\tilde{\gamma}$ as a function of $r$, with
$n_{th}=0,\,0.05,\,0.1,\,0.15$ (same plot style as for the fidelities).}
\end{figure}

In fig.~(\ref{fig:FidTelepMixedNG}); we plot the optimal fidelities for the mixed
non-Gaussian resources as a function of squeezing for various choices of the thermal
parameters $n_{th,A}=n_{th,B}=n_{th}$. Along, we plot the fidelities for a Gaussian
two-mode squeezed vacuum, used as a benchmark. We observe that, as expected, the
fidelity decreases for increasing $n_{th}$: the thermal noise sensibly reduces the
success probability of teleportation, as it reduces the entanglement content of the
resources. We can also observe that for similar noise $n_{th}$ the asymptote for
large $r$ is the same for all resources: the asymptote indicates the convergence of
all resources to a mixed two-mode squeezed vacuum at high squeezing, with the
"mixedness" putting an upper bound on teleportation fidelity even for this nearly
ideal resource.

An important observation is that in our plot, for the chosen realistic values for
$n_{th}$; the fidelity associated with both non-Gaussian mixed resources never drops
below the threshold of classical teleportation~\cite{CriteriaCVTelep} with maximal
fidelity $\mathcal{F}_{cls}^{max} =1/2$. The ability of the resource to keep the
fidelity above this benchmark value of $1/2$ will be made a practical measure of
resilience in presence of noise; in the next section.

At fixed squeezing and thermal parameters, the non-Gaussian resources always perform
better than the Gaussian two-mode squeezed vacuum. Furthermore, mixed squeezed
Bell-like states have a higher performance than the mixed squeezed cat-like states.
We remark that the better performance of the mixed Bell-like states over the mixed
cat-like states has been demonstrated for this \emph{simple} analysis of
environmental noise. Other sources of noise (and decoherence) have not been studied.

We have chosen the mixed Gaussian state to be a reference resource (see
subsection~\ref{NGau:SqueBell:OptBell}) for the evaluation of the performance of the
squeezed Bell-like state in teleportation. To develop this relationship between
resources further; we define the relative fidelity (see eq.~(\ref{eq:Deltafidelity}))
given for mixed squeezed Bell-like state and referring to the mixed Gaussian state
\begin{equation}\label{eq:Deltafidnoisyres}
\Delta\mathcal{F}^{(c)}_{opt} (r,n_{th}) \,=\,
\frac{\mathcal{F}^{(c)}(r,n_{th},n_{th},\delta_{max}^{(c,th)}) -
\mathcal{F}^{(c)}(r,n_{th},n_{th},0)}{\mathcal{F}^{(c)}(r,n_{th},n_{th},0)}
\end{equation}
where $\mathcal{F}^{(c)}(r,n_{th},n_{th},0)$ is the fidelity for the mixed Gaussian
resource and $\mathcal{F}^{(c)}(r,n_{th},n_{th},\delta_{max}^{(c,th)})$ is the
fidelity for the optimized noisy Bell-like resource (eq.~(\ref{eq:FidMixedSBell})).

\begin{figure}[ht] \label{fig:NoisyDeltaFidTelep}
\begin{centering}
\includegraphics[width=12cm]{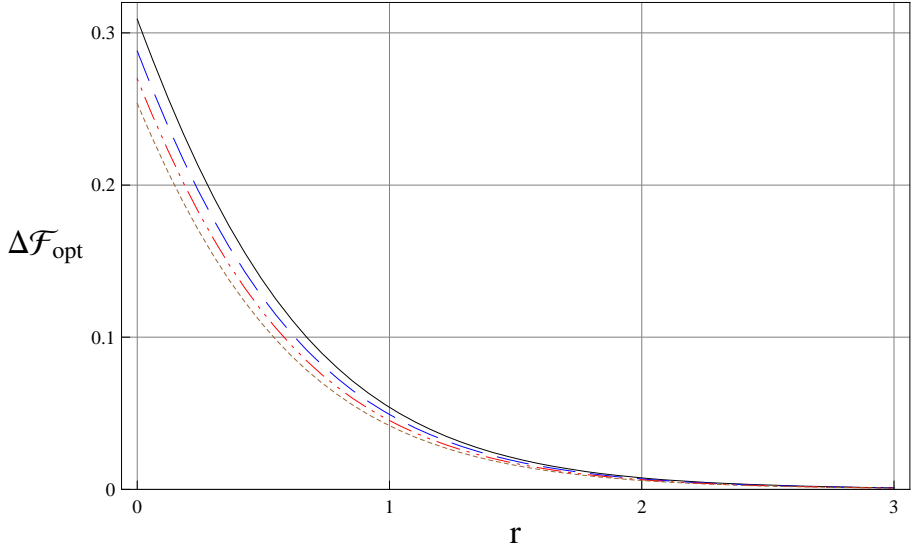}
\end{centering}
\caption[Relative fidelity for mixed Bell-like resources with respect to mixed
Gaussian resources for coherent state inputs]{Relative fidelity for mixed Bell-like
resources with respect to mixed Gaussian resources for coherent state inputs;
$\Delta\mathcal{F}^{(c)}_{opt} (r,n_{th})$ as a function of the squeezing $r$, for
several choices of the thermal parameter $n_{th,A}=n_{th,B}=n_{th}$: $n_{th}=0$ (full
black line), $n_{th}=0.05$ (long-dashed line), $n_{th}=0.10$ (double-dot dashed
line), $n_{th}=0.15$ (dotted line).}
\end{figure}

In fig.~(\ref{fig:NoisyDeltaFidTelep}) we show the behavior of
$\Delta\mathcal{F}^{(c)}_{opt} (r,n_{th})$ as a function of $r$. We see that the
percent gain in the fidelity for $n_{th}=0$ essentially coincides with the one
defined in the absence of thermal noise (fig.~(\ref{fig:DeltaFid})). It is also
evident that for ever increasing $n_{th}$, both the resources will converge
asymptotically to $\Delta\mathcal{F}^{(c)}_{opt} (r,n_{th})=0$. This signals a
similar and \emph{very poor} performance in teleportation for both resources;
corresponding to a separable, mixed, thermal state of two-modes.

We have shown that the mixed non-Gaussian resources; although with reduced fidelity
with respect to the ideal resources, are better choices than the mixed Gaussian
two-mode squeezed resource. Both the non-Gaussian resources studied here also provide
acceptable (always better than the Gaussian) values of the fidelity for technically
feasible squeezing $r$ and realistic values of thermal mean photon number $n_{th}$.

\section{Inseparability criteria and classical teleportation:
Bell-like, cat-like and Gaussian resources} \label{Noisy:Insepar}

In ref.~\cite{TelepNonGauss} it has been shown that the entanglement, that is a
fundamental requirement for the efficient implementation of nonclassical
teleportation protocols, is remarkably enhanced in the Bell-like state when compared
to the Gaussian squeezed vacuum. Even though the pure squeezed Bell-like state has
entanglement for $r=0$; the introduction of noise (and of a mixed nature) induces a
lowering of the amount of entanglement in the Bell-like state. It is then necessary
to check that, for fixed values of the thermal parameter $n_{th}$, a sufficient
amount of entanglement still survives \emph{and} is useful for teleportation, above a
classical threshold value.

In order to avoid the not-yet-solved problem of computing the amount of entanglement
by means of a measure appropriate for all mixed states; we can exploit an
inseparability criterion, based on the condition of positivity under partial
transposition (\textbf{PPT}
criterion)~\cite{InsepDGCZ,InsepSimon,InsepHZ,InsepShchukin,InsepDellAnno}. For the
purpose of knowing where~(for which values of $r$ and $n_{th}$) a state becomes
inseparable and \emph{possibly} useful for quantum teleportation, we assume this
criterion to be just as useful as a proper entanglement measure.

In our case, the PPT criterion can be expressed through an inequality for a
\textit{inseparability parameter} $\Delta$ involving only second order statistical
moments:
\begin{equation}
\Delta \,=\, \langle\: \hat{a}_{A}^{\dag}\,\hat{a}_{A}\:\rangle \: \langle
\:\hat{a}_{B}^{\dag}\,\hat{a}_{B}\:\rangle \;-\; |\,\langle\: \hat{a}_{A}\,
\hat{a}_{B}\: \rangle\,|^{2} \, <\,0 \label{eq:InsepSU11}
\end{equation}
which are straightforward to calculate, as the characteristic function of a quantum
state is the generating function for the statistical moments of the state (see
eq.(\ref{eq:CharGenMom})).

\begin{figure}[ht] \label{fig:NoisyInsepTest}
\begin{centering}
\includegraphics[width=12cm]{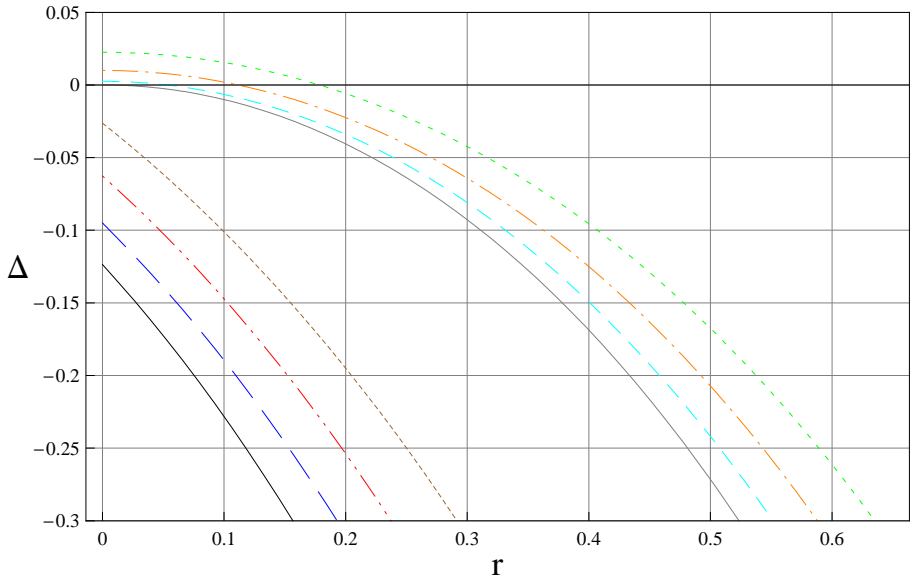}
\end{centering}
\caption[The inseparability parameter $\Delta$ for mixed squeezed Bell-like
states]{Behavior of the inseparability parameter $\Delta$ as a function of the
squeezing $r$, for several choices of the thermal parameters
$n_{th,A}=n_{th,B}=n_{th}$. Plots are for the mixed squeezed Bell-like state, with
$n_{th}=0$ (full black line), $n_{th}=0.05$ (long-dashed line), $n_{th}=0.10$
(double-dot dashed line), $n_{th}=0.15$ (dotted line); and for the noisy squeezed
vacuum state, with $n_{th}=0$ (full gray line), $n_{th}=0.05$ (dashed line),
$n_{th}=0.10$ (dot-dashed line), $n_{th}=0.15$ (long-dotted line).}
\end{figure}

Let us recall that the inequality in eq.~(\ref{eq:InsepSU11}) is a sufficient
inseparability condition  for non-Gaussian states, being necessary and sufficient for
Gaussian states. In fig.~(\ref{fig:NoisyInsepTest}); we plot the behavior of $\Delta$
as a function of $r$ for several choices of the thermal parameters
$n_{th,A}=n_{th,B}=n_{th}$, both for the mixed squeezed Bell-like state and for the
mixed squeezed vacuum state.

Recall again that the pure Bell-like state is already entangled for $r=0$ and
$\delta\neq\, 0,\pi/2$ (see fig.~(\ref{fig:vonNeumSBell})). For realistic values of
$n_{th}$ ($0$ to $0.15$) we see the mixed Bell-like state exhibiting entanglement. On
the other hand, for $n_{th}>\,0$, the mixed squeezed vacuum state has $\Delta \geq 0$
at sufficiently low values of $r$; i.e. it becomes separable. Specifically, for the
mixed Gaussian state the threshold value $n_{th}^{(sep)}$ for separability is
$n_{th}^{(sep)}(r)=\frac{1\,-\,e^{\,-\,2\,r}}{2}$.

Though we have defined no measure for entanglement, and in particular, no measure of
entanglement useful for \textbf{CV} teleportation; we have in the two-mode squeezed
vacuum a \emph{practical reference resource},with a "maximum classical fidelity"
$\mathcal{F}_{cls}^{max} = 0.5$ that is widely accepted as a practical
threshold~\cite{CriteriaCVTelep,CriteriaCVTelep2} for \textbf{CV} teleportation of
coherent states. We can assume for practical purposes that quantum teleportation
becomes impractical when thermal noise causes fidelity to go under this value.

Therefore, for a resource that teleports coherent states, we define the classical
threshold value of the thermal parameter $n_{th}^{(cls)}(r)$: At a fixed $r$ and for
$n_{th}=n_{th}^{(cls)}(r)$, the optimal fidelity $\mathcal{F}_{opt} =
\mathcal{F}_{cls}^{max}=1/2$. The optimal fidelity goes above the $1/2$ threshold in
a smooth manner for $n_{th}<n_{th}^{(cls)}(r)$, and goes smoothly below the threshold
$1/2$ for $n_{th}>n_{th}^{(cls)}(r)$. For this it is assumed (and found for the
resources used here) that the optimal fidelity is an analytical function of $r$ and
$n_{th}$.

\begin{figure}[ht] \label{fig:NoisynthClassicalFid}
\begin{centering}
\includegraphics[width=12cm]{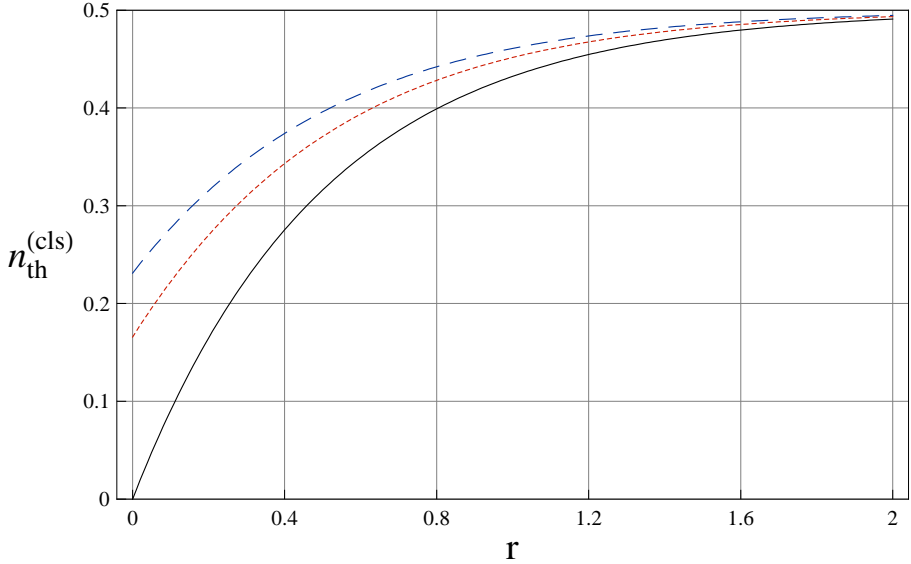}
\end{centering}
\caption[The threshold thermal parameter $n_{th}^{(cls)}$ of the mixed Bell-like and
the mixed cat-like resource states]{Behavior of the classical threshold value
$n_{th}^{(cls)}$ of the thermal parameter as a function of $r$, for mixed squeezed
Bell-like resources (dashed line), mixed squeezed cat-like resources (dotted line),
and mixed squeezed vacuum, shown as a reference (full line).}
\end{figure}

In fig.~\ref{fig:NoisynthClassicalFid} we plot $n_{th}^{(cls)}$ as a function of $r$
for the optimized, mixed squeezed Bell-like resource, the optimized mixed squeezed
cat-like resource and the mixed squeezed Gaussian resource, shown as a reference.

We see that the threshold value $n_{th}^{(cls)}$ associated to non-Gaussian resources
is larger than that associated to Gaussian ones, and in any case sensibly larger than
the reasonable realistic values ($n_{th}\leq\,0.15$) we have considered.  Even for
$r=0$ the non-Gaussian resources, always under realistic conditions, show
entanglement allowing teleportation fidelities greater than the classical threshold.
On the contrary, the fidelity associated with the mixed squeezed vacuum state, at low
values of $r$ falls below the classical threshold, unless optimization is performed
on the local degrees of squeezing, without modifying the Braunstein-Kimble
protocol~\cite{Telepoppy}.

It is remarkable that the squeezed Bell-like state has a consistently higher value of
$n_{th}^{(cls)}$ than the squeezed cat-like state. Taking this result at face value,
it can be said that the Bell-like state is more resilient in noisy environments than
the squeezed cat-like state, at equal $r$. Note lastly that for mixed squeezed
vacuums, $n_{th}^{(cls)}(r)$ coincides with the threshold value for separability,
that is $n_{th}^{(cls)}(r)=n_{th}^{(sep)}(r)=\frac{1\,-\,e^{\,-\,2\,r}}{2}$.

Compare fig.~(\ref{fig:NoisyDeltaFidTelep}) with
fig.~(\ref{fig:NoisynthClassicalFid}) for the squeezed Bell-like resource. In the
former, increasing $n_{th}$ makes for a marked decrease in
$\Delta\mathcal{F}^{(c)}_{opt} (r,n_{th})$ for all $r$. In the latter, the asymptotic
limit of $n_{th}^{(cls)}(r)$, for both Gaussian and non-Gaussian resources is the
same. Therefore, we have a value of $n_{th}\approx\,0.5$ beyond which no
\emph{reasonable} amount of two-mode squeezing $r$ will avail to produce a
useful~\footnote{More useful than a separable pure Gaussian state in any case.}
resource for quantum teleportation out of either a Gaussian resource or a Bell-like
resource of any kind. Fortunately, this value of $n_{th}$ is considered \emph{huge}
for a realistic experimental setting.

\chapter{Conclusions}
\addcontentsline{toc}{chapter}{Conclusions} \label{CONCLU}

We have developed a Wigner's characteristic function based formalism that allows for
the intuitive representation of the formalism for \textbf{CV} quantum teleportation
for any combination of resource and input states for which these characteristic
functions exist. The formalism is not confined to the original protocol; furthermore,
it is possible to introduce further and more complicated steps to the protocol;
measurements different from homodyne detection of \textbf{EPR} states and the mixing
in of external modes representing noise, photon addition and photon subtraction. We
have further shown that an analysis of the expression for the teleportation fidelity
makes for a trivial derivation of results for resource states that are simple
mixtures of pure states. The characteristic function formalism, as it stands, can
easily accommodate most of the conceivable operations and measurements involved in
\textbf{CV} teleportation, together with simple modifications to the protocol. It is
then, possible, to use characteristic functions to "model" other quantum information
protocols in a convenient manner.

We have presented a thorough comparison, with regard to the performance in
continuous-variable quantum teleportation, between standard Gaussian, wholly
non-Gaussian (two-mode squeezed Fock state) and degaussified resources (such as
photon-added and photon-subtracted squeezed states) and a new type of
\emph{sculptured} resource that interpolates between these states and can be
optimized because it depends on an extra, relative-phase, independent free parameter
in addition to squeezing. These sculptured non-Gaussian resources we have named
\emph{squeezed Bell-like states}: They hybridize discrete single-photon pumping,
coherent superposition of Bell two-qubit eigenstates, and \textbf{CV} squeezing; thus
including the above mentioned resources as "special cases". The maximization of the
teleportation fidelity (an analytical expression which is valid for a given input
state) is made with respect to the free parameter. Therefore we have produced a
fidelity that is optimal for the class of Bell-like states including, most
importantly, all the "special case" resources. Understanding the enhancement yielded
by sculptured squeezed Bell-like resources in teleportation success is possible when
certain properties of the resources used are studied and compared jointly. The
optimized squeezed Bell-like states are those states that are as close as possible to
the simultaneous maximization of entanglement, non-Gaussianity, and affinity to the
two-mode squeezed vacuum.

We have proposed a method for the experimental generation of squeezed Bell-like
states using a cascading setup of second and third-order non-linear crystals. Given
the nature of the squeezed Bell-like states, this method is to be additionally
regarded as an alternative method for the generation of degaussified resources.

We have further generalized the Bell-like states by including an additional
four-photon term in the superposition, producing squeezed symmetric superpositions of
Fock (\textbf{SSSF}) states. These resources, with an added dimension for
sculpturing, can be utilized to an optimal fidelity of teleportation that is a
notable improvement even over the optimal Bell-like state fidelity. We have
considered the Bell-like states as first-order truncations of a $r+s$ squeezing
Gaussian state where $r$ is genuine two-mode squeezing and $s$ depends on the
character of the Bell-like state; we found, remarkably, that the optimal
\textbf{SSSF} resource is to be formulated as the second-order truncation of the same
$r+s$ squeezing Gaussian states. We have argued that a Gaussian "truncation" should
be compared to the $r+s$ squeezing Gaussian instead of the $r$ squeezing Gaussian.
The analysis of the properties for these states, the Bell-like and the \textbf{SSSF}
shows that the optimal teleportation resource comes close to simultaneously
maximizing entanglement and affinity to the two-mode squeezed vacuum. Not so for the
non-Gaussianity.

We have introduced, and optimized for teleportation a class of two-mode squeezed
cat-like states; superpositions of coherent states that are then two-mode squeezed.
These states have been optimized for teleportation over their coherent amplitudes,
obtaining performances and entanglement values that are higher than those for
Gaussian resources; but lower than those for Bell-like states. These entangled states
are interesting because of their, in principle, greater
feasibility~\cite{GrangierCats} in comparison with Bell-like and \textbf{SSSF}
states, which involve superpositions of few-photon Fock states.

Given the above results, we state that the optimization of the ideal \textbf{CV}
teleportation protocol with non-Gaussian resources necessitates only the formulation
of more general and more complicated non-Gaussian resources. The only difficulty to
be incurred in the generalization of non-Gaussian resources lies is in the analytic
calculation of the  fidelities for the purpose of optimization of the same.

Further optimization is in principle possible with respect to the local parts of the
resource states, in analogy to the case of standard Gaussian
resources~\cite{Telepoppy}. One could think of extending the sculpturing to the
entire basis of Bell states, to generate entangled non-Gaussian resources that can
never be reduced to proper truncations of Gaussian squeezed resources. Such
\emph{fully sculptured} resources might allow for the further enhancement of the
teleportation success due to the presence of a larger number of experimentally
adjustable free parameters in addition to squeezing. Fully sculptured states could be
applied to hybrid schemes of teleportation combining continuous-variable inputs with
discrete-variable resources and viceversa. In this framework, a particularly
appealing line of research would be to look for modified schemes of teleportation
beyond the standard Braunstein-Kimble protocol, to be realized by generalized
measurements in combination with state-control enhancing unitary operations.

We have studied the efficiency of \textbf{CV} teleportation of input coherent states
using, as resources, squeezed Bell-like and squeezed cat-like states, prepared,
\emph{superimposed} over initial thermal states that represent thermal noise. Thus we
have defined general non-Gaussian, realistically mixed entangled states. We have
shown that, although the thermal noise strongly affects the success probability of
teleportation, the resource provided by the optimized mixed extensions of the
non-Gaussian squeezed Bell-like and squeezed cat-like states guarantee a sufficiently
high fidelity, for realistic values of the average thermal photon number $n_{th}$ and
of the squeezing $r$. A better performance is assured always, with respect to the
mixed extension of the Gaussian twin beam.

We have calculated some simple expressions for the characteristic function of the
teleportation output for different models of nonideal Bell measurement. The analysis
of "noisy" teleportation that we have performed can be generalized using the above
mentioned results, while taking into account the decoherence induced by the
propagation of the resource in noisy channels.

Finally, the present discussion could be extended to other types of quantum
information tasks and processes besides teleportation. For instance, it would be
interesting to investigate the comparative effects of non-Gaussian inputs and
non-Gaussian resources in schemes for the generation of macroscopic and mesoscopic
optomechanical entanglement~\cite{Aspelmeyer}. A further goal will be the study of
optimal teleportation with non-Gaussian resources of two-mode and multimode
states~\cite{twomodeinputTelep}.

\clearpage{\pagestyle{empty}\cleardoublepage}

\appendix
\noappendicestocpagenum
\chapter{Analytic expressions of Teleportation fidelities}
\label{APPA} \markboth{\appendixname\ \thechapter\ \quad }{}

In this appendix we report the explicit, analytic expressions for the teleportation
fidelities calculated in chapters~\ref{NGau} and~\ref{Noisy}. We have deemed these
expressions both too cumbersome and unnecessary for the purposes of our exposition
regarding the \textbf{CV} teleportation protocol with Non-Gaussian resources; relying
instead on the plots accompanying the exposition for the graphical representation of
said fidelities.

\section{The fidelities: noisy squeezed Bell-like states with coherent state inputs}
\label{APPA:NoisySqBellCoh}

The fidelity $\mathcal{F}^{(c)}(r\,,\,n_{th,A}\,,\,n_{th,B}\,,\,\delta)$
(superposition phase $\theta=0$; squeezing angle $\phi=\pi$) for the \textbf{CV}
teleportation of a coherent state input of arbitrary amplitude, using as a resource
the mixed squeezed Bell-like state (see eq.~(\ref{eq:MixResCharNoise} and
eq.~(\ref{eq:CharSqueBell}) reads
\begin{equation}
\mathcal{F}^{(c)}(\,r\,,\,n_{th,A}\,,\,n_{th,B}\,,\,\delta) \,=\,
\frac{1+e^{\,2\,r}f_{th}\:+\:e^{\,4\,r}f_{th}^{2}\:+\:e^{\,2\,r}f_{th}\,\cos(2\delta)\:+\:[1\,+\,e^{\,2\,r}f_{th}]\sin
(2\,\delta)}{e^{\,-2\,r}[1\,+\,e^{\,2\,r}f_{th}]^{3}} \label{eq:FidMixedSBell}
\end{equation}
where
\begin{equation}\label{eq:fth}
  f_{th}\equiv\,1\,+\,n_{th,A}\,+\,n_{th,B}
\end{equation}

The fidelities for the states that are special cases of the Squeezed Bell-like state
(see section~\ref{NGau:Chzation}) can be obtained by the substitution the appropriate
values of the $\delta$ parameter in eq.~(\ref{eq:FidMixedSBell}) and the appropriate
levels of noise $n_{th,A}$ and $n_{th,B}$. For the case in which $\delta=0$ and
$n_{th,A}=n_{th,B}$=0, the fidelity equals the well-known result
\begin{equation}
\mathcal{F}_{TwB}(r)\,=\, \frac{1}{1\,+\,e^{\,-\,2\,r}} \label{eq:TwBFid}
\end{equation}
holding for pure two-mode squeezed vacuum resources and coherent state
inputs~\cite{CriteriaCVTelep}.

In the absence of thermal fields, for $n_{th,A}=n_{th,B}=0$; we have $f_{th}=1$, and
the fidelity reducing to the pure resource fidelities studied in chapter~\ref{NGau}.
Notice that, in the limit of large thermal parameter, the fidelity of teleportation
in eq.~(\ref{eq:FidMixedSBell}) is strongly suppressed and vanishes asymptotically.

\section{The fidelities: noisy squeezed cat-like states with coherent state inputs}
\label{APPA:NoisySqCatCoh}

We report the analytical expressions for the fidelities of teleportation of input
coherent states using pure and mixed squeezed cat-like states as resources. By
putting, as usual, the phases $\theta=0$ and $\phi=\pi$, the fidelity for the
squeezed cat-like state (eq.~(\ref{eq:squeezCat})) reads
\begin{equation}
\mathcal{F}_{SC}'(r\,,\,\delta\,,\,\gamma) \,=\, \frac{\cos^{2}(\delta) \,+\,
e^{\,\frac{(\gamma\,-\,\gamma^{*})^{2}}{1\,+\,e^{\,2\,r}}}\,\sin^{2}(\delta)\:+\:
e^{\,-|\,\gamma\,|^{2}}\:(e^{\,\frac{\gamma^{2}}{1\,+\,e^{\,2\,r}}}+e^{\frac{\gamma^{*2}}{1\,+\,e^{\,2\,r}}})
\sin(\delta)\:\cos(\delta)}{(1\,+\,e^{\,-2\,r})(1\,+\,e^{-|\,\gamma\,|^{2}}\sin
(2\,\delta))} \label{eq:FidelitySqueezKat}
\end{equation}

It is worth noticing that for $\delta\rightarrow 0$ and/or $\gamma\rightarrow 0$, the
fidelity in eq.~(\ref{eq:FidelitySqueezKat}) reduces to the well known expression of
eq.~(\ref{eq:TwBFid}). A preliminary numerical optimization procedure for
eq.~(\ref{eq:FidelitySqueezKat}) allows us to fix the parameters $\arg\gamma$ and
$\delta$ to the values $\arg\gamma = 0$ and $\delta=\frac{\pi}{4}$, leading to the
simplification of the above fidelity to the expression
\begin{equation}
\mathcal{F}_{SC}\left(r\,,\,|\,\gamma\,| \right) \,=\, \frac{1\,+\,
e^{\,\frac{-|\,\gamma\,|^{2}}{1\,+\,e^{\,-2\,r}}}}{(1\,+\,e^{\,-2\,r})(1\,+\,e^{-|\,\gamma\,|^{2}})}
\label{eq:FidelitySqueezKat2}
\end{equation}

Thus, at fixed squeezing $r$ the maximization can be carried out with the real
amplitude $|\,\gamma\,|$ being the only free parameter. We have
$\mathcal{F}_{opt}\,=\,\max_{|\,\gamma\,|}\mathcal{F}_{SC}\,\left(r\,,\,|\,\gamma\,|\right)$.
In the limit of zero squeezing ($r=0$), the value of the optimal fidelity becomes
$\mathcal{F}_{opt}\,=\,[4\:(\sqrt{2}\,-\,1)]^{-1}\simeq\,0.6035$, for
$|\,\gamma\,|=\ln^{1/2}(\sqrt{2}\,-\,1)^{-2}\simeq\, 1.3276$.

Lastly; consider the "noisy", mixed squeezed cat-like resource with fixed angle
$\delta=\frac{\pi}{4}$ and phases $\phi=\pi$, $\theta=0$, $\arg\gamma =0$. The
fidelity $\mathcal{F}_{SC}^{(th)}(r\,,\,n_{th,1}\,,\,n_{th,2},|\,\gamma\,|)$ for the
mixed squeezed cat-like state is given by
\begin{equation}
\mathcal{F}_{SC}^{(th)}(r\,,\,n_{th,1}\,,\,n_{th,2},|\,\gamma\,|) \,=\, \frac{1 \,+\,
e^{\,-|\,\gamma\,|^{2}}e^{\,\frac{|\,\gamma\,|^{2}}{(1 \,+\, e^{\,2\,r}
\,f_{th})}}}{e^{\,-\,2\,r}(1 \,+\, e^{\,2\,r}\,
f_{th})(1\,+\,e^{\,-\,|\,\gamma\,|^{2}})} \label{eq:FidelitySCMixed}
\end{equation}
where $f_{th}$ is given by eq.~(\ref{eq:fth}).

\backmatter

\bibliographystyle{abbrv}

\end{document}